\documentstyle[aps,eqsecnum,floats]{revtex}
\begin{document}
\draft
\title{Scalar, electromagnetic, and gravitational self-forces 
       in weakly curved spacetimes}    
\author{Michael J.~Pfenning and Eric Poisson}
\address{Department of Physics, University of Guelph, Guelph,
         Ontario, Canada N1G 2W1}
\date{Revised draft, September 12, 2001} 
\maketitle
\begin{abstract}
We calculate the self-force experienced by a point scalar charge $q$,
a point electric charge $e$, and a point mass $m$ moving in a weakly
curved spacetime characterized by a time-independent Newtonian
potential $\Phi$. We assume that the matter distribution responsible
for this potential is bounded, so that $\Phi \sim -M/r$ at large
distances $r$ from the matter, whose total mass is $M$; otherwise, the
Newtonian potential is left unspecified. (We use units in which
$G=c=1$.) The self-forces are calculated by first computing the
retarded Green's functions for scalar, electromagnetic, and
(linearized) gravitational fields in the weakly curved spacetime, and
then evaluating an integral over the particle's past world line. The
self-force typically contains both a conservative and a
nonconservative (radiation-reaction) part. For the scalar charge, the
conservative part of the self-force is equal to $2\xi q^2 M
\bbox{\hat{r}}/r^3$, where $\xi$ is a dimensionless constant measuring
the coupling of the scalar field to the spacetime curvature, and
$\bbox{\hat{r}}$ is a unit vector pointing in the radial
direction. For the electric charge, the conservative part of the
self-force is $e^2 M \bbox{\hat{r}}/r^3$. For the massive particle,
the conservative force vanishes. For the scalar charge, the
radiation-reaction force is $\frac{1}{3} q^2 d\bbox{g}/dt$, where
$\bbox{g} = -\bbox{\nabla} \Phi$ is the Newtonian gravitational
field. For the electric charge, the radiation-reaction force is
$\frac{2}{3} e^2 d\bbox{g}/dt$. For the massive particle, the
radiation-reaction force is $-\frac{11}{3} m^2 d\bbox{g}/dt$. Our
result for the gravitational self-force is disturbing: a
radiation-reaction force should not appear in the equations of motion
at this level of approximation, and it should certainly not give rise
to radiation antidamping. In the last section of the paper we prove
that while a massive particle in a vacuum spacetime is subjected only 
to its self-force, it is also subjected to a matter-mediated force
when it moves in a spacetime that contains matter; this force
originates from the changes in the matter distribution that are
induced by the presence of the particle. We show that the
matter-mediated force contains a radiation-damping term that precisely
cancels out the antidamping contribution from the gravitational
self-force. When both forces are combined, the equations of motion are 
conservative, and they agree with the appropriate limit of the
standard post-Newtonian equations of motion.  
\end{abstract}

\section{Introduction and summary}  

\subsection*{Motion of massive bodies in strong gravitational
       fields} 

The problem of determining the motion of $N$ bodies subjected to their
mutual gravitational interactions has been a center piece of general
relativity from its inception. Work on this started shortly after the
introduction of the theory, and in 1938, a firm formulation of the
equations of motion was given by Einstein, Infeld, and Hoffmann
\cite{1}, who provided post-Newtonian corrections to the Newtonian
equations of motion. Higher-order corrections were later added
\cite{2,3,4,5}, and work continues today, with Damour, Jaranowski \&
Sch\"afer \cite{6,6.2}, de Andrade, Blanchet \& Faye \cite{7,7.2},
and Pati \& Will \cite{8} currently computing corrections of third
post-Newtonian order. A technical review of this field of research, as
of 1987, can be found in Ref.~\cite{9}; the early history was
described by Havas \cite{10}.     

While the equations of motion for slowly moving bodies in a weak
gravitational field are now very well understood, the same cannot be
said of the fast motion of a massive body in a strong gravitational
field. The prototypical problem here is to determine the motion of a
structureless particle of mass $m$ in the gravitational field of a
much more massive black hole. While it is known that the motion
is geodesic in the limit $m \to 0$, the issue is to determine the
corrections to geodesic motion that appear when $m$ in
nonvanishing. An important effect that must be introduced is the loss
of orbital energy and angular momentum to the gravitational radiation
emitted by the moving particle; this is effected by an appropriate
radiation-reaction force. Another important effect is a conservative
correction to the equations of motion, which disappears when $m \to
0$. Schematically, therefore, the equations of motion will take the
form  
\begin{equation} 
m u^\alpha_{\ ;\beta} u^\beta = f^\alpha_{\rm self}, 
\label{1.1}
\end{equation}
where $u^\alpha$ is the particle's four-velocity in the background of   
the massive black hole, $u^\alpha_{\ ;\beta} u^\beta$ its acceleration
(the semicolon denotes covariant differentiation in the black-hole
metric), and $f^\alpha_{\rm self}$ is the self-force. This contains
both a conservative and a dissipative (radiation-reaction) component,
and it scales as $m^2$ in the small-mass limit. (Throughout the paper
we work in relativistic units, with $G$ and $c$ both set equal to
unity.)  

A useful way to look at Eq.~(\ref{1.1}) is to imagine that the motion
is actually geodesic in a spacetime that is not the background
spacetime of the massive black hole, but the perturbed spacetime that
contains the particle as well. The geodesic motion in the perturbed
spacetime can be expressed as a forced motion in the background
spacetime, and this gives rise to Eq.~(\ref{1.1}). The
implementation of this idea, however, presents some difficulty: If the
particle is pointlike, the perturbation diverges at the position of
the particle, and the geodesic equation is not defined on the world
line. The perturbation must then be decomposed into a part that is
singular but does not influence the motion of the particle, and a
smooth remainder that is entirely responsible for the
self-force \cite{11}. This decomposition requires great care. 

In simple situations, the orbital evolution of a particle emitting
gravitational waves can be determined without the involvement of a
self-force. If the black hole is nonrotating, the orbital
evolution is determined by energy and angular-momentum balance
\cite{12,13}: The rate at which the radiation carries energy and
angular momentum both to infinity and across the event horizon matches
the rate at which orbital energy and angular momentum is lost by the
particle. This information alone is sufficient to solve for the
motion, at least when the time scale for orbital evolution is long
compared with the orbital period. If the black hole is rotating, the
statement continues to be true provided that the orbit is either
equatorial or circular \cite{14,15,16,17,18,19}. When dealing with
generic orbits around a rotating black hole, however, the
loss of orbital energy and angular momentum no longer constitutes
sufficient information --- the rate of change of the ``Carter
constant'' \cite{20} must also be determined, and this requires the
involvement of a self-force \cite{18}.  

\subsection*{Sources of gravitational waves for LISA} 

There is a practical necessity for the computation of the
gravitational self-force. Solar-mass compact objects in highly
eccentric, nonequatorial orbits around rapidly rotating, massive black
holes (in the range between $10^3$ and $10^8$ solar masses) emit
gravitational waves that will be measured by eventual spaceborne
interferometric detectors \cite{21}, which operate in the
low-frequency band (in the range between $10^{-4}$ and 1 Hz). One such
detector, the Laser Interferometer Space Antenna (LISA) \cite{22}, has
been selected as one of three ``Cornerstone missions'' in the
``Horizon 2000+'' program of the European Space Agency. A possible
involvement by the National Aeronautics and Space Administration ---
now appearing likely after the publication of the Taylor-McKee decadal
survey \cite{23}, in which LISA is listed as a high-priority,
moderate-size mission --- would increase the likelihood
that this detector will be deployed in the not-too-distant future. A
realistic projection places the launch in the neighborhood of 2010.      

The detailed modeling of these sources of gravitational waves for
LISA, to the extent that templates could be provided for data
analysis, will require a detailed computation of the orbital
evolution. This, in turn, requires that Eq.~(\ref{1.1}) be evaluated
and solved for generic orbits around a Kerr black hole. For this we
need a practical way of computing the gravitational self-force.    

\subsection*{Gravitational self-force} 

This issue was taken up recently by Mino, Sasaki, and Tanaka
\cite{24}, who calculated the self-force acting on a point particle of
mass $m$ moving in an arbitrary background spacetime with metric
$g_{\alpha\beta}$; they assume that the metric satisfies the Einstein
field equations in vacuum. Their calculation is based on a careful
analysis of the perturbation field in the vicinity of the particle,
the (gravitational) perturbation being created by the particle
itself. It involves a careful decomposition of the perturbation into a
part that is singular at the particle's location but does not affect
its motion, and a part that is well-behaved and governs the 
motion. Their expression for the self-force was then reproduced by
Quinn and Wald \cite{25}, on the basis of a different approach
involving a comparison between self-forces acting in two different
spacetimes. Both teams found the following equations of motion:    
\begin{equation} 
m u^\alpha_{\ ;\beta} u^\beta = f^\alpha_{\rm ext} - \frac{11}{3} m   
\bigl(\delta^\alpha_{\ \beta} + u^\alpha u_\beta \bigr)
\dot{f}^\beta_{\rm ext} + f^{\alpha}_{\rm grav}.  
\label{1.2}
\end{equation}
Here, $u^\alpha$ is the particle's four-velocity in the background
spacetime, $f^\alpha_{\rm ext}$ is an external force acting on the
particle, and an overdot indicates differentiation with respect to
proper time $\tau$. The gravitational self-force is given by 
\begin{equation} 
f^{\alpha}_{\rm grav} = - 2m^2 \int_{-\infty}^{\tau^-} \Bigl( 
2 G^\alpha_{\ \beta\mu'\nu';\gamma} 
- G_{\beta\gamma\mu'\nu'}^{\ \ \ \ \ \ \ ;\alpha} 
+ u^\alpha G_{\beta\gamma\mu'\nu';\delta} u^\delta \Bigr) 
u^\beta u^\gamma u^{\mu'} u^{\nu'}\, d\tau'.  
\label{1.3}
\end{equation} 
The four-index object $G^{\alpha\beta}_{\ \ \gamma'\delta'}(x,x')$
appearing inside the integral is the retarded gravitational Green's
function \cite{26}, whose precise definition is given by
Eq.~(\ref{3.24}) below (our normalization differs from the
conventional choice by a factor of 4); the unprimed indices refer to
the field point $x$ (which is identified in the integral with the
current position of the particle), while the primed indices refer to
the source point $x'$ (identified with the particle's past
position). The integral extends over the past world line of the
particle, from $\tau' = -\infty$ to (almost) the current time, $\tau'
= \tau^- \equiv \tau - \epsilon$, where $\epsilon$ is very small and
positive \cite{25}. The integration is cut short to avoid the singular
behavior of the Green's function as $x'$ approaches $x$; it involves
only the smooth part of the Green's function, which is often referred
to as its ``tail part'' \cite{27}. In a situation in which the
particle is subjected only to its self-force, the equations of motion
reduce to $m u^\alpha_{\ ;\beta} u^\beta = f^{\alpha}_{\rm
grav}$. Because the self-force scales like $m^2$, we recover geodesic
motion in the limit $m \to 0$. 

The difficulty of evaluating Eq.~(\ref{1.3}) resides with the 
computation of the gravitational Green's function, which would be very
hard to carry out for an arbitrary spacetime. There is hope for
progress, however, if the spacetime possesses symmetries, such as
staticity and spherical symmetry in the case of a Schwarzschild black
hole, or stationarity and axial symmetry in the case of a Kerr black
hole. In such cases, a plausible method of computation would be based
on a separation-of-variables approach, and the {\it modes} of the
Green's function could be obtained fairly easily \cite{28}. But
this approach is not free of difficulties: While the individual modes
of the Green's function stay finite and continuous as $x$ approaches
$x'$ (though they are discontinuous in their first derivatives), the  
sum over modes does not converge. Essentially, this is because the
modes do not distinguish between the singular and smooth parts of the
Green's function; they contribute to both, and the singular behavior
of the Green's function gives rise to a divergent sum. Recently, Ori,
Burko, and Barack \cite{29,30,31,32,33,34,34.2} have devised
a way to regularize the mode sums, so as to extract from them a
meaningful expression for the self-force. They have applied their
technique to a number of simple situations involving scalar,
electromagnetic, and gravitational radiation. Similar
regularization methods were proposed by Lousto \cite{35}, as well as 
Nakano, Mino, and Sasaki \cite{35.2,35.4}. It appears likely that in 
the near future, this method will be used successfully to solve more
complicated problems, including the prototypical problem mentioned
previously.   

\subsection*{Electromagnetic self-force} 

The derivation of Eqs.~(\ref{1.2}) and (\ref{1.3}) by Mino 
{\it et al.} \cite{24}, and Quinn \& Wald \cite{25}, is based on
methods previously introduced by DeWitt and Brehme \cite{36}. These
authors calculated the self-force acting on an {\it electrically
charged} particle moving in an arbitrary spacetime with metric
$g_{\alpha\beta}$. As corrected by Hobbs \cite{37}, the equations of
motion of a charged particle are 
\begin{equation} 
m u^\alpha_{\ ;\beta} u^\beta = f^\alpha_{\rm ext} 
+ \frac{2}{3} \frac{e^2}{m}   
\bigl(\delta^\alpha_{\ \beta} + u^\alpha u_\beta \bigr)
\dot{f}^\beta_{\rm ext} + \frac{1}{3} e^2 \bigl( R^\alpha_{\ \beta} 
u^\beta + u^\alpha R_{\beta\gamma} u^\beta u^\gamma \bigr)
+ f^{\alpha}_{\rm em},   
\label{1.4}
\end{equation}
where $e$ is the particle's electric charge, $f^\alpha_{\rm ext}$ an
external force acting on the particle, $R_{\alpha\beta}$ the
spacetime's Ricci tensor, and  
\begin{equation} 
f^\alpha_{\rm em} = - e^2 \int_{-\infty}^{\tau^-} \bigl( 
G^\alpha_{\ \gamma';\beta} - G_{\beta\gamma'}^{\ \ \ ;\alpha} \bigr)  
u^\beta u^{\gamma'}\, d\tau'  
\label{1.5}
\end{equation}
is the electromagnetic self-force. The two-index object 
$G^\alpha_{\ \beta'}(x,x')$ is the retarded electromagnetic Green's
function \cite{36}, whose precise definition is given by
Eq.~(\ref{3.16}) below. In the absence of an external force, and in a
region of spacetime empty of matter, the equations of motion reduce to  
$m u^\alpha_{\ ;\beta} u^\beta = f^{\alpha}_{\rm em}$. In
flat spacetime, on the other hand, $f^{\alpha}_{\rm em} = 0$ because
the Green's function contains only a singular part; the smooth, or
tail, part vanishes. In flat spacetime, therefore, we recover the
Abrahams-Lorentz-Dirac equation \cite{38,39,40}, 
\begin{equation}
m u^\alpha_{\ ;\beta} u^\beta = f^\alpha_{\rm ext} 
+ \frac{2}{3} \frac{e^2}{m} \bigl(\delta^\alpha_{\ \beta} 
+ u^\alpha u_\beta \bigr) \dot{f}^\beta_{\rm ext}. 
\label{1.6}
\end{equation} 
The practical computation of the electromagnetic self-force presents
the same technical difficulties as in the gravitational case.    

\subsection*{Scalar self-force} 

The self-force acting on a particle with {\it scalar} charge $q$ was
recently calculated by Quinn \cite{41}. In this case, the equations of
motion are  
\begin{equation} 
m u^\alpha_{\ ;\beta} u^\beta = f^\alpha_{\rm ext} 
+ \frac{1}{3} \frac{q^2}{m}  
\bigl(\delta^\alpha_{\ \beta} + u^\alpha u_\beta \bigr)
\dot{f}^\beta_{\rm ext} + \frac{1}{6} q^2 \bigl( R^\alpha_{\ \beta}  
u^\beta + u^\alpha R_{\beta\gamma} u^\beta u^\gamma \bigr)
+ f^{\alpha}_{\rm scalar},   
\label{1.7}
\end{equation}
where the self-force is now given by 
\begin{equation}
f^\alpha_{\rm scalar} = q^2 \bigl(g^{\alpha\beta} + u^\alpha 
u^\beta \bigr) \int_{-\infty}^{\tau^-} G_{,\beta}\, d\tau', 
\label{1.8}
\end{equation}
in which the retarded Green's function $G(x,x')$ is a scalar; this
is defined by Eq.~(\ref{3.10}) below. In the absence of an external
force, the equations of motion reduce to $m u^\alpha_{\ ;\beta}
u^\beta = f^{\alpha}_{\rm scalar}$, and this is the simplest
realization of the class of equations that govern the orbital
evolution of a particle moving in a strong gravitational field. The
computation of the scalar, electromagnetic, or gravitational
self-force involves solving for a retarded Green's function, and then
performing an integration over the particle's past world line. Because 
there is only one component to the scalar Green's function (as opposed
to 16 for the electromagnetic function, or 100 for the gravitational
case), Eq.~(\ref{1.8}) captures the essence of the problem and avoids
many technical complications. For this reason, it has been the
starting point of many recent investigations
\cite{30,31,32,33,34,34.2,35.2,35.4,42}.       

\subsection*{Weak-field limit} 

The computation of the various Green's functions is a very difficult
undertaking even when the spacetime possesses many symmetries. This
is, however, a very tractable task when the spacetime is only weakly 
curved. In such a situation, the metric itself can be calculated
perturbatively as an expansion about flat spacetime, and the same
technique can be employed to find the Green's functions and compute
the self-forces. Our purpose with this paper is to do just that. We
will consider spacetimes for which the metric can be expressed as  
\begin{equation}
ds^2 = -(1+2\Phi)\, dt^2 + (1-2\Phi)(dx^2+dy^2+dz^2), 
\label{1.9}
\end{equation}
in which $\Phi(x,y,z) \ll 1$ is a generic Newtonian potential
satisfying Poisson's equation, $\nabla^2 \Phi = 4\pi \rho$, where
$\rho$ is the mass density. We will keep $\Phi$ unspecified throughout
the paper, but assume that it is small everywhere. We shall also
assume that the mass distribution is bounded, so that $\Phi$ behaves
as $-M/r$ at large distances $r$ from the center of mass; $M = \int
\rho\, d^3x$ is the total mass, and $r^2 = x^2+y^2+z^2$. Various
aspects of those spacetimes are discussed in Sec.~II. In Sec.~III and
IV we calculate the scalar, electromagnetic, and gravitational Green's
functions to first order in $\Phi$. In Sec.~V we use the Green's
functions to calculate the scalar, electromagnetic, and gravitational
self-forces.   

Our motivation for carrying out such a (long) computation is to
provide a useful check on the formalism, especially in its
gravitational formulation. Admittedly, the weak-field limit is not by
itself very interesting, but understanding this limit may well be a
necessary step toward understanding the strong-field behavior of
the gravitational self-force. So while we cannot hope to learn
anything new from such a weak-field computation, we can use our 
knowledge of the post-Newtonian equations of motion \cite{3,4,5,9} to
produce a nontrivial check of the self-force method; the results found
here should of course match the results from the literature. As we
shall see, we have been successful at producing this check: the method
works.  

\subsection*{DeWitt and DeWitt} 

We are not the first to perform such a check on the formalism. In
1964, DeWitt and DeWitt \cite{44} calculated the electromagnetic
self-force, as given by Eq.~(\ref{1.5}), for the weakly curved
spacetime of Eq.~(\ref{1.9}). They considered only the special case
$\Phi(\bbox{x}) = -M/r$, where $r = |\bbox{x}|$, but noted that their
results generalize to arbitrary potentials by superposition. (We use a 
bold-faced symbol to denote a three-dimensional vector living in flat
space.) They found that to leading order in a weak-field, slow-motion
approximation, the spatial components of the self-force are  
\begin{equation}
\bbox{f}_{\! \rm em}(\bbox{x}) = e^2 \frac{M}{r^3}\, \bbox{\hat{r}}  
+ \frac{2}{3}\, e^2 \frac{d \bbox{g}}{dt},   
\label{1.10}
\end{equation} 
where $\bbox{\hat{r}} = \bbox{x}/r$ and $\bbox{g}(\bbox{x}) =
-\bbox{\nabla} \Phi$ is the Newtonian gravitational field. In this
limit, the equations of motion $m u^\alpha_{\ ;\beta} u^\beta 
= f^{\alpha}_{\rm em}$ become 
\begin{equation} 
m \frac{d^2 \bbox{z}}{dt^2} = m \bbox{g}(\bbox{z}) 
+ \bbox{f}_{\! \rm em}(\bbox{z}),
\label{1.11} 
\end{equation}
where $\bbox{z}(t)$ is the trajectory of the charged particle. The
first term on the right-hand side of Eq.~(\ref{1.10}) is a repulsive 
correction to the local force of gravity; it agrees with the
weak-field limit of a result by Smith and Will \cite{45}, who
calculated the force required to hold a charged particle in place in
the (exact) field of a Schwarzschild black hole. The second term is
the usual expression for the radiation-reaction force experienced by a
charged particle subjected to an external force $m \bbox{g}$
\cite{38,39}.    

The result of Eq.~(\ref{1.10}) is remarkable because this entirely
local expression for the self-force derives from a nonlocal
formulation involving the entire past history of the charged
particle; in the weak-field limit, any trace of nonlocality is
lost. The result is remarkable also because while the expression for
the radiation-reaction component of the self-force is the expected
$\frac{2}{3} e^2 d \bbox{g}/dt$, it is very hard to see how such a
simple result could ever follow from such a complicated expression as
Eq.~(\ref{1.5}). We notice that this result for the radiation-reaction
force can be obtained from two very different approaches. In a
flat-spacetime point of view, the equations of motion for a charged
particle are $m u^\alpha_{\ ;\beta} u^\beta =
f^\alpha_{\rm ext} + \frac{2}{3}(e^2/m) (\delta^\alpha_{\ \beta} +
u^\alpha u_\beta ) \dot{f}^\beta_{\rm ext}$; in the slow-motion limit,
the external force is $m \bbox{g}$, and we recover the usual
result. In a curved-spacetime point of view, on the other hand, there
is no external force and the equations of motion are 
$m u^\alpha_{\ ;\beta} u^\beta = f^{\alpha}_{\rm em}$; 
in the weak-field, slow-motion limit, we also recover the usual
result. This agreement is necessary on physical grounds, but it could
hardly have been anticipated on the basis of a quick inspection of
Eq.~(\ref{1.5}); the calculations required to produce Eq.~(\ref{1.10}) 
are very involved.  

In their paper \cite{44}, DeWitt and DeWitt gave the following
physical picture for the self-force of Eq.~(\ref{1.10}). This picture
relies heavily on their expression for the Newtonian potential, $\Phi
= -M/r$, which corresponds to a point mass $M$ located at the origin
of the coordinate system. 

Because the fundamental expression for the self-force,
Eq.~(\ref{1.5}), involves only the smooth, or tail, part of the
electromagnetic Green's function, there is a priori no {\it local}
contribution to this force coming from the world-line
integral. Indeed, DeWitt and DeWitt found that the quantity inside 
this integral vanishes in the immediate past of the event
$(t,\bbox{x})$, which corresponds to the particle's current
position. In fact, it vanishes until the time delay, $t-t'$, becomes
equal to $r+r'$, the distance between the central mass $M$ and the
particle's current position $\bbox{x}$, added to the distance to the
particle's old position $\bbox{x'}$. At $t - t' = r + r'$, the
electromagnetic Green's function undergoes an abrupt change of
behavior, and the self-force is entirely due to this sudden
transition. (We expand our discussion of this point in Appendix B.)
The physical effect responsible for the force, in DeWitt and DeWitt's
view \cite{44}, is a signal that originates from the particle at an 
earlier time $t'$, propagates toward the central mass $M$ at the
speed of light, bounces off the central mass, and comes
back to the particle at the current time $t$. Although the self-force
is nonlocal, Eq.~(\ref{1.10}) involves the conditions at the current
time only. This is because the time delay in not noticeable at the
level of approximation maintained in the calculation. To leading order
in a weak-field, slow-motion approximation, the electromagnetic
self-force appears to be entirely local. 

The physical picture described in the preceding paragraph appeared to  
us to be slightly suspicious. The sudden change in the behavior of
the electromagnetic Green's function at $t - t' = r + r'$ is dictated
by the presence of the point mass $M$ at $\bbox{x} = 0$, which
mediates the interaction between the charged particle and its earlier
self. The region of spacetime near the central mass is therefore seen
to have an important effect on the Green's function, and it is this
effect --- the sudden change --- which apparently gives rise to
the self-force. But $\Phi = -M/r$ cannot be considered to be small in
this region of spacetime, and the perturbative method of calculation
of the Green's function must come into question. While there is little
reason to doubt the validity of Eq.~(\ref{1.10}), it appeared to us
that its derivation by DeWitt and DeWitt \cite{44} left room for
improvement.   

\subsection*{This paper} 

Part of this paper is concerned with providing a derivation of
Eq.~(\ref{1.10}) that is free of this criticism. Throughout the paper
we assume that $\Phi(\bbox{x})$ is everywhere much smaller than unity,
and reduces to $-M/r$ only far away from the mass 
distribution. Building upon DeWitt and DeWitt's work \cite{44}, we
introduce (in Secs.~III and IV) techniques that allow us to compute
the electromagnetic Green's function for such generic Newtonian
potentials. In Sec.~V we recalculate the electromagnetic self-force
for these potentials, and we reproduce Eq.~(\ref{1.10}). While our
result is compatible with the physical picture suggested by DeWitt and
DeWitt, our derivation shows very clearly that the result is largely
independent of the conditions near the mass distribution. We consider
this improvement on the original derivation to be a significant
contribution to this field. Our derivation is also more transparent,
in the sense that the computational labor involved is much reduced
compared with the original calculations of DeWitt and DeWitt. 

The techniques introduced in Secs.~III and IV allow us to calculate
all three types of Green's functions --- scalar, electromagnetic, and
gravitational. Those are used in Sec.~V to calculate the respective
self-forces. Part of the paper is therefore devoted to the calculation
of the scalar and gravitational self-forces in the weakly curved
spacetimes of Eq.~(\ref{1.9}). To the best of our knowledge, such
calculations have not yet been presented in the literature.   

\subsection*{Scalar self-force} 

The computations are simplest for the case of a scalar charge, and we 
obtain   
\begin{equation}
\bbox{f}_{\! \rm scalar} = 2 \xi q^2 \frac{M}{r^3}\, \bbox{\hat{r}}      
+ \frac{1}{3}\, q^2 \frac{d \bbox{g}}{dt}, 
\label{1.12} 
\end{equation}
where $\xi$ is a constant measuring the coupling of the scalar
field to the spacetime curvature; a precise definition is given by
Eq.~(\ref{2.7}) below. The equations of motion for the scalar charge
$q$ are identical to Eq.~(\ref{1.11}), but with the scalar
self-force replacing the electromagnetic self-force. For minimal
coupling ($\xi = 0$), the self-force is entirely dissipative, and we 
recover the expected result, $\frac{1}{3} q^2 d\bbox{g}/dt$; this is
the radiation-reaction force experienced by a scalar charge subjected 
to an external force $m \bbox{g}$. For minimal coupling, the
self-force acting on a stationary charge vanishes; this agrees with
Wiseman's result \cite{42} for the force required to hold a scalar
charge in place in the exact field of a Schwarzschild black hole. 
If $\xi > 0$, the conservative part of the self-force is repulsive. If
$\xi < 0$, it is attractive. This result for the scalar self-force is
very similar to Eq.~(\ref{1.10}), and we notice that the highly
nonlocal expression of Eq.~(\ref{1.8}) has managed to produce the
expected local result. Once more the dual point of view holds: We can
adopt a flat-spacetime point of view, set $f^\alpha_{\rm scalar}$ to
zero in Eq.~(\ref{1.7}), and get the correct result by equating the
external force to $m \bbox{g}$; or we can adopt a curved-spacetime
point of view, set $f^\alpha_{\rm ext}$ to zero in Eq.~(\ref{1.7}),
and get the correct result by evaluating the world-line integral of
Eq.~(\ref{1.8}).  

\subsection*{Gravitational self-force and matter-mediated force}   

The computations required for the gravitational self-force are the
most involved, but here we also face a serious technical problem. We
have stressed the importance of working with a Newtonian potential
that is {\it everywhere} small. In fact, the perturbative calculation
of the Green's function requires that the deviations of the metric
$g_{\alpha\beta}$ with respect to the Minkowski values
$\eta_{\alpha\beta}$ be everywhere small. To accommodate this
requirement, it is necessary that the spacetime contain matter: all
vacuum solutions to the Einstein field equations representing an
isolated massive object necessarily possess regions of 
strong curvature. Such strong-field regions would affect the Green's
function in a way that cannot be predicted by a perturbative
calculation; the entire method would fail, even if the Green's
function were to be evaluated only in the weak-field region of the
spacetime. Thus, the global character of the Green's function,
together with the weak-field limitations of our calculational methods,
dictate that we work with spacetimes that contain matter. 

The problem resides with the fact that the equations of motion for a  
massive particle, as given by Eqs.~(\ref{1.2}) and (\ref{1.3}), are
restricted to {\it vacuum} spacetimes. The
Mino-Sasaki-Tanaka-Quinn-Wald equations of motion \cite{24,25} are
therefore not directly suited to a weak-field calculation. An
extension to spacetimes containing matter must be produced. We do this 
in Sec.~VI, proceeding in two steps.      

First, we incorporate the modifications to the gravitational Green's 
function that come from the presence of Ricci-tensor terms in Green's 
equation. These modifications affect the gravitational
self-force, but they do not alter Eq.~(\ref{1.3}) if the particle is
restricted to move in a vacuum region of the spacetime. Second, we
consider how the presence of matter modifies the
equations of motion. Because the background stress-energy tensor
$T^{\alpha\beta}$ depends on the metric, it necessarily suffers a
perturbation when the massive particle is inserted in the
spacetime. Physically, this corresponds to the fact that while the
object of mass $M$ --- the star --- is at rest when the particle is
absent, it must move in the presence of the particle: both objects 
move around a fixed center of mass \cite{45.5}. This perturbative
motion of the star produces an additional metric perturbation, over
and above the perturbation directly associated with the particle's own
stress-energy tensor. This must be accounted for, and we shall see
that its effect is to modify the right-hand side of Eq.~(\ref{1.2}) by
a term $f^\alpha_{\rm mm}$ which we call the ``matter-mediated''
force. While the self-force can be thought of as a direct action of
the particle on itself, this additional force can be thought of as an
indirect action mediated by the presence of matter: the particle acts 
on the star, and the star acts back on the particle.        

The gravitational self-force is computed in Sec.~V, and this
calculation incorporates the effects of the matter on the retarded
Green's function. We obtain  
\begin{equation}
\bbox{f}_{\! \rm grav} = - \frac{11}{3}\, m^2 \frac{d \bbox{g}}{dt},     
\label{1.13}
\end{equation} 
and according to the naive equations of motion $m d^2 \bbox{z}/dt^2 
= m \bbox{g} + \bbox{f}_{\! \rm grav}$, the gravitational self-force
does work on the particle at an average rate $dW/dt = 
+\frac{11}{3} m^2 |\bbox{g}|^2$; it therefore gives rise to radiation  
{\it antidamping} \cite{46}. Notice that the self-force represents a
correction of 1.5{\sc pn} (post-Newtonian) order to the Newtonian
equations of motion \cite{PN}. Notice also that the dual point of view
seems at work also here: our expression for the ``radiation-reaction
force'' could be calculated on the basis of Eq.~(\ref{1.2}) by
adopting a ``flat-spacetime point of view'' in which 
$f^\alpha_{\rm grav}$ would be set to zero and the ``external force''
equated to $m \bbox{g}$.    

The remarkable conspiracy that makes the tail integral reproduce the
external-force term in the equations of motion is seen to be at play
in all three cases. While the agreement between the flat-spacetime and 
curved-spacetime points of view is quite necessary in the case of the 
scalar and electromagnetic self-forces, it is decidedly disturbing in
the gravitational case. How can we understand this result? 

The answer obviously comes from the matter-mediated force, which we
define and compute in Sec.~VI. For this we obtain 
\begin{equation} 
\bbox{f}_{\! \rm mm} = m \delta \bbox{g} + \mbox{1{\sc pn}}  
+ \frac{11}{3}\, m^2 \frac{d \bbox{g}}{dt}, 
\label{1.14}
\end{equation}
where the first term represents the change in the star's Newtonian
gravitational field associated with its motion around the fixed center
of mass, the second term is a post-Newtonian correction to the
Newtonian force $m \bbox{g}$, and the third term is a
radiation-damping term that precisely cancels out the antidamping
force of Eq.~(\ref{1.13}). (Such a cancellation was noticed a long
time ago by Carmeli \cite{51}, in the context of a very different
formulation of the equations of motion.) 

A precise expression for the matter-mediated force can be found in
Eq.~(\ref{6.54}) below. When it is substituted, together with the
self-force, into the equations of motion, we find that they take the
form   
\begin{equation}
\frac{d^2 \bbox{z}}{dt^2} = -\frac{M}{\rho^3}\, \biggl[   
\biggl( 1 + \bbox{v}^2 - \frac{5m}{\rho} - 4 \bbox{v} \cdot \bbox{V}
\biggr) \bbox{\rho} -  \bbox{\rho} \cdot (4 \bbox{v} - 3\bbox{V}) 
\bbox{v} + 4 (\bbox{\rho} \cdot \bbox{v}) \bbox{V} \biggr]
+ O(2{\sc pn},m^2,M^2),  
\label{1.15}
\end{equation}  
which contains no trace of a radiation-reaction force. Here,
$\bbox{z}(t)$ gives the position of the particle, and $\bbox{v}(t) = 
d\bbox{z}/dt$ is its velocity; the star moves on a 
trajectory $\bbox{Z}(t)$ with a velocity $\bbox{V}(t) =
d\bbox{Z}/dt$. We use $\bbox{\rho} = \bbox{z} - \bbox{Z}$ to designate 
the separation between the two objects, and $\rho \equiv |\bbox{z} - 
\bbox{Z}|$. The equations of motion for the star are 
\begin{equation} 
\frac{d^2 \bbox{Z}}{dt^2} = \frac{m}{\rho^3}\, \Biggl\{   
\Biggl[ 1 + 2 \bbox{v}^2 - \frac{3}{2} \biggl( 
\frac{\bbox{\rho} \cdot \bbox{v}}{\rho} \biggr)^2 \Biggr] \bbox{\rho}  
- 3 (\bbox{\rho} \cdot \bbox{v}) \bbox{v} \Biggr\} 
+ O(2{\sc pn},m^2,mM).    
\label{1.16}
\end{equation}
Our calculations cannot reproduce {\it all} the terms in the 
Einstein-Infeld-Hoffmann equations of motion \cite{1,55.5}:
Eq.~(\ref{1.15}) omits terms that are quadratic in $m$ and $M$,
while Eq.~(\ref{1.16}) neglects terms of order $m^2$ and $mM$. The
formalism we use in this paper is based on linear perturbation
theory, and it would be incapable of producing corrections of order
$m^2$. On the other hand, the corrections of order $mM$ and $M^2$ are
within the reach of the formalism, but in order to produce them we
would need to go beyond the weak-field approximation considered in
this paper. Within these limitations, however, we have complete
agreement between the calculations presented in this paper and the
standard post-Newtonian treatment of the two-body problem.   

\subsection*{Organization} 

The technical part of the paper begins in Sec.~II with a detailed 
discussion of the metric of Eq.~(\ref{1.9}), and a derivation of the 
scalar, electromagnetic, and gravitational wave equations for the
weakly curved spacetimes described by this metric. In Sec.~III we
introduce the two-point functions $A(x,x')$ and $B(x,x')$, and show
how the scalar, electromagnetic, and gravitational Green's functions
can be obtained from them by acting with differential operators. In
Sec.~IV we introduce methods to compute the two-point functions, and
evaluate them in interesting limiting cases. [In Appendix A we expand 
our discussion of $A(x,x')$, and in Appendix B we provide an explicit
computation of the two-point functions for the special case $\Phi =
-M/r$.] In Sec.~V we use our preceding results for the Green's
functions to compute the scalar, electromagnetic, and gravitational
self-forces for the spacetimes of Eq.~(\ref{1.9}); our results were 
quoted in Eqs.~(\ref{1.12}), (\ref{1.10}), and (\ref{1.13}), 
respectively. Finally, in Sec.~VI we introduce and compute the
gravitational matter-mediated force, and show that it cancels out the 
antidamping force calculated in Sec.~V. 

\section{Weakly curved spacetime} 

\subsection{The spacetime and its geometric quantities} 

The spacetimes considered in this paper have a metric given by 
\begin{equation}
ds^2 = -(1 + 2\Phi)\, dt^2 + (1-2\Phi)(dx^2 + dy^2 + dz^2), 
\label{2.1}
\end{equation}
in which $\Phi(\bbox{x})$ is a Newtonian potential, a function of the
spatial coordinates $\bbox{x}$ satisfying Poisson's equation, 
\begin{equation}
\nabla^2 \Phi = 4\pi \rho,
\label{2.2}
\end{equation}
where $\rho(\bbox{x})$ is the mass density. By virtue of
Eq.~(\ref{2.2}) and the fact that $\Phi \ll 1$ everywhere in the
spacetime, the metric of Eq.~(\ref{2.1}) satisfies the linearized
Einstein field equations. Throughout the paper we will work
consistently to first order in $\Phi$. If the metric were representing
the gravitational field of a point mass $M$ located
at the origin of the coordinate system, then $\rho = M
\delta(\bbox{x})$ and $\Phi = -M/r$, where $r = | \bbox{x} |$. We will
not need to adopt this particular form for the Newtonian potential
(which would violate the condition that $\Phi$ be small everywhere),
and we shall work with generic potentials satisfying
Eq.~(\ref{2.2}). We will, however, assume that $\Phi$ becomes equal to 
$-M/r$, with $M = \int \rho\, d^3 x$, far from the matter
distribution. Notice that we do not allow $\Phi$ to depend on time.  

Introducing $t^\alpha = \partial x^\alpha/\partial t$ as the timelike
Killing vector of the spacetime, we write the metric tensor as 
\begin{equation}
g_{\alpha\beta} = \eta_{\alpha\beta} - 2\Phi \chi_{\alpha\beta},
\qquad 
\chi_{\alpha\beta} \equiv \eta_{\alpha\beta} + 2 t_\alpha t_\beta, 
\label{2.3}
\end{equation}
where $\eta_{\alpha\beta} = \mbox{diag}(-1,1,1,1)$ is the Minkowski 
metric, which was used to lower the index on $t^\alpha$. Throughout
this section we will continue to lower and raise indices with the
Minkowski metric, unless it is otherwise indicated. Notice that
$\chi_{\alpha\beta} = \mbox{diag}(1,1,1,1)$. From Eq.~(\ref{2.3}) it
is easy to derive $\sqrt{-g} = 1 - 2\Phi$ and $g^{\alpha\beta} =
\eta^{\alpha\beta} + 2\Phi \chi^{\alpha\beta}$. These results hold to 
first order in $\Phi$.   

The metric of Eq.~(\ref{2.3}) comes with the following Christoffel
symbols: 
\begin{equation}
\Gamma^\mu_{\, \alpha\beta} = -\chi^\mu_{\ \alpha} \Phi_{,\beta} -
\chi^\mu_{\ \beta} \Phi_{,\alpha} + \chi_{\alpha\beta} \Phi^{,\mu}. 
\label{2.4}
\end{equation}
As a consequence of the relation $\Phi_{,\alpha} t^\alpha = 0$, they 
satisfy $\eta^{\alpha\beta} \Gamma^\mu_{\, \alpha\beta} = 0$. The
Riemann tensor is given by 
\begin{equation}
R^\mu_{\ \alpha\beta\gamma} = -\chi^\mu_{\ \gamma} \Phi_{,\alpha\beta} 
+ \chi^\mu_{\ \beta} \Phi_{,\alpha\gamma} - \chi_{\alpha\beta}
\Phi^{,\mu}_{\ \gamma} + \chi_{\alpha\gamma} \Phi^{,\mu}_{\ \beta}. 
\label{2.5}
\end{equation} 
Contracting over the first and third indices gives the Ricci tensor: 
\begin{equation}
R_{\alpha\beta} = \chi_{\alpha\beta} \Box \Phi, 
\label{2.6}
\end{equation}
where $\Box \Phi \equiv \eta^{\alpha\beta}
\Phi_{,\alpha\beta}$. Because $\Phi$ does not depend on time, 
$\Box \Phi = \nabla^2 \Phi \equiv \delta^{ab} \Phi_{,ab}$, where $x^a$
denotes the spatial coordinates. The Ricci scalar is 
$R = 2 \Box \Phi$, and the Einstein tensor is 
$G^{\alpha \beta} = 2 t^\alpha t^\beta \Box \Phi$.
Equation (\ref{2.2}) follows from the Einstein field equations with
$T^{\alpha\beta} = \rho t^\alpha t^\beta$; this represents a fluid of
mass density $\rho$ and negligible pressure at rest in the spacetime.   

\subsection{Wave equations} 

The field equation for a massless scalar field $\psi$ in a curved
spacetime with metric $g_{\alpha\beta}$ is 
\begin{equation} 
g^{\alpha\beta} \psi_{;\alpha\beta} - \xi R \psi = -4\pi \mu,
\label{2.7}
\end{equation}
where the semicolon designates covariant differentiation compatible
with the metric tensor, $\xi$ is an arbitrary constant measuring the 
coupling to curvature, and $\mu$ is a given source term. It will be
convenient to work with a densitized form for the field equation,
\begin{equation}
E[\psi] = -4\pi \sqrt{-g}\, \mu,
\label{2.8}
\end{equation}
where $E[\psi]$ stands for the left-hand side of Eq.~(\ref{2.7})
multiplied by $\sqrt{-g}$. For the weakly curved spacetime of
Eq.~(\ref{2.3}), we find that 
\begin{equation}
E[\psi] = \Box \psi + 4\Phi t^\alpha t^\beta \psi_{,\alpha\beta} -
2\xi (\Box \Phi) \psi, 
\label{2.9}
\end{equation}
where $\Box \equiv \eta^{\alpha\beta} \partial_\alpha \partial_\beta$
is the flat-spacetime wave operator and, as was pointed out
previously, $\Box \Phi = \nabla^2 \Phi = 4\pi \rho$.      

An electromagnetic field is represented by a vector potential
$A^\alpha$ which is here assumed to satisfy the Lorentz-gauge
condition, $A^\alpha_{\ ;\alpha} = 0$. Its field equations are  
\begin{equation}
g^{\mu\nu} A^\alpha_{\ ;\mu\nu} - R^\alpha_{\ \mu} 
A^\mu = -4\pi j^\alpha,
\label{2.10}
\end{equation}
where $j^\alpha$ is a given current density. The wave equation in 
densitized form is 
\begin{equation}
E^\alpha[A] = -4\pi \sqrt{-g}\, j^\alpha,
\label{2.11}
\end{equation}
where $E^\alpha[A]$ stands for the left-hand side of Eq.~(\ref{2.10}) 
multiplied by $\sqrt{-g}$. A straightforward computation reveals that
for the weakly curved spacetime of Eq.~(\ref{2.3}),  
\begin{equation}
E^\alpha[A] = \Box A^\alpha 
+ 4\Phi t^\mu t^\nu A^\alpha_{\ ,\mu\nu} 
- 2 \bigl(\chi^\alpha_{\ \mu} \Phi_{,\nu}
+ \chi^\alpha_{\ \nu} \Phi_{,\mu} - \chi_{\mu\nu}
  \Phi^{,\alpha} \bigr) A^{\mu,\nu} 
- \bigl[ \chi^{\alpha\mu} \Phi_{,\mu\nu} 
+ 2 \chi^\alpha_{\ \nu} \Box \Phi 
- \chi^\mu_{\ \nu} \Phi^{,\alpha}_{\ \mu} \bigr] A^\nu.  
\label{2.12}
\end{equation}
Here, $\Box$ acting on the vector field $A^\alpha$ still stands for
the scalar wave operator of flat spacetime.      

A gravitational perturbation on a background metric $g_{\alpha\beta}$
is described by a tensor $\gamma_{\alpha\beta}$, and the full
metric is $\hat{g}_{\alpha\beta} = g_{\alpha\beta} +
\gamma_{\alpha\beta}$. By linearizing the Einstein field equations
about the background, one obtains a wave equation for the
perturbation. This equation takes a simpler form if it is expressed in
terms of a ``trace-reversed'' field $\bar{\gamma}_{\alpha\beta}
\equiv \gamma_{\alpha\beta} - \frac{1}{2} (g^{\mu\nu} \gamma_{\mu\nu}) 
g_{\alpha\beta}$, which is then chosen to satisfy a Lorentz-gauge
condition: $\bar{\gamma}^{\alpha\beta}_{\ \ \ \, ;\beta} = 0$. (Here
and below, indices on the perturbation field are raised and lowered
with the background metric, and a semicolon designates covariant
differentiation on the background.) With these choices, the Einstein 
field equations take the form of a wave equation for
$\bar{\gamma}^{\alpha\beta}$ \cite{26}:   
\begin{equation} 
g^{\mu\nu} \bar{\gamma}^{\alpha\beta}_{\ \ \, ;\mu\nu} 
+ 2 R_{\mu\ \nu}^{\ \alpha\ \beta} \bar{\gamma}^{\mu\nu} 
+ S_{\mu\ \nu}^{\ \alpha\ \beta} \bar{\gamma}^{\mu\nu}
= -16\pi \delta T^{\alpha\beta}, 
\label{2.13}
\end{equation}
where $\delta T^{\alpha\beta}$ is the perturbation in the
stress-energy tensor (so that the full stress-energy tensor is
$\hat{T}^{\alpha\beta} = T^{\alpha\beta} + \delta T^{\alpha\beta}$,
with the first term denoting the background's stress-energy tensor),  
$R_{\mu\alpha\nu\beta}$ is the Riemann tensor of the background
spacetime, and $S_{\mu\alpha\nu\beta}$ is constructed from the
background's Ricci and Einstein tensors,  
\begin{equation}
S_{\mu\alpha\nu\beta} = 2 R_{\mu(\alpha} g_{\beta)\nu}  
- R_{\mu\nu} g_{\alpha\beta} - R g_{\mu(\alpha} g_{\beta)\nu} 
- 2 g_{\mu\nu} G_{\alpha\beta}.  
\label{2.14}
\end{equation} 
The parentheses around tensor indices indicate symmetrization with
respect to these indices. Once Eq.~(\ref{2.13}) has been solved for
the trace-reversed perturbation, the metric is recovered from the
relation $\gamma_{\alpha\beta} = 
\bar{\gamma}_{\alpha\beta} - \frac{1}{2} g_{\alpha\beta}
\bar{\gamma}$, where $\bar{\gamma} = g^{\alpha\beta}
\bar{\gamma}_{\alpha\beta}$. 

The densitized wave equation is 
\begin{equation}
E^{\alpha\beta}[\bar{\gamma}] = 
-16\pi \sqrt{-g}\, \delta T^{\alpha\beta}, 
\label{2.15}
\end{equation}
where $E^{\alpha\beta}[\bar{\gamma}]$ stands for the left-hand side of 
Eq.~(\ref{2.13}) multiplied by $\sqrt{-g}$. A rather long but
straightforward computation reveals that in a weakly curved
spacetime with metric (\ref{2.3}),  
\begin{eqnarray}
E^{\alpha\beta}[\bar{\gamma}] &=& \Box \bar{\gamma}^{\alpha\beta} 
+ 4\Phi t^\mu t^\nu \bar{\gamma}^{\alpha\beta}_{\ \ ,\mu\nu}  
- 4 \bigl(\chi^{(\alpha}_{\ \, \mu} \Phi_{,\nu}
+ \chi^{(\alpha}_{\ \, \nu} \Phi_{,\mu} - \chi_{\mu\nu}
  \Phi^{(,\alpha} \bigr) \bar{\gamma}^{\beta)\mu,\nu} 
+ 2 \bigl( \chi_{\mu\nu} \Phi^{,\alpha\beta} 
- 2 \chi^{(\alpha}_{\ \mu} \Phi^{,\beta)}_{\ \ \nu} 
+ \chi^{\alpha\beta} \Phi_{,\mu\nu} \bigr) \bar{\gamma}^{\mu\nu} 
\nonumber \\ & & \mbox{} 
- \Box \Phi \Bigl( \eta^{\alpha\beta} \chi_{\mu\nu}
                    \bar{\gamma}^{\mu\nu} 
       + 2 \bar{\gamma}^{\alpha\beta}
       + 4 t^\alpha t^\beta \eta_{\mu\nu} \bar{\gamma}^{\mu\nu} 
            \Bigr).  
\label{2.16}
\end{eqnarray} 
In the first term on the right-hand side, $\Box = \eta^{\mu\nu}
\partial_\mu \partial_\nu$ is the scalar wave operator.   

\section{Green's functions} 

\subsection{Generating two-point functions} 

The task before us in this section is the computation of retarded
Green's functions for the wave equations considered in Sec.~II --- 
Eqs.~(\ref{2.8}), (\ref{2.11}), and (\ref{2.15}). As we shall see,
these Green's functions can be constructed by acting with differential  
operators on two generating two-point functions, $A(x,x')$ and
$B(x,x')$, defined by 
\begin{equation}
A(x,x') = \frac{1}{2\pi} \int G^{\rm flat}(x,x'') \Phi(x'') 
G^{\rm flat}(x'',x')\, d^4 x''
\label{3.1}
\end{equation}
and 
\begin{equation}
B(x,x') = \int G^{\rm flat}(x,x'') \rho(x'') 
G^{\rm flat}(x'',x')\, d^4 x''. 
\label{3.2}
\end{equation}
Here, $\Phi$ is the Newtonian potential of Eq.~(\ref{2.1}), $\rho$ is
the mass density of Eq.~(\ref{2.2}), and $G^{\rm flat}(x,x')$ is the
retarded Green's function of the flat-spacetime wave operator, 
\begin{equation}
G^{\rm flat}(x,x') = \frac{\delta(t-t'-|\bbox{x} - \bbox{x'}|)}
{|\bbox{x} - \bbox{x'}|}; 
\label{3.3}
\end{equation}
this satisfies $\Box G^{\rm flat}(x,x') = -4\pi \delta_4(x-x')$, where 
$\delta_4(x-x')$ is a four-dimensional $\delta$-function, equal
to $\delta(t-t') \delta_3(\bbox{x} - \bbox{x'})$. This method to
calculate the Green's function originated in the work of DeWitt and
DeWitt \cite{44}, who used it to compute the scalar and
electromagnetic Green's functions for the special case $\Phi =
-M/r$. Here we generalize their method to arbitrary potentials, and to
the gravitational Green's function.   

Inside the integrals of Eqs.~(\ref{3.1}) and (\ref{3.2}), the function  
$G^{\rm flat}(x,x'')$ picks out the past light cone of the point $x$,
while $G^{\rm flat}(x'',x')$ picks out the future light cone of the
point $x'$. (These light cones are those of the flat background
spacetime.) The intersection of the two light cones defines a closed  
two-surface $\cal S$, and we see that $A(x,x')$ is the average
of the Newtonian potential over $\cal S$, while $B(x,x')$ is the
average of the mass density. Both two-point functions are zero if $x$
and $x'$ are spacelike related, because the surface of intersection
then disappears. 

Averages of derivatives of the Newtonian potential can be obtained by 
differentiating $A(x,x')$. For example, 
\begin{equation}
\frac{1}{2\pi} \int G^{\rm flat}(x,x'') 
\bigl[\partial_{\alpha''} \Phi(x'')\bigr]   
G^{\rm flat}(x'',x')\, d^4 x'' = 
(\partial_\alpha + \partial_{\alpha'}) A(x,x') 
\label{3.4}
\end{equation} 
and
\begin{equation}
\frac{1}{2\pi} \int G^{\rm flat}(x,x'') 
\bigl[ \partial_{\alpha''\beta''} \Phi(x'') \bigr] 
G^{\rm flat}(x'',x')\, d^4 x'' = 
(\partial_{\alpha\beta} + \partial_{\alpha'\beta} +
\partial_{\alpha\beta'} + \partial_{\alpha'\beta'}) A(x,x'). 
\label{3.5}
\end{equation} 
We use the notation $\partial_{\alpha'} f = \partial f/\partial
x^{\alpha'}$, $\partial_{\alpha\beta'} f = \partial^2 f/\partial
x^\alpha \partial x^{\beta'}$, etc., in which the tensor index
(either $\alpha$, $\alpha'$, or $\alpha''$) indicates with respect to
which variable (between $x$, $x'$, and $x''$) the function $f$ is
being differentiated. 

Equations (\ref{3.4}) and (\ref{3.5}) are easy to establish. (In the
following we keep the notation simple and drop the factor of
$1/2\pi$ in front of the integrals). First we write
$\partial_\alpha A = \int (\partial_\alpha G) \Phi G\, dx''$, and use
the fact that $G^{\rm flat}(x,x'')$ depends only on the difference
between $x$ and $x''$. This allows us to express the right-hand side
as $-\int (\partial_{\alpha''} G) \Phi G\, dx''$. Integrating by
parts, we obtain $+\int G (\partial_{\alpha''} \Phi) G\, dx'' + \int G
\Phi (\partial_{\alpha''} G)\, dx''$. In the second term, the
derivative operator can be switched to $-\partial_{\alpha'}$, which
can be taken outside the integral. The second term becomes 
$-\partial_{\alpha'} A$, and Eq.~(\ref{3.4}) follows. To derive
Eq.~(\ref{3.5}) we differentiate Eq.~(\ref{3.4}) with respect to
$x^\beta$, and go through the same procedure. 

Because the Newtonian potential does not depend on time,
Eq.~(\ref{3.4}) implies 
\begin{equation}
\partial_t A(x,x') + \partial_{t'} A(x,x') = 0. 
\label{3.6}
\end{equation} 
Another useful identity results from taking the trace of
Eq.~(\ref{3.5}). Using Poisson's equation $\Box \Phi = 
\nabla^2 \Phi = 4\pi \rho$ and Eq.~(\ref{3.2}), we obtain 
\begin{equation}
\frac{1}{2} ( \Box + \Box' ) A(x,x') 
+ \partial^{\alpha'}_{\ \alpha} A(x,x') = B(x,x'), 
\label{3.7}
\end{equation}
where $\Box' \equiv \eta^{\mu\nu} \partial_{\mu'} \partial_{\nu'}$;
summation over $\alpha'$ and $\alpha$ in the third term is
understood. By virtue of Eq.~(\ref{3.6}), all time derivatives drop
out of Eq.~(\ref{3.7}), which can then be written as $\frac{1}{2}
( \nabla^2 + \nabla^{\prime 2} ) A(x,x') + \partial^{a'}_{\ a} A(x,x') 
= B(x,x')$. The left-hand side of Eq.~(\ref{3.7}) can be simplified
differently. The wave operator acting on $A$ can be taken inside the
integral, where it acts on $G^{\rm flat}$. This returns a
$\delta$-function, and the integral can be evaluated. This gives 
\begin{equation}
\Box A(x,x') = -2 \Phi(x) G^{\rm flat}(x,x'), \qquad
\Box' A(x,x') = -2 \Phi(x') G^{\rm flat}(x,x'), 
\label{3.8}
\end{equation} 
and these expressions can be substituted into Eq.~(\ref{3.7}). 

\subsection{Scalar Green's function} 

We are looking for a function $G(x,x')$ that solves \cite{36} 
\begin{equation}
E[G] = -4\pi \delta_4(x-x')
\label{3.9}
\end{equation}
to first order in the Newtonian potential $\Phi$. Here,
$\delta_4(x-x')$ is the four-dimensional $\delta$-function, and $E[G]$
stands for the left-hand side of Eq.~(\ref{2.7}) multiplied by
$\sqrt{-g}$, with $G(x,x')$ taking the place of $\psi(x)$. In terms of
this Green's function, the solution to Eq.~(\ref{2.8}) is 
\begin{equation}
\psi(x) = \int G(x,x') \mu(x') \sqrt{-g'}\, d^4 x', 
\label{3.10}
\end{equation} 
where $g'$ is the metric determinant evaluated at $x'$. 

To find this Green's function, we write 
\begin{equation}
G(x,x') = G^{\rm flat}(x,x') + \dot{G}(x,x') + O(\Phi^2), 
\label{3.11}
\end{equation}
in which $G^{\rm flat}(x,x')$ is flat-spacetime solution given by
Eq.~(\ref{3.3}), and $\dot{G}(x,x')$ is the first-order correction
(linear in the Newtonian potential). [The overdot notation 
suggests that $G(x,x')$ should be viewed as a smooth function of a
smallness parameter $\epsilon$, and that Eq.~(\ref{3.11}) gives the
first two terms of its Taylor expansion about $\epsilon = 0$.] By
virtue of Eq.~(\ref{2.9}), in which we substitute $\Box \Phi 
= 4\pi \rho$, we find that $\dot{G}(x,x')$ satisfies 
\begin{equation}
\Box \dot{G}(x,x') = -4 \Phi(x) \partial_{tt} G^{\rm flat}(x,x') +
8\pi \xi \rho(x) G^{\rm flat}(x,x').
\label{3.12}
\end{equation} 
The solution is
\begin{equation}
\dot{G}(x,x') = \frac{1}{\pi} \int G^{\rm flat}(x,x'') \Phi(x'')
\partial_{t''t''} G^{\rm flat}(x'',x')\, d^4 x'' 
- 2\xi \int G^{\rm flat}(x,x'') \rho(x'') G^{\rm flat}(x'',x')\, 
d^4 x''.
\label{3.13}
\end{equation} 
In the first integral, the derivative operator can be switched to
$\partial_{t't'}$ because the flat-spacetime Green's function depends
only on $x''-x'$. The derivatives can then be taken outside the
integral. Using Eqs.~(\ref{3.1}) and (\ref{3.2}), we
obtain $\dot{G} = 2 \partial_{t't'} A - 2\xi B$. Using
Eq.~(\ref{3.6}), we express the final result as  
\begin{equation}
\dot{G}(x,x') = -2 \partial_{t t'} A(x,x') - 2\xi B(x,x'). 
\label{3.14}
\end{equation}
Equations (\ref{3.11}) and (\ref{3.14}) give the scalar Green's
function to first order in $\Phi$. As was claimed previously,
$G(x,x')$ can be expressed in terms of the generating two-point
functions introduced in Sec.~III A. For the special case $\xi = 0$,
our Green's function agrees with the one obtained by DeWitt and
DeWitt \cite{44}.     

\subsection{Electromagnetic Green's function} 

The electromagnetic Green's function $G^\alpha_{\ \beta'}(x,x')$ is a 
solution to \cite{36}  
\begin{equation}
E^\alpha_{\ \beta'}[G] = -4\pi \delta^\alpha_{\ \beta'}
\delta_4(x-x'),  
\label{3.15}
\end{equation}
where $E^\alpha_{\ \beta'}[G]$ stands for the left-hand side of
Eq.~(\ref{2.10}) multiplied by $\sqrt{-g}$, with 
$G^\alpha_{\ \beta'}(x,x')$ taking the place of
$A^\alpha(x)$. Here and below, the unprimed tensor indices refer to
the point $x$, while the primed indices refer to $x'$. In terms of
this Green's function, the solution to Eq.~(\ref{2.11}) is 
\begin{equation}
A^\alpha(x) = \int G^\alpha_{\ \beta'}(x,x') j^{\beta'}(x')
\sqrt{-g'}\, d^4 x'. 
\label{3.16}
\end{equation}
We seek to determine the Green's function to first order in $\Phi$. 

For this purpose, we write
\begin{equation}
G^\alpha_{\ \beta'}(x,x') = G^{\rm flat}(x,x') \delta^\alpha_{\ \beta'}
+ \dot{G}^\alpha_{\ \beta'}(x,x') + O(\Phi^2) 
\label{3.17}
\end{equation}
and substitute into Eq.~(\ref{3.15}). Here, $G^{\rm flat}(x,x')$ is
the scalar Green's function of Eq.~(\ref{3.3}), and 
$\dot{G}^\alpha_{\ \beta'}(x,x')$ is the first-order correction to the
electromagnetic Green's function. We find that this must satisfy 
\begin{equation}
\Box \dot{G}^\alpha_{\ \beta'} = -4\Phi \partial_{tt} G^{\rm flat} 
\delta^\alpha_{\ \beta} + 2 \bigl( \chi^\alpha_{\ \mu} \Phi_{,\beta} +
\chi^\alpha_{\ \beta} \Phi_{,\mu} - \chi_{\beta\mu} \Phi^{,\alpha}
\bigr) G^{{\rm flat},\mu} + 8\pi \rho \chi^\alpha_{\ \beta} 
G^{\rm flat}. 
\label{3.18}
\end{equation}
We have used the fact that $\Phi$ does not depend
on time, so that, for example, $\chi^{\alpha\mu} \Phi_{,\mu\beta} =   
\eta^{\alpha\mu} \Phi_{,\mu\beta} = \Phi^{,\alpha}_{\ \beta}$; such
terms end up canceling out. We have also, on the right-hand side,
dropped the distinction between $\beta$ and $\beta'$; this distinction 
is not necessary when dealing with a constant tensor such as
$\chi^\alpha_{\ \beta}$. 

Equation (\ref{3.18}) is solved by once again invoking the
flat-spacetime Green's function. The solution is 
\begin{equation}
\dot{G}^\alpha_{\ \beta'}(x,x') = - \frac{1}{4\pi} \int  
G^{\rm flat}(x,x'') \mbox{\sc rhs}(x'')\, d^4 x'',
\end{equation}
where $\mbox{\sc rhs}(x'')$ stands
for the right-hand side of Eq.~(\ref{3.18}) evaluated at
$x = x''$. The derivatives appearing in Eq.~(\ref{3.18}) are now taken 
with respect to $x''$. When $\partial_{\mu''}$ (say) is acting on 
$G^{\rm flat}(x'',x')$, it can be replaced by $-\partial_{\mu'}$ and
taken outside the integral. This leaves us with integrals involving 
$G^{\rm flat}(x,x'')$, $G^{\rm flat}(x'',x')$, as well as
$\Phi(x'')$ and its derivatives. Those integrals can all be expressed
in terms of the two-point functions $A(x,x')$ and $B(x,x')$ defined in
Eqs.~(\ref{3.1}) and (\ref{3.2}), thanks to the relations (\ref{3.4})
and (\ref{3.5}). The results can then be simplified by using 
Eqs.~(\ref{3.6})--(\ref{3.8}), as well as Eq.~(\ref{2.3}) for
$\chi_{\alpha\beta}$. The end result is 
\begin{equation}
\dot{G}^\alpha_{\ \beta'} = - 2 \partial_{tt'} A 
\delta^\alpha_{\ \beta} + \bigl( \partial^{\alpha'}_{\ \beta} 
- \partial^{\alpha}_{\ \beta'} \bigr) A + 2 t^\alpha
(\partial_{t'\beta} - \partial_{t\beta'} ) A + 2t_\beta   
\bigl( \partial^{\alpha'}_{\ t} 
- \partial^{\alpha}_{\ t'} \bigr) A + \chi^\alpha_{\ \beta} \bigl
( \Delta \Phi G^{\rm flat} - B ), 
\label{3.19}
\end{equation} 
where $\Delta \Phi \equiv \Phi(x) - \Phi(x')$. Equations (\ref{3.17})
and (\ref{3.19}) give $G^\alpha_{\ \beta'}(x,x')$, the electromagnetic 
Green's function, to first order in $\Phi$. 

Equation (\ref{3.19}) is easier to deal with if we distinguish between
its temporal and spatial components. From it we read off 
\begin{eqnarray} 
\dot{G}^t_{\ t'} &=& - \Delta \Phi G^{\rm flat} - 2 \partial_{tt'} A
+ B, \nonumber \\
\dot{G}^t_{\ a'} &=& ( \partial_{t'a} - \partial_{ta'} ) A, 
\nonumber \\ 
& & \label{3.20} \\
\dot{G}^a_{\ t'} &=& \bigl( \partial^a_{\ t'} - \partial^{a'}_{\ t}
\bigr) A,   
\nonumber \\
\dot{G}^a_{\ b'} &=& \delta^a_{\ b} \bigl( \Delta \Phi G^{\rm flat}
- 2 \partial_{tt'} A  - B \bigr) + 
\bigl( \partial^{a'}_{\ b} - \partial^a_{\ b'} \bigr) A. \nonumber 
\end{eqnarray} 
We notice that in Eqs.~(\ref{3.19}) and (\ref{3.20}), the part of 
$\dot{G}^{\alpha}_{\ \beta'}(x,x')$ that involves $G^{\rm flat}(x,x')$
has support only on the past light cone of the point $x$. The
remaining part, that involving the two-point functions 
$A(x,x')$ and $B(x,x')$, have support inside the light cone as well;
this is the ``tail part'' of the Green's function \cite{27}. Our
electromagnetic Green's function agrees with DeWitt and DeWitt
\cite{44}. 

\subsection{Gravitational Green's function} 

The trace-reversed gravitational Green's function 
${\cal G}^{\alpha\beta}_{\ \ \gamma'\delta'}(x,x')$ satisfies the 
equation \cite{26} 
\begin{equation}
E^{\alpha\beta}_{\ \ \gamma'\delta'}[{\cal G}] = -4\pi   
\delta^{(\alpha}_{\ \gamma'} \delta^{\beta)}_{\ \delta'}   
\delta_4(x-x'), 
\label{3.21}
\end{equation}
where $E^{\alpha\beta}_{\ \ \gamma'\delta'}[{\cal G}]$ stands for the  
left-hand side of Eq.~(\ref{2.13}) multiplied by $\sqrt{-g}$, but with  
${\cal G}^{\alpha\beta}_{\ \ \gamma'\delta'}(x,x')$ replacing
$\bar{\gamma}^{\alpha\beta}(x)$. In terms of this Green's function,
the solution to Eq.~(\ref{2.15}) is 
\begin{equation}
\bar{\gamma}^{\alpha\beta}(x) = 4 \int 
{\cal G}^{\alpha\beta}_{\ \ \gamma'\delta'}(x,x') 
\delta T^{\gamma'\delta'}(x') \sqrt{-g'}\, d^4 x'. 
\label{3.22}
\end{equation}
Our goal here is again to calculate the Green's function to first
order in the Newtonian potential. The steps are virtually identical to
what was done for the scalar and electromagnetic cases. The only
difference is that the expressions are longer.  

At the end of the calculation we will have to ``trace-reverse'' the
trace-reversed Green's function to obtain 
$G^{\alpha\beta}_{\ \ \gamma'\delta'}(x,x')$, the Green's function
directly associated with the metric perturbation
$\gamma_{\alpha\beta}$. This is given by 
\begin{equation} 
G^{\alpha\beta}_{\ \ \gamma'\delta'}(x,x') = 
{\cal G}^{\alpha\beta}_{\ \ \gamma'\delta'}(x,x') - \frac{1}{2}\,  
g^{\alpha\beta}(x) g_{\mu\nu}(x) 
{\cal G}^{\mu\nu}_{\ \ \gamma'\delta'}(x,x'). 
\label{3.23}
\end{equation}
In terms of this, the metric perturbation is recovered by integrating 
\begin{equation}
\gamma^{\alpha\beta}(x) = 4 \int 
G^{\alpha\beta}_{\ \ \gamma'\delta'}(x,x') 
\delta T^{\gamma'\delta'}(x') \sqrt{-g'}\, d^4 x'. 
\label{3.24}
\end{equation}

To find the trace-reversed Green's function, we write 
\begin{equation}
{\cal G}^{\alpha\beta}_{\ \ \gamma'\delta'}(x,x') = 
G^{\rm flat}(x,x') \delta^{(\alpha}_{\ \gamma'} 
                   \delta^{\beta)}_{\ \delta'} 
+ \dot{\cal G}^{\alpha\beta}_{\ \ \gamma'\delta'}(x,x') 
+ O(\Phi^2), 
\label{3.25}
\end{equation}
where $G^{\rm flat}(x,x')$ is the flat-spacetime scalar Green's
function, and $\dot{\cal G}^{\alpha\beta}_{\ \ \gamma'\delta'}(x,x')$ 
the correction of order $\Phi$. By substituting Eq.~(\ref{3.25})
into Eq.~(\ref{3.21}), we find that this must satisfy 
\begin{eqnarray}
\Box \dot{\cal G}^{\alpha\beta}_{\ \ \gamma'\delta'} &=& 
- 4\Phi \partial_{tt} G^{\rm flat} \delta^{(\alpha}_{\ \gamma} 
  \delta^{\beta)}_{\ \delta}
+ 4 \delta^{(\alpha}_{\ (\gamma} \Bigl( 
     \chi^{\beta)}_{\ \delta)} \Phi_{,\mu} 
   + \Phi_{,\delta)} \chi^{\beta)}_{\ \mu} 
   - \chi_{\delta)\mu} \Phi^{,\beta)} \Bigr) G^{{\rm flat},\mu} 
\nonumber \\ & & \mbox{}
- 2 \Bigl( \chi_{\gamma\delta} \Phi^{,\alpha\beta} 
   - 2\chi^{(\alpha}_{\ (\gamma} \Phi^{,\beta)}_{\ \, \delta)} 
   + \chi^{\alpha\beta} \Phi_{,\gamma\delta} \Bigr) G^{\rm flat}
 + 4\pi \Bigl( \eta^{\alpha\beta} \chi_{\gamma\delta} 
   + 2 \delta^{(\alpha}_{\ \gamma} \delta^{\beta)}_{\ \delta}
   + 4 t^\alpha t^\beta \eta_{\gamma\delta} 
        \Bigr) \rho\, G^{\rm flat}, 
\label{3.26}
\end{eqnarray} 
where we have gone through the same simplification steps as in the 
electromagnetic case. The remaining steps are also similar: We
integrate Eq.~(\ref{3.26}) with the flat-spacetime Green's
function, and we simplify the resulting integrals as was described in
the paragraph preceding Eq.~(\ref{3.19}). After a rather long
computation, we obtain 
\begin{eqnarray} 
\dot{\cal G}^{\alpha\beta}_{\ \ \gamma'\delta'} &=& 
- 2 \partial_{tt'} A \delta^{(\alpha}_{\ \gamma} 
                     \delta^{\beta)}_{\ \delta} 
+ 2 \delta^{(\alpha}_{\ (\gamma} \Bigl[ 
    \partial^{\beta')}_{\ \delta)} - \partial^{\beta)}_{\ \delta')}  
   + 2 t^{\beta)} \bigl( \partial_{\delta)t'} - \partial_{\delta')t}
   \bigr) - 2 t_{\delta)} \bigl(\partial^{\beta)}_{\ t'} 
           - \partial^{\beta')}_{\ t}\bigr) \Bigr] A 
+ \chi_{\gamma\delta} \bigl(\partial^\alpha + \partial^{\alpha'}
  \bigr) \bigl(\partial^\beta + \partial^{\beta'}\bigr) A  
\nonumber \\ & & \mbox{}
- 2 \chi^{(\alpha}_{\ (\gamma} \bigl( \partial^{\beta)} +
    \partial^{\beta')} \bigr) \bigl(\partial_{\delta)} +
    \partial_{\delta')} \bigr) A 
+ \chi^{\alpha\beta} \bigl( \partial_\gamma + \partial_{\gamma'}
  \bigr) \bigl(\partial_\delta + \partial_{\delta'} \bigr) A    
+ 2 \chi^{(\alpha}_{\ (\gamma} \delta^{\beta)}_{\ \delta)} 
  \Delta \Phi G^{\rm flat}
\nonumber \\ & & \mbox{}
+  \Bigl( 2 \chi^{(\alpha}_{\ (\gamma} \delta^{\beta)}_{\ \delta)}  
  - \eta^{\alpha\beta} \chi_{\gamma\delta} 
  - 2 \delta^{(\alpha}_{\ \gamma} 
      \delta^{\beta)}_{\ \delta} 
  - 4 t^\alpha t^\beta \eta_{\gamma\delta} \Bigr) B,  
\label{3.27}
\end{eqnarray} 
where $\Delta \Phi = \Phi(x) - \Phi(x')$. Equations (\ref{3.25}) and
(\ref{3.27}) give ${\cal G}^{\alpha\beta}_{\ \ \gamma'\delta'}(x,x')$, 
the trace-reversed gravitational Green's function, to first order in
$\Phi$.  

From Eq.~(\ref{3.27}) we extract the following components: 
\begin{eqnarray}
\dot{\cal G}^{tt}_{\ \ t't'} &=& - 2 \Delta \Phi G^{\rm flat} 
- 2 \partial_{tt'} A + B, \nonumber \\
\dot{\cal G}^{tt}_{\ \ t'a'} &=& (\partial_{t'a} - \partial_{ta'}) A, 
\nonumber \\
\dot{\cal G}^{tt}_{\ \ a'b'} &=& (\partial_a + \partial_{a'}) 
(\partial_b + \partial_{b'}) A - 3 \delta_{ab} B, \nonumber \\
& & \nonumber \\
\dot{\cal G}^{ta}_{\ \ t't'} &=& \bigl(\partial^a_{\ t'} -
    \partial^{a'}_{\ t}\bigr) A, \nonumber \\
\dot{\cal G}^{ta}_{\ \ t'b'} &=& -\delta^a_{\ b}( \partial_{tt'} A  
  + B) + \frac{1}{2} \bigl(\partial^a_{\ b} + 2\partial^{a'}_{\ b}  
  + \partial^{a'}_{\ b'} \bigr) A,
\label{3.28} \\
\dot{\cal G}^{ta}_{\ \ b'c'} &=& \delta^a_{\ (b} 
\bigl( \partial_{c)t'} - \partial_{c')t} \bigr) A, \nonumber \\
& & \nonumber \\
\dot{\cal G}^{ab}_{\ \ t't'} &=& \bigl( \partial^a + \partial^{a'}
\bigr) \bigl(\partial^b + \partial^{b'} \bigr) A 
- \delta^{ab} B, \nonumber \\
\dot{\cal G}^{ab}_{\ \ t'c'} &=& \delta^{(a}_{\ c} 
\bigl(\partial^{b)}_{\ t'} - \partial^{b')}_{\ t} \bigr) A, 
\nonumber \\  
\dot{\cal G}^{ab}_{\ \ c'd'} &=& 2 \delta^{(a}_{\ c} 
   \delta^{b)}_{\ d} \bigl( \Delta \Phi G^{\rm flat} 
- \partial_{tt'} A \bigr) 
+ \delta^{ab} \bigl(\partial_c + \partial_{c'}
\bigr) \bigl(\partial_d + \partial_{d'} \bigr) A 
- 2\delta^{(a}_{\ (c} \Bigl( \partial^{b)}_{\ d)} 
+ 2 \partial^{b)}_{\ d')} + \partial^{b')}_{\ d')} \Bigr) A
\nonumber \\ & & \mbox{} 
 + \delta_{cd} \bigl(\partial^a + \partial^{a'}\bigr)
\bigl(\partial^b + \partial^{b'}\bigr) A 
- \delta^{ab} \delta_{cd} B. \nonumber
\end{eqnarray} 

To ``trace-reverse'' the Green's function according to
Eq.~(\ref{3.23}) is a straightforward operation, but it is important 
to keep in mind that the metric involved in this computation is not
$\eta^{\alpha\beta}$, but $\eta^{\alpha\beta} + 2\Phi
\chi^{\alpha\beta}$. We find that the gravitational Green's function
is given by 
\begin{equation} 
G^{\alpha\beta}_{\ \ \gamma'\delta'}(x,x') = \Bigl( 
\delta^{(\alpha}_{\ \gamma'} \delta^{\beta)}_{\ \delta'} 
- \frac{1}{2} \eta^{\alpha\beta} \eta_{\gamma'\delta'} \Bigr)  
G^{\rm flat}(x,x') 
+ \dot{G}^{\alpha\beta}_{\ \ \gamma'\delta'}(x,x') + O(\Phi)^2, 
\label{3.29}
\end{equation}
where 
\begin{equation}
\dot{G}^{\alpha\beta}_{\ \ \gamma'\delta'}(x,x') = 
\dot{\cal G}^{\alpha\beta}_{\ \ \gamma'\delta'}(x,x')  
- \frac{1}{2} \eta^{\alpha\beta} \eta_{\mu\nu} 
\dot{\cal G}^{\mu\nu}_{\ \ \gamma'\delta'}(x,x')  
- 2\bigl( t^\alpha t^\beta \eta_{\gamma'\delta'} 
- \eta^{\alpha\beta} t_{\gamma'} t_{\delta'} \bigr) 
\Phi(x) G^{\rm flat}(x,x'). 
\label{3.30}
\end{equation}
The complete listing of components is 
\begin{eqnarray}
\dot{G}^{tt}_{\ \ t't'} &=& - (\Delta \Phi G^{\rm flat} 
+ \partial_{tt'} A), \nonumber \\
\dot{G}^{tt}_{\ \ t'a'} &=& (\partial_{t'a} - \partial_{ta'}) A, 
\nonumber \\
\dot{G}^{tt}_{\ \ a'b'} &=& -\delta_{ab} \bigl[ 
\langle \Phi \rangle G^{\rm flat} 
+ \partial_{tt'} A + 2B \bigr] 
+ (\partial_a + \partial_{a'}) (\partial_b + \partial_{b'}) A, 
\nonumber \\
& & \nonumber \\
\dot{G}^{ta}_{\ \ t't'} &=& \bigl(\partial^a_{\ t'} -
    \partial^{a'}_{\ t} \bigr) A, \nonumber \\
\dot{G}^{ta}_{\ \ t'b'} &=& -\delta^a_{\ b}( \partial_{tt'} A  
  + B) + \frac{1}{2} \bigl(\partial^a_{\ b} 
  + 2\partial^{a'}_{\ b} + \partial^{a'}_{\ b'} \bigr) A, 
\label{3.31} \\
\dot{G}^{ta}_{\ \ b'c'} &=& \delta^a_{\ (b} \bigl(\partial_{c)t'} 
  - \partial_{c')t} \bigr) A, \nonumber \\
& & \nonumber \\
\dot{G}^{ab}_{\ \ t't'} &=& \delta^{ab} \bigl[ 
\langle \Phi \rangle G^{\rm flat} - \partial_{tt'} A \bigr] 
+ \bigl(\partial^a + \partial^{a'} \bigr)
  \bigl(\partial_b + \partial_{b'} \bigr) A, 
\nonumber \\
\dot{G}^{ab}_{\ \ t'c'} &=& \delta^{(a}_{\ c} 
\bigl(\partial^{b)}_{\ t'} - \partial^{b')}_{\ t} \bigr) A, 
\nonumber \\  
\dot{G}^{ab}_{\ \ c'd'} &=& \bigl( 2\delta^{(a}_{\ c} 
   \delta^{b)}_{\ d} - \delta^{ab} \delta_{cd} \bigr) 
   \bigl(\Delta \Phi G^{\rm flat} - \partial_{tt'} A \bigr) 
+ \delta^{ab}(\partial_c + \partial_{c'})(\partial_d 
+ \partial_{d'}) A  
\nonumber \\ & & \mbox{} 
- 2\delta^{(a}_{\ (c} \Bigl( \partial^{b)}_{\ d)} 
+ 2 \partial^{b)}_{\ d')} + \partial^{b')}_{\ d')} \Bigr) A 
+ \delta_{cd} \bigl(\partial^a + \partial^{a'} \bigr)
  \bigl(\partial^b + \partial^{b'}\bigr)A 
- 2 \delta^{ab} \delta_{cd} B. \nonumber    
\end{eqnarray} 
Here we use the notation $\langle \Phi \rangle \equiv \Phi(x) +
\Phi(x')$, as well as $\Delta \Phi = \Phi(x) - \Phi(x')$. It should be
noted that in Eq.~(\ref{3.31}), the part of 
$\dot{G}^{\alpha\beta}_{\ \ \gamma'\delta'}(x,x')$ that involves 
$G^{\rm flat}(x,x')$ has support only on the past light cone 
of the point $x$. The remaining, tail part has support inside the
light cone as well. We believe that these results for the
gravitational Green's function are new.     

\section{Evaluation of two-point functions} 

The scalar, electromagnetic, and gravitational Green's functions can
all be expressed in terms of $A(x,x')$ and $B(x,x')$, the two-point
functions introduced in Sec.~III A. Our task in this section is to
evaluate these two-point functions. We will keep $\Phi(x)$ completely
generic, demanding only that far from the matter distribution, it
becomes equal to $-M/r$, where $r = |\bbox{x}|$ and $M$ is the total
mass. This distinguishes our work from the earlier work of DeWitt and
DeWitt \cite{44}, who considered only the special case $\Phi =
-M/r$. A consequence of keeping $\Phi$ generic is that we will obtain
only partial information regarding the two-point functions. This,
however, will be sufficient for the computation of the self-forces
presented in Sec.~V. The results specifically required for this
computation are derived in this section. Additional results are
presented in Appendix A, which gives an extended discussion of the
two-point function $A(x,x')$. Complete expressions for the two-point
functions are derived in Appendix B for the special case $\Phi =
-M/r$; these reproduce the earlier results of DeWitt and DeWitt.   

\subsection{Ellipsoidal coordinates} 

When Eq.~(\ref{3.3}) is substituted into Eqs.~(\ref{3.1}) and
(\ref{3.2}), we find that the $dt''$ integration can be carried out
immediately, and we obtain 
\begin{equation}
A(x,x') = \frac{1}{2\pi} \int 
\frac{\Phi(\bbox{x''})}{|\bbox{x}-\bbox{x''}| 
|\bbox{x''}-\bbox{x'}|}\, \delta \bigl(\Delta t -
|\bbox{x}-\bbox{x''}| - |\bbox{x''}-\bbox{x'}| \bigr)\, d^3 x''
\label{4.1}
\end{equation} 
and 
\begin{equation}
B(x,x') = \int \frac{\rho(\bbox{x''})}{|\bbox{x}-\bbox{x''}|  
|\bbox{x''}-\bbox{x'}|}\, \delta \bigl(\Delta t -
|\bbox{x}-\bbox{x''}| - |\bbox{x''}-\bbox{x'}| \bigr)\, d^3 x'', 
\label{4.2}
\end{equation} 
where $\Delta t \equiv t - t'$. These are the integrals we will
attempt to evaluate. The method of calculation presented below comes
from DeWitt and DeWitt \cite{44}, but this particular implementation
was suggested to us by Alan Wiseman \cite{47}.  
 
The $\delta$-function in Eqs.~(\ref{4.1}) and (\ref{4.2}) enforces the
condition $|\bbox{x}-\bbox{x''}| + |\bbox{x''}-\bbox{x'}| = \Delta t$,
which defines the closed, two-dimensional surface $\cal S$ formed by
the intersection of $x$'s past light cone with $x'$'s future light
cone. This surface is an ellipsoid of revolution centered at
$\bbox{x}_0 \equiv \frac{1}{2}(\bbox{x}+\bbox{x'})$, of semi-major 
axis $\frac{1}{2}\Delta t$ and ellipticity $\frac{1}{2}|\bbox{x} -
\bbox{x'}|$. To integrate Eqs.~(\ref{4.1}) and (\ref{4.2}), we will
adopt ellipsoidal coordinates adapted to this geometry.  

We first summarize our notation: 
\begin{equation}
\Delta t = t - t', \qquad
\bbox{x}_0 = \frac{1}{2} (\bbox{x} + \bbox{x'}), \qquad
\bbox{R} = \bbox{x} - \bbox{x'}, \qquad 
R = |\bbox{x} - \bbox{x'}|, \qquad  
\bbox{\hat{n}} = \frac{\bbox{R}}{R}, \qquad 
e = \frac{1}{2} R. 
\end{equation} 
The diagram of Fig.~1 illustrates the situation. 

\begin{figure}[t]
\special{hscale=40 vscale=40 hoffset=130.0 voffset=-270.0
         angle=0 psfile=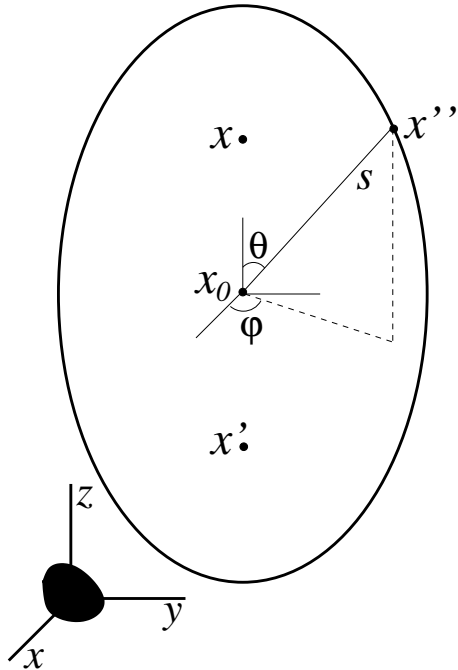}
\vspace*{3.5in}
\caption{The diagram shows the matter distribution located near the
origin of the Cartesian coordinate system, the points $\bbox{x}$,
$\bbox{x'}$, and $\bbox{x}_0$, to which the ellipsoidal coordinates
$(s,\theta,\phi)$ are attached. The diagram also shows one
representative ellipsoid of constant $s$, and a point $\bbox{x''} =
\bbox{x}_0 + \bbox{\eta}(s,\theta,\phi)$ lying on this ellipsoid.} 
\end{figure} 

For convenience we orient our coordinate axes such that the unit
vector $\bbox{\hat{n}}$ points in the same direction as the $z$, or
third, axis; there is no loss of generality involved in this
choice. We then represent the vector $\bbox{x''}$ in terms of
ellipsoidal coordinates $(s,\theta,\phi)$. The coordinate
transformation is  
\begin{equation}
\bbox{x''} = \bbox{x}_0 + \bbox{\eta}(s,\theta,\phi),  
\label{4.4}
\end{equation}
where 
\begin{equation}
\eta^1 = \sqrt{s^2-e^2} \sin\theta \cos\phi, \qquad
\eta^2 = \sqrt{s^2-e^2} \sin\theta \sin\phi, \qquad
\eta^3 = s \cos\theta. 
\label{4.5}
\end{equation} 
The parameter $e \equiv \frac{1}{2} R$ is the ellipticity of the new
coordinate system. It is easy to check that in these coordinates, 
$|\bbox{x}-\bbox{x''}| = s-e\cos\theta$ and 
$|\bbox{x''}-\bbox{x'}| = s+e\cos\theta$. The Jacobian of the 
transformation (\ref{4.4}) is $(s^2-e^2 \cos^2\theta) \sin\theta$. 

With these results, we find that Eq.~(\ref{4.1}) reduces to $A(x,x') = 
(2\pi)^{-1} \int \Phi(\bbox{x}_0 + \bbox{\eta}) \delta(\Delta t - 2s)\,
ds d\Omega$, where $d\Omega = \sin\theta\, d\theta d\phi$ is the
element of solid angle. The integration over $s$ is immediate, and we
obtain  
\begin{equation}
A(x,x') = \frac{1}{4\pi} \int \Phi\bigl(\bbox{x}_0 +
\bbox{\eta}(\theta,\phi) \bigr)\, d\Omega, 
\label{4.6}
\end{equation}
where the vector $\bbox{\eta}(\theta,\phi)$ is now given by
Eq.~(\ref{4.5}) with $s$ set equal to $\frac{1}{2} \Delta t$. Thus,
$A(x,x')$ is the average of $\Phi$ over the ellipsoid $s = \frac{1}{2}
\Delta t$, which is just the two-surface $\cal S$ introduced
previously. Similarly,  
\begin{equation}
B(x,x') = \frac{1}{2} \int \rho\bigl(\bbox{x}_0 +
\bbox{\eta}(\theta,\phi) \bigr)\, d\Omega 
\label{4.7}
\end{equation}
is the average of $2\pi \rho$ over the same ellipsoid. It is important
to remember that in Eqs.~(\ref{4.5})--(\ref{4.7}), the third
coordinate axis has been aligned with the vector $\bbox{R} \equiv
\bbox{x} - \bbox{x'}$.   

It should be noted that $A(x,x')$ and $B(x,x')$ are nonzero if and
only if $x$ is in the causal future of $x'$. To properly reflect this,
the integrals of Eqs.~(\ref{4.6}) and (\ref{4.7}) should be multiplied
by the step function $\theta(\Delta t - R)$. To keep the notation
simple, however, we choose to leave this factor implicit in our
expressions; for the rest of this section, $\Delta t$ will be
understood to always be larger than $R$.         

\subsection{$A(x,x')$ near coincidence}   

The integral of Eq.~(\ref{4.6}) can readily be evaluated, for an
arbitrary Newtonian potential $\Phi$, if the points $x$ and $x'$ are
close together in spacetime. This is the near-coincidence
approximation. Because these points must be timelike related for
$A(x,x')$ to be nonvanishing, we are looking at evaluating
Eq.~(\ref{4.6}) in a limit where $R$ and $\Delta t$ are both small,
but the ratio $\Delta t/R$ is maintained larger than unity. In this
limit, the vector $\bbox{\eta}(\theta,\phi)$ defined in Eq.~(\ref{4.5})
is small, and $\Phi$ can be expanded in Taylor series about the point
$\bbox{x}_0$; this is the basis for our approximation. For
concreteness, we will assume that $\bbox{x}_0$ lies outside the matter
distribution, so that $\rho(\bbox{x}_0) = 0$. 

Expanding $\Phi$ in powers of $\bbox{\eta}$ in Eq.~(\ref{4.6}) gives 
\begin{equation}
4\pi A = \Phi(\bbox{x_0}) \int d\Omega   
+ \Phi_{,a}(\bbox{x_0}) \int \eta^a\, d\Omega  
+ \frac{1}{2} \Phi_{,ab}(\bbox{x_0}) \int \eta^a \eta^b\, d\Omega  
+ \cdots.
\end{equation}
The first integral, with a factor of $(4\pi)^{-1}$ in
front, gives unity. The average of $\eta^a$ on the sphere is zero, and
the average of $\eta^a \eta^b$ is the tensor $q^{ab} = 
\frac{1}{3} (s^2 - e^2) \delta^{ab} + \frac{1}{3} e^2 \hat{n}^a
\hat{n}^b$, where $s=\frac{1}{2} \Delta t$ and $e=\frac{1}{2} R$. The
part of $q^{ab}$ that is proportional to $\delta^{ab}$ multiplies
$\Phi_{,ab}(\bbox{x}_0)$ and returns something proportional to
$\nabla^2 \Phi$ evaluated at $\bbox{x}_0$; because this point is 
outside the matter, this is zero. We are left with  
\begin{equation}
A = \Phi(\bbox{x}_0) + \frac{1}{24}\, R^2 
\Phi_{,ab}(\bbox{x_0}) \hat{n}^a \hat{n}^b + \cdots. 
\label{4.8}
\end{equation}
Notice that the right-hand side of this equation no longer depends on
$\Delta t$: the two-point function $A(x,x')$ is time-independent near 
coincidence.      

It is convenient to shift the reference point from $\bbox{x}_0$ to
$\bbox{x}$. To order $R^2$, this means re-expressing
$\Phi(\bbox{x}_0)$ as 
\begin{equation}
\Phi(\bbox{x} - {\textstyle \frac{1}{2}} R \bbox{\hat{n}}) =  
\Phi(\bbox{x}) - \frac{1}{2}\, R \Phi_{,a}(\bbox{x}) \hat{n}^a 
+ \frac{1}{8}\, R^2 \Phi_{,ab}(\bbox{x}) \hat{n}^a \hat{n}^b + \cdots.  
\label{4.9}
\end{equation} 
Substituting this into Eq.~(\ref{4.8}), we arrive at 
\begin{equation}
A(x,x') = \Phi(\bbox{x}) - \frac{1}{2}\, \Phi_{,a}(\bbox{x}) R^a 
+ \frac{1}{6}\, \Phi_{,ab}(\bbox{x}) R^a R^b + O(R^3). 
\label{4.10}
\end{equation} 
This is our final expression for the two-point function in the
near-coincidence approximation. We recall that $\bbox{R} = 
\bbox{x} - \bbox{x'}$.  

In the sequel we will need expressions for various derivatives of
$A(x,x')$ evaluated at coincidence, $x = x'$. It is straightforward to
differentiate Eq.~(\ref{4.10}) with respect to either $x$ or $x'$, to
obtain    
\begin{equation} 
\lim_{x' \to x} \partial_t A = \lim_{x' \to x} \partial_{t'} A = 0
\label{4.11}
\end{equation}
and 
\begin{equation}
\lim_{x' \to x} \partial_{ab} A = 2 \lim_{x' \to x} \partial_{ab'} A 
= \lim_{x' \to x} \partial_{a'b'} A = \frac{1}{3}\,
\Phi_{,ab}(\bbox{x}).  
\label{4.12}
\end{equation} 
These results hold for arbitrary Newtonian potentials, provided that
the point $\bbox{x}_0$ lies outside the matter distribution. 

\subsection{$A(x,x')$ for long delays} 

We can also evaluate $A(x,x')$ for arbitrary potentials if we assume
that $x$ and $x'$ are widely separated in spacetime. More precisely,
we now take $\Delta t$ to be extremely large, and in particular,
much larger than $R = |\bbox{x}-\bbox{x'}|$, $r=|\bbox{x}|$, and 
$r' = |\bbox{x'}|$. This is the long-delay approximation. In this
limit, $s$ is much larger than $e$, and $\bbox{\eta} \simeq 
(s\sin\theta\cos\phi, s\sin\theta\sin\phi, s\cos\theta)$. Because $s$
is very large, we will neglect $\bbox{x}_0$ in front of $\bbox{\eta}$
in Eq.~(\ref{4.6}), so that $A \simeq (4\pi)^{-1} \int
\Phi(\bbox{\eta})\, d\Omega$. Since $\bbox{\eta}$ is located well
outside the matter distribution, we can approximate the Newtonian
potential by $-M/|\bbox{\eta}|$, its first term in a multipole
expansion. (Here, $M = \int \rho\, d^3x$ is the total mass of the
matter distribution.) Because $|\bbox{\eta}| \simeq s = \frac{1}{2}
\Delta t$ in the long-delay limit, we arrive at 
\begin{equation} 
A(x,x') \simeq - \frac{2M}{\Delta t}. 
\label{4.13}
\end{equation} 
We recall that $\Delta t = t - t'$. Equation (\ref{4.13}) shows that
$A(x,x')$ and all its derivatives vanish in the limit 
$t' \to -\infty$.     

\subsection{$B(x,x')$ at large distances} 

We will evaluate $B(x,x')$ in the case where $x$ and $x'$ are both 
located well outside the matter distribution. For this calculation it
is easier to go back to Eq.~(\ref{4.2}), in which we assume that
$r \equiv |\bbox{x}|$ and $r' \equiv |\bbox{x'}|$ are both very
large. (A more precise criterion will be given below.) These
conditions will allow us to develop a multipole expansion for this
two-point function.  

Equations (\ref{4.2}) and (\ref{4.7}) indicate that $B(x,x')$ is the
average of the mass density $\rho$ over the ellipsoid $s = 
\frac{1}{2} \Delta t$. The diagram of Fig.~1 makes it clear that
unless $\Delta t$ falls within the appropriate interval, the ellipsoid 
will fail to intersect the matter distribution. Thus, $B(x,x')$ is
nonzero only when $\Delta t$ lies within this interval, which we might   
call the intersection window. When $r$ and $r'$ are both large, the
ellipsoid is also very large, and the intersection window becomes
comparatively short. In this limit, $B(x,x')$ is ``on'' for a very
short time, and its behavior suggests that of a
$\delta$-function. This expectation is borne out by an explicit
calculation. 

In the denominator of Eq.~(\ref{4.2}), we approximate
$|\bbox{x}-\bbox{x''}|$ by $r$ and $|\bbox{x''}-\bbox{x'}|$ by $r'$,
taking $|\bbox{x''}|$ to be much smaller than both $r$ and $r'$. In
the $\delta$-function we go to the next order of approximation and
write $|\bbox{x}-\bbox{x''}| \simeq r - \bbox{x} \cdot \bbox{x''}/r$
and $|\bbox{x''}-\bbox{x'}| \simeq r' - \bbox{x'} \cdot
\bbox{x''}/r'$. If we introduce the vector 
\begin{equation}
\bbox{m} = \frac{\bbox{x}}{r} + \frac{\bbox{x'}}{r'}, 
\label{4.14}
\end{equation}
then the $\delta$-function becomes $\delta(u + \bbox{m} \cdot
\bbox{x''})$, where $u \equiv \Delta t - r - r'$. This we expand in
powers of $\bbox{x''}$ and express as $\delta(u) + \delta'(u)
(\bbox{m} \cdot \bbox{x''}) + \frac{1}{2} \delta''(u) (\bbox{m} \cdot
\bbox{x''})^2 + \cdots$, where a prime denotes differentiation with
respect to the argument. Substituting all this inside the integral of 
Eq.~(\ref{4.2}), we obtain 
\begin{equation} 
B \simeq \frac{1}{rr'} \biggl[ \delta(u) 
\int \rho(\bbox{x''})\, d^3x'' + \delta'(u) m_a 
\int \rho(\bbox{x''}) x^{a''}\, d^3 x'' + \frac{1}{2} 
\delta''(u) m_a m_b 
\int \rho(\bbox{x''}) x^{a''} x^{b''} \, d^3 x'' + \cdots \biggr].  
\label{4.15}
\end{equation} 
The first integral gives the total mass $M$. The second integral gives
the dipole moment of the mass distribution; this vanishes if, as we
assume, the origin of the coordinate system is attached to the center
of mass. The third integral is $Q^{ab}$, the quadrupole moment of the
mass distribution. We have therefore obtained 
\begin{equation}
B(x,x') \simeq \frac{1}{rr'} \biggl[ M \delta(u) 
+ \frac{1}{2} Q_{ab} m^a m^b \delta''(u) + \cdots
\biggr],  
\label{4.16}
\end{equation}
where, we recall, $u = \Delta t - r - r'$. This equation displays the 
first two terms in the multipole expansion of the two-point function
$B(x,x')$. Notice that this expansion is analogous to a multipole
expansion of a radiative field in the wave zone; this is not a
near-zone expansion of a quasi-stationary field. 

In the sequel we will approximate $B(x,x')$ by 
\begin{equation}
B(x,x') = \frac{M}{rr'}\, \delta(\Delta t - r - r'), 
\label{4.17}
\end{equation}
its leading term in the multipole expansion. In this
approximation, $B(x,x')$ is nonzero only when $\Delta t = r + r'$. The
time delay corresponds to a signal propagating at the speed of light
from $\bbox{x'}$ to the center of the mass distribution, and then on
to $\bbox{x}$. The approximation has made the intersection window 
infinitely short.  

To estimate the error involved in going from Eq.~(\ref{4.16}) to
Eq.~(\ref{4.17}), we integrate $B(x,x')$ over a smooth test function
$f(t)$. Setting $t'=0$ for simplicity, Eq.~(\ref{4.16}) gives 
\begin{equation}
\int B f(t)\, dt \simeq \frac{M}{rr'} f(r+r') \biggr( 1 
+ \frac{Q}{M} \frac{f''}{f} + \cdots \biggr), 
\label{4.18}
\end{equation}
where $Q \equiv \frac{1}{2} Q_{ab} m^a m^b$. The factor in front the
large brackets is what would be returned by the approximation of
Eq.~(\ref{4.17}). We wish to show that in typical situations, the
second term inside the brackets is much smaller than unity. Let $\ell$
be the characteristic size of the matter distribution, and let $\tau$
be the characteristic time over which the function $f(t)$
changes. Then $Q/M$ is of order $\ell^2$, and $f''/f$ is of order
$1/\tau^2$. The correction term is therefore of order
$(\ell/\tau)^2$. In a typical situation, $\tau$ would be associated
with the orbital time scale of an object located at $\bbox{x}$ or
$\bbox{x'}$. Assuming for concreteness that $r$ is the shortest
distance, we have $\tau^2 \sim r^3/M$. With this, we find that the
correction term in Eq.~(\ref{4.18}) is of the order of $(\ell/r)^2
(M/r)$. So long as $r$ and $r'$ are both much larger than $\ell$ and
$M$, Eq.~(\ref{4.17}) makes an excellent approximation.        

\section{Self-forces} 

\subsection{Scalar self-force} 

In the absence of an external force, a point particle with scalar
charge $q$ experiences a self-force \cite{41} 
\begin{equation}
f^\alpha(\tau) = q^2 \bigr[ g^{\alpha\beta} + u^\alpha(\tau)
u^\beta(\tau) \bigr] \int_{-\infty}^{\tau^-} G_{,\beta}(\tau,\tau')\,
d\tau',
\label{5.1}
\end{equation}
so that its equations of motion are $m u^\alpha_{\ ;\beta} u^\beta =
f^\alpha$. The particle's world line is represented by the relations
$z^\alpha(\tau)$, where $\tau$ is proper time, and $u^\alpha(\tau) = 
dz^\alpha/d\tau$. The self-force is evaluated at the current position
of the particle, and it involves an integration over its past
history. Inside the integral, the retarded Green's function is first
written as $G(x,z(\tau'))$, in terms of an arbitrary field point $x$ 
and the past position $z(\tau')$ of the particle. The Green's
function is then differentiated with respect to $x^\beta$, and the
result is evaluated at $x = z(\tau)$, the particle's current
position. This is what we have denoted by
$G_{,\beta}(\tau,\tau')$ in Eq.~(\ref{5.1}). The upper limit of
integration is $\tau^-\equiv \tau - \epsilon$, where $\epsilon$ is a
small, positive number that is taken to zero at the end of the
calculation. The purpose of this cutoff is to remove the singular,
$\delta$-function part of the Green's function which has support on
the past light cone of the point $z(\tau)$. What survives is the
``tail part'' of the Green's function, which has support inside the
light cone.  

This expression for the self-force was derived by Quinn \cite{41}, who
assumed that the scalar field was minimally coupled to curvature ---
he considered the special case $\xi = 0$. There is no difficulty,
however, in extending Quinn's result to arbitrary couplings to
curvature. So long as the particle moves in a region of spacetime in
which the Ricci scalar is zero, his results carry over directly. The
only effect that $\xi$ has on the motion is through the Green's
function, which is sensitive to a nonzero Ricci scalar 
{\it somewhere} in spacetime. 

We will calculate $f^\alpha$ to first order in the Newtonian potential
$\Phi$ in a weak-field approximation; the particle therefore moves in
the weakly curved spacetime of Sec.~II. We assume that the particle is
gravitationally bound to the matter distribution, and infer from the
virial theorem that $v^2$, the square of the particle's velocity, is
of the same order of magnitude as $\Phi$. We therefore add a
slow-motion assumption to our weak-field assumption. Accordingly, we
will neglect in all expressions terms involving $\Phi^2$, $\Phi v^2$,
and $v^4$. Within this approximation, we have that the spatial
components of the self-force are given by 
\begin{equation}
f^a = q^2 \int_{-\infty}^t (\dot{G}_{,a} + v^a \dot{G}_{,t})\, dt',  
\label{5.2}
\end{equation}
where $\bbox{v} = d\bbox{z}(t)/dt$ is the current velocity of the 
particle, which does not depend on $t'$; the past velocity will be
denoted $\bbox{v'} = d\bbox{z}(t')/dt'$, and this does depend on
$t'$. We have made $t'$ the new variable of integration,
and neglected all $O(v^2,vv',v^{\prime 2})$ terms inside the integral, 
which through the Green's function is already linear in the Newtonian
potential. This fact allows us to freely alter the position of all
latin indices, implicitly using $\delta_{ab}$, the metric of flat
space. And since we will substitute only the tail part of the Green's
function inside the integral, we have made $t$ the upper limit of
integration (instead of the redundant $t-\epsilon$).   

From Eq.~(\ref{3.14}) we find that the tail part of the scalar Green's 
function is given by 
\begin{equation} 
\dot{G}(x,z') = -2 \partial_{tt'} A(x,z') - 2\xi B(x,z').   
\label{5.3}
\end{equation}
Here $z' \equiv (t',\bbox{z}(t'))$, and $A(x,z')$, $B(x,z')$ are the
two-point functions introduced in Sec.~III A and computed in
Sec.~IV; we will set $x=z \equiv (t,\bbox{z}(t))$ after taking the 
derivatives. Substituting Eq.~(\ref{5.3}) into Eq.~(\ref{5.2}) gives
$f^a = f^a_A + f^a_B$, where 
\begin{equation}
f^a_A = -2 q^2 \int_{-\infty}^t ( A_{,tt'a} + v^a A_{,tt't} )\, dt' 
\label{5.4}
\end{equation}
and 
\begin{equation}
f^a_B = -2 \xi q^2 \int_{-\infty}^t ( B_{,a} + v^a B_{,t} )\, dt'.  
\label{5.5}
\end{equation} 
We will first work on $f^a_A$, and then move on to $f^a_B$. 

The first term inside the integral of Eq.~(\ref{5.4}) can be expressed
as $\partial_{t'} A_{,ta} = d A_{,ta} /dt' - A_{,tab'} v^{b'}$,
allowing part of the integration to be carried out. Similarly, we
write the second term inside the integral as $v^a \partial_{t'}
A_{,tt} = v^a d A_{,tt}/dt' + O(vv')$, and Eq.~(\ref{5.4}) becomes  
\begin{equation}
f^a_A = -2q^2 \Bigl[ A_{,ta} + v^a A_{,tt} \Bigr]^t_{-\infty} 
+ 2q^2 \int^t_{-\infty} A_{,tab'} v^{b'}\, dt'. 
\label{5.6}
\end{equation} 
We now use Eq.~(\ref{3.6}) and replace $A_{,tab'}$ by $-\partial_{t'}
A_{,ab'} = -d A_{,ab'} /dt' + O(v')$. After integrating by parts,
we obtain  
\begin{equation}
f^a_A = -2q^2 \Bigl[ A_{,ta} + v^a A_{,tt} + A_{,ab'} v^{b'}
\Bigr]^t_{-\infty} + 2q^2 \int^t_{-\infty} A_{,ab'} a^{b'}\, dt',  
\label{5.7}
\end{equation} 
where $\bbox{a'} = d\bbox{v'}/dt'$ is the past acceleration of
the particle. Because the particle is not subjected to any force except
for gravity, its acceleration is linear in the Newtonian potential 
$\Phi$. Since $A(x,z')$ is already linear in $\Phi$, the integral of
Eq.~(\ref{5.7}) is $O(\Phi^2)$ and it can be neglected. The first
part of the self-force can therefore be computed without having to
evaluate a single integral! And Eq.~(\ref{5.7}) shows that the
self-force depends on the conditions at the current time, and in the
infinite past. At this stage, the nonlocal character of the force has
(almost) disappeared.  

In Sec.~IV C we learned that as $t'$ approaches $-\infty$, $A(x,z')$  
behaves as $2M/t'$. This shows that in Eq.~(\ref{5.7}), there are
actually no contributions from the infinite past. Any nonvanishing
contribution must therefore come from the current time, $t' = t$, at
which the points $\bbox{z}(t')$ and $\bbox{z}(t)$ are
coincident. Equation (\ref{4.11}) tells us that the time derivative of 
$A(x,z')$ is zero at coincidence. Setting $v^{b'} = v^b$, we find that 
the sole surviving contribution to $f^a_A$ is $-2 q^2 A_{,ab'} v^b$,
in which the derivatives of the two-point function must be evaluated
at coincidence. This was worked out in Eq.~(\ref{4.12}), and we arrive
at   
\begin{equation}
f^a_A = -\frac{1}{3}\, q^2 \Phi_{,ab} v^b. 
\label{5.8}
\end{equation} 
We may express this in the alternative form 
\begin{equation}
\bbox{f}_{\! A} = \frac{1}{3}\, q^2 \frac{d \bbox{g}}{dt},  
\label{5.9}
\end{equation}    
where $\bbox{g} = -\bbox{\nabla} \Phi$ is the Newtonian gravitational
field. In components, the right-hand side of Eq.~(\ref{5.9}) involves
$dg^a/dt = -d\Phi_{,a}/dt = - \Phi_{,ab} v^b$, where we have used
the fact that the Newtonian potential does not explicitly depend on 
time.  

We now turn to the second part of the self-force, given by 
Eq.~(\ref{5.5}) in terms of the two-point function
$B(x,z')$. According to Eq.~(\ref{4.17}), this is 
\begin{equation}
B(x,z') = \frac{M}{rr'}\, \delta(u), 
\label{5.10}
\end{equation}
where $u = t-t'-r-r'$, $r = |\bbox{x}|$, and $r' =
|\bbox{z}(t')|$. As before, the field point $x = (t,\bbox{x})$ will be 
identified with $z \equiv (t,\bbox{z}(t))$ after differentiation. It is 
important to keep in mind that in Eq.~(\ref{5.10}), $r'$ depends on
$t'$. We will use the notation $v_r' = dr'/dt'$. 

We first show that in Eq.~(\ref{5.5}), the integral involving the time
derivative of $B(x,z')$ is negligible. According to Eq.~(\ref{5.10}),
$B_{,t} = M \delta'(u) / (rr')$, where a prime indicates
differentiation with respect to $u$. Changing the variable of
integration from $t'$ to $u$, using $du = -(1+v'_r)\, dt'$, we have
that $\int B_{,t}\, dt' = (M/r) \int (r')^{-1}[1 + O(v')]
\delta'(u)\, du$. Here the lower limit of integration is $-(r+r')$,
the upper limit is $\infty$, and we neglect terms of order $v'$
because the integral comes with a factor of $v^a$ in
front. Integration yields $M v_r'/(r r^{\prime 2})$ evaluated at 
$u=0$. We conclude that the second term of Eq.~(\ref{5.5}) is of
order $(q/r)^2 (M/r) v^2$ and therefore negligible.      

Differentiation of Eq.~(\ref{5.10}) with respect to the spatial
variables $x^a$ yields
\begin{equation}
B_{,a} = -\frac{M r_{,a}}{r^2 r'} \Bigl[ \delta(u) + r \delta'(u)
\Bigr], 
\label{5.11}
\end{equation}
where a prime still indicates differentiation with respect to
$u$. Changing variables of integration from $t'$ to $u$, we find that
the first integral of Eq.~(\ref{5.5}) becomes 
\begin{eqnarray} 
\int_{-\infty}^t B_{,a}\, dt' &=& -\frac{M r_{,a}}{r^2} 
\int_{-(r+r')}^\infty \frac{\delta(u) + r\delta'(u)}{r'(1+v_r')}\, du 
\nonumber \\ 
&=& -\frac{M r_{,a}}{r^2}\, \biggl[ \frac{1}{r'(1+v_r')} - r
\frac{d}{du}\, \frac{1}{r'(1+v_r')} \biggr] \biggr|_{u=0}  
\nonumber \\
&=& -\frac{M r_{,a}}{r^2 r'}\, \Bigl[ 1 - \bigl(1 + r/r'\bigr) v_r' +
O\bigl(v^{\prime 2},a'\bigr) \Bigr] \biggr|_{u=0},   
\label{5.12}
\end{eqnarray} 
where, as indicated, we neglect terms that are of the second order in
the velocity, and terms that involve the particle's acceleration.  

To evaluate the right-hand side of Eq.~(\ref{5.12}) is made difficult
by the fact that the relation $u = t-t'-r-r'(t')$ cannot be inverted
for $t'$, because the function $r'(t')$ is not known. For the purposes
of this calculation, however, it is sufficient to give an approximate
inversion. We let the radial part of the particle's trajectory be
described by the relation $r = R(t)$, so that $r' = R(t') 
= R(t-\Delta t)$. This we approximate by $R(t) - \dot{R}(t) \Delta t + 
\frac{1}{2} \ddot{R}(t) \Delta t^2 + \cdots$, where overdots indicate 
differentiation with respect to $t$. Neglecting the acceleration term,
we have $r' = r - v_r \Delta t + O(a)$, where $v_r \equiv \dot{R}(t)$.
Similarly, we obtain $v_r' \equiv \dot{R}(t') = v_r + O(a)$. When
$u=0$, we find that $\Delta t = 2r/(1+v_r) + O(a) = 2r(1-v_r) 
+ O(v^2,a)$, giving $r' = r(1-2 v_r) + O(v^2,a)$. Substituting this
into Eq.~(\ref{5.12}), we find
\begin{equation}
\int_{-\infty}^t B_{,a}\, dt' = - \frac{M r_{,a}}{r^3}\, 
\Bigl[ 1 + O(v^2,a) \Bigr]. 
\label{5.13}
\end{equation} 
Finally, going back to Eq.~(\ref{5.5}), we arrive at 
\begin{equation}
\bbox{f}_{\! B} = 2 \xi q^2 \frac{M}{r^3}\, \bbox{\hat{r}},  
\label{5.14}
\end{equation} 
where $\bbox{\hat{r}} = \bbox{x}/r$ is a unit vector pointing in the
radial direction; its components are equal to $x^a/r = r_{,a}$. 

The total self-force acting on a point particle with scalar charge $q$
is given by the sum of Eqs.~(\ref{5.9}) and (\ref{5.14}). This is  
\begin{equation}
\bbox{f} = 2 \xi q^2 \frac{M}{r^3}\, \bbox{\hat{r}}    
+ \frac{1}{3}\, q^2 \frac{d \bbox{g}}{dt}, 
\label{5.15}
\end{equation} 
where, we recall, $\bbox{g} = -\bbox{\nabla} \Phi$ is the Newtonian 
gravitational field. The equations of motion for the particle are then
$d^2\bbox{z}/dt^2 = \bbox{g} + \bbox{f}/m$, where $m$ is its
mass. The first term on the right-hand side of Eq.~(\ref{5.15}) is a
conservative correction to the local gravity. This is a repulsive
force if $\xi > 0$, an attractive force if $\xi < 0$, and this part of
the self-force disappears altogether if the scalar field is minimally
coupled to curvature. The second term is nonconservative; this is the
well-known expression for the radiation-reaction force experienced by
a scalar charge moving under the influence of an external force 
$m \bbox{g}$.   

\subsection{Electromagnetic self-force} 

The steps required to calculate the electromagnetic self-force are
very similar to those presented in the preceding subsection. Our
discussion here will therefore be brief. We employ the same
notation, and work within the same set of approximations.  

Our starting point is the DeWitt-Brehme expression for the self-force
experienced by a point particle with an electric charge $e$ \cite{36}:  
\begin{equation}
f^\alpha = - e^2 \int_{-\infty}^{\tau^-} \bigl( 
G^\alpha_{\ \gamma';\beta} - G_{\beta\gamma'}^{\ \ \ ;\alpha} \bigr)  
u^\beta u^{\gamma'}\, d\tau',  
\label{5.16}
\end{equation}
where $G^\alpha_{\ \beta'}(\tau,\tau')$ is the retarded Green's
function of the electromagnetic field, $u^\alpha =
dz^\alpha(\tau)/d\tau$ is the current four-velocity of the charged
particle, and $u^{\alpha'} = dz^\alpha(\tau')/d\tau'$ its past
four-velocity. With the approximations introduced in Sec.~V A, we find 
that the spatial components of the self-force are given by 
\begin{equation}
f^a = -e^2 \int_{-\infty}^t \Bigl[ (\dot{G}_{at',t} - \dot{G}_{tt',a})
+ (\dot{G}_{at',b} - \dot{G}_{bt',a}) v^b + (\dot{G}_{ab',t} 
- \dot{G}_{tb',a}) v^{b'} \Bigr]\, dt';  
\label{5.17}
\end{equation}
only the tail part of the Green's function appears under the
integral. Using Eq.~(\ref{3.20}), we find that the self-force can
again be written as a sum of two parts, $\bbox{f} = \bbox{f}_{\! A} 
+ \bbox{f}_{\! B}$, where  
\begin{equation}   
f^a_A = e^2 \int_{-\infty}^t \Bigl[ (A_{,tt'a} + A_{,tta'}) +
(A_{,ta'b} - A_{,tab'}) v^b + (2\delta_{ab} A_{,tt't} - A_{,ta'b} +
2A_{,tab'} - A_{,t'ab}) v^{b'} \Bigr]\, dt' 
\label{5.18}
\end{equation}
and 
\begin{equation}
f^a_B = - e^2 \int_{-\infty}^t \bigl( B_{,a} - B_{,t} v^{a'} 
\bigr)\, dt'.  
\label{5.19}
\end{equation} 
The evaluation of $\bbox{f}_{\! B}$ proceeds as in Sec.~V A, using
Eq.~(\ref{5.13}) and the fact that the second term does not contribute
to the force at this level of approximation. The result is $f^a_B =
 e^2 (M/r^3) r_{,a}$. The evaluation of $\bbox{f}_{\! A}$ also
involves familiar steps, which we now describe.   

The first group of terms inside the integral of Eq.~(\ref{5.18}) are 
expressed as 
\begin{equation}
\partial_{t'} \bigl(A_{,ta} - A_{,ta'} \bigr) = 
\frac{d}{dt'} \bigl( A_{,ta} - A_{,ta'} \bigr) 
- \bigl( A_{,tab'} - A_{,ta'b'} \bigr) v^{b'}. 
\end{equation}
The first term is
immediately integrated, and by virtue of Eq.~(\ref{4.11}), this gives
no contribution to the self-force. The remaining terms are absorbed
into the third group of terms in Eq.~(\ref{5.18}), those which are
proportional to $v^{b'}$. In the second group of terms, the quantity
in front of $v^b$ is expressed as 
\begin{equation}
-\partial_{t'} \bigl( A_{,a'b} - A_{,ab'} \bigr)
= -\frac{d}{dt'} \bigl( A_{,a'b} - A_{,ab'} \bigr) + O(v'). 
\end{equation}
This is also immediately
integrated. Equation (\ref{4.12}) shows that this also gives no
contribution to the self-force. We are therefore left with the group
of terms proportional to $v^{b'}$. This takes the form of 
$(d C_{ab}/dt') v^{b'}$, where 
\begin{equation}
C_{ab} = 2\delta_{ab} A_{,tt} 
- A_{,ab} - A_{,ab'} + A_{,a'b} - A_{,a'b'}.
\end{equation}
The time derivative is
then transferred to $v^{b'}$ by integration by parts. This produces an
integral that can be neglected, and boundary terms at $t' = t$ (the 
boundary terms at $t' = -\infty$ all vanish). Those give 
$C_{ab} v^{b}$ evaluated at coincidence. Using Eqs.~(\ref{4.11}) and
(\ref{4.12}) we find that Eq.~(\ref{5.18}) becomes 
$f^a_A = -\frac{2}{3} e^2 \Phi_{,ab} v^b$.   

We have found that the self-force acting on a point particle with
electric charge $e$ is given by 
\begin{equation}
\bbox{f} = e^2 \frac{M}{r^3}\, \bbox{\hat{r}}    
+ \frac{2}{3}\, e^2 \frac{d \bbox{g}}{dt},   
\label{5.20}
\end{equation} 
where $\bbox{\hat{r}} = \bbox{x}/r$ and $\bbox{g} = -\bbox{\nabla}
\Phi$ is the Newtonian gravitational field. The first term on the
right-hand side of Eq.~(\ref{5.20}) represents a repulsive,
conservative force. The second term gives the well-known expression
for the radiation-reaction force experienced by an electric charge
moving under the influence of an external force $m \bbox{g}$
\cite{38,39}.   

Equation (\ref{5.20}) implies that the force required to keep an 
electric charge static in the gravitational field of an isolated
object of mass $M$ must be smaller than $mg$ by $e^2 M/r^3$. Our
derivation indicates that this correction is caused by an interaction
between the electromagnetic field and the matter distribution; this
interpretation is suggested by the fact that this contribution to the
self-force can be traced back to the two-point function $B(x,x')$,
which is directly associated with $\rho$, the mass density. While this
is a valid interpretation, it is interesting to note that the
correction persists even in the total absence of matter, although this
cannot be revealed in a weak-field approach. The force required to
hold an electric charge in place in the exact field of a Schwarzschild
black hole was calculated by Smith and Will \cite{45}. In the
weak-field limit, these authors recover the conservative term in
Eq.~(\ref{5.20}). In this case, clearly, the effect has nothing to do
with the presence of matter. It must instead be attributed to an
interaction between the electromagnetic field and the hole's event
horizon. That the conservative self-force should be so insensitive to
the source of the gravitational field is quite interesting. This
robustness, however, doesn't seem to apply to scalar charges: The
extra force required to keep a scalar charge in place in a
Schwarzschild spacetime is always zero, irrespective of the coupling
to curvature \cite{42,47,48,52,53}. This is quite different from what
was obtained in Sec.~V A, where we found that in the presence of
matter, there is a nonzero self-force for nonminimal coupling.       

\subsection{Gravitational self-force} 

The self-force experienced by a point particle of mass $m$ moving in
a curved spacetime is given by \cite{24,25} 
\begin{equation}
f^\alpha = -2m^2 \int_{-\infty}^{\tau^-} \Bigl( 
2 G^\alpha_{\ \beta\mu'\nu';\gamma} 
- G_{\beta\gamma\mu'\nu'}^{\ \ \ \ \ \ \ ;\alpha} 
+ u^\alpha G_{\beta\gamma\mu'\nu';\delta} u^\delta \Bigr) 
u^\beta u^\gamma u^{\mu'} u^{\nu'}\, d\tau', 
\label{5.21}
\end{equation}
where $G^{\alpha\beta}_{\ \ \mu'\nu'}(\tau,\tau')$ is the
gravitational Green's function, $u^\alpha$ the current four-velocity
of the particle, and $u^{\mu'}$ its past four-velocity. This result
was derived for the first time by Mino, Sasaki, and Tanaka \cite{24},
although an incomplete attempt was made earlier by Morette-DeWitt and
Ging \cite{54}. The gravitational self-force was later recovered by
Quinn and Wald \cite{25}, who gave a more transparent
derivation. Their expression is technically wrong, however, because it
incorrectly puts  
${\cal G}^{\alpha\beta}_{\ \ \mu'\nu'}$, the trace-reversed Green's
function (see Sec.~III D), in place of $G^{\alpha\beta}_{\ \ \mu'\nu'}$ 
in Eq.~(\ref{5.21}); this slight oversight hardly diminishes the value
of their paper. 

Our expression for the gravitational self-force differs from Mino 
{\it et al.} \cite{24} in two ways. First, we use a different
normalization for the Green's function, and Eq.~(\ref{5.21}) is valid
in the normalization imposed by Eq.~(\ref{3.24}). Second, and more
importantly, both Mino {\it et al.} and Quinn \& Wald \cite{25} assume
that the background metric $g_{\alpha\beta}$ is a solution to the
Einstein field equations {\it in vacuum}, a condition that is
incompatible with the weak-field approximation adopted in this 
paper. [Please refer back to the discussion presented in the
paragraphs before Eq.~(\ref{1.13}).] As we have already indicated, the 
extension of the Mino-Sasaki-Tanaka-Quinn-Wald equations of motion to
spacetimes that contain matter is not entirely trivial. In this
subsection we consider only the straightforward modifications to the
Green's function that come from the presence of Ricci-tensor terms in
Eq.~(\ref{2.13}). In other words, we still define the gravitational
self-force by Eq.~(\ref{5.21}), but we substitute in it the Green's
function computed in Sec.~III D to account for the presence of
matter. In Sec.~VI we shall prove that this substitution is
appropriate, and derive additional (matter-mediated) corrections to
the equations of motion.   

Under the weak-field, slow-motion approximation, we find that the
spatial components of the self-force are given by 
\begin{eqnarray}
f^a &=& -2m^2 \int_{-\infty}^t \Bigl[ (2\dot{G}_{att't',t} 
- \dot{G}_{ttt't',a}) + (\delta_{ab} \dot{G}_{ttt't',t} 
+ 2 \dot{G}_{abt't',t} + 2 \dot{G}_{att't',b} 
- 2 \dot{G}_{btt't',a}) v^b 
\nonumber \\ & & \mbox{} 
+ 2(2 \dot{G}_{atb't',t} - \dot{G}_{ttb't',a}) v^{b'} \Bigr]\, dt'.  
\label{5.22}
\end{eqnarray} 
With the tail part of the Green's function extracted from
Eq.~(\ref{3.31}), we have $\bbox{f} = \bbox{f}_{\! A} 
+ \bbox{f}_{\! B}$, where  
\begin{eqnarray}
f^a_A &=& -2m^2 \int_{-\infty}^t \Bigl[ (2A_{,tta'} - A_{,tt'a}) 
+(-3\delta_{ab} A_{,tt't} + 2A_{,tab} + 4A_{,ta'b} + 2A_{,ta'b'})v^b 
\nonumber \\ & & \mbox{} 
+ 2(2\delta_{ab} A_{,tt't} - 2A_{,ta'b} - A_{,ta'b'} 
+ A_{,tab'})v^{b'} \Bigr]\, dt' 
\label{5.23}
\end{eqnarray}
and
\begin{equation} 
f^a_B = -8m^2 \int_{-\infty}^t B_{,t} v^{a'}\, dt'. 
\label{5.24}
\end{equation} 
Referring back to the discussion preceding Eq.~(\ref{5.11}), we see
that $\bbox{f}_{\! B}$ is of the order of $(mM/r^2)(m/r)v^2$, which is
a correction of order $\Phi v^2$ relative to Newtonian gravity. This
gives a second post-Newtonian (2{\sc pn}) correction \cite{PN} to 
the Newtonian equations of motion, and according to our approximation
rules, this must be neglected.   

The evaluation of $\bbox{f}_{\! A}$ is lengthier, but the steps are
familiar. The first group of terms inside the integral of
Eq.~(\ref{5.23}) are 
\begin{equation}
-\partial_{t'} \bigl(2 A_{,ta'} + A_{,ta}\bigr) = 
-\frac{d}{dt'} \bigl( 2 A_{,ta'} + A_{,ta} \bigr) +
\bigl( 2 A_{,ta'b'} + A_{,tab'} \bigr) v^{b'}.
\end{equation}
The total derivatives contributes
nothing to the self-force, and the remaining terms are absorbed into
the third group. In the second group, the quantity in front of $v^b$
is 
\begin{equation}
-\frac{d}{dt'} \bigl( -3 \delta_{ab}A_{,tt'} + 2A_{,ab} + 4A_{,a'b} 
+ 2A_{,a'b'} \bigr) + O(v'), 
\end{equation}
and this group contributes $4m^2 \Phi_{,ab} v^b$ to the
self-force. Finally, the third group of terms is
\begin{equation}
2 v^{b'} \frac{d}{dt'} \bigl( 2\delta_{ab} A_{,tt} + 2A_{,a'b} 
- \frac{3}{2}A_{,ab'} \bigr) + O(v^{\prime 2}),
\end{equation}
and after integrating 
by parts, we find that this contributes $-\frac{1}{3} m^2 \Phi_{,ab}
v^b$ to the self-force. Summing the contributions gives 
$f^a_A = \frac{11}{3} m^2 \Phi_{,ab} v^b$.  

The gravitational self-force acting on a point particle of mass $m$ is
therefore 
\begin{equation}
\bbox{f} = - \frac{11}{3}\, m^2 \frac{d \bbox{g}}{dt},  
\label{5.25}
\end{equation} 
where $\bbox{g} = -\bbox{\nabla} \Phi$. This gives a correction of
1.5{\sc pn} order to the equations of motion, and appears to represent
radiation reaction. According to the naive equations of motion $d^2
\bbox{z}/dt^2 = \bbox{g} + \bbox{f}/m$, this term does work on the
particle at an average rate $dW/dt = + \frac{11}{3} m^2
|\bbox{g}|^2$. We see that the self-force of Eq.~(\ref{5.25}) seems to
bring energy to the particle.    

There are two major problems of interpretation associated with
Eq.~(\ref{5.25}). First, the self-force seems to give rise to
radiation {\it antidamping} instead of radiation damping
\cite{46}. Second, radiation reaction seems to occur at 1.5{\sc pn}
order, while it is known that the effect should appear only at 
2.5{\sc pn} order in a post-Newtonian expansion of the relativistic
equations of motion \cite{3,4,5,9}. In the next section we shall see 
that these problems disappear once we properly incorporate additional 
(matter-mediated) corrections to the equations of motion.    

\section{Massive particle in a spacetime containing matter} 

Our task in this section is to produce a proper extension of the 
Mino-Sasaki-Tanaka-Quinn-Wald equations of motion \cite{24,25} to
spacetimes that contain matter. First (Secs.~VI A and B), we 
establish that in this context, the gravitational self-force
$f^\alpha_{\rm self}$ is still given by Eq.~(\ref{1.3}), but that the
retarded Green's function must be a solution to Eq.~(\ref{3.21}),
which includes Ricci-tensor terms generated by the matter. That this
does indeed give the correct expression for the self-force was
presented as an assumption in the preceding sections of this
paper. Second (Sec.~VI C), we derive the existence of an
additional term in the equations of motion, which now take the form  
\begin{equation}
m u^\alpha_{\ ;\beta} u^\beta = f^\alpha_{\rm self} 
+ f^\alpha_{\rm mm},
\label{6.1}
\end{equation}
where $f^\alpha_{\rm mm}$ is the matter-mediated force alluded to 
previously. This force originates from the change in the background
metric associated with the motion of the mass $M$ around the system's
center of mass; this motion is caused by the gravitational action of
the particle on the central mass. Third (Secs.~VI D to H), we
compute the matter-mediated force, and show that it contains a
radiation-damping term that precisely cancels the antidamping term in
the self-force. Such a cancellation was noticed a long time ago by
Carmeli \cite{51}, but in the context of a very different formulation
of the equations of motion. We will see (Sec.~VI I) that within the
weak-field, slow-motion approximation considered here, Eq.~(\ref{6.1})
reduces to the appropriate limit of the standard post-Newtonian
equations of motion \cite{3,4,5,9}.    

\subsection{Background spacetime and test particle} 

Let us consider the background spacetime first. We are given a metric
$g_{\alpha\beta}$ that satisfies the Einstein field equations in the
presence of matter. We assume that the matter distribution is bounded
and describes an isolated object of mass $M$; throughout this section
we will refer to this object as ``the star''. We write the field
equations as $G^{\alpha\beta}[g] = 8\pi T^{\alpha\beta}[g]$, and
indicate explicitly that the stress-energy tensor is a functional of
the metric. For the moment we take $g_{\alpha\beta}$ to be an
{\it exact} solution to the field equations; the weak-field
approximation will be incorporated at a later stage. And for the
moment we avoid adopting a specific phenomenology for the
stress-energy tensor; the star can always be thought of as a ball of
perfect fluid with a specified equation of state.  

In this spacetime we insert a point particle of mass $m$. We assume
that this particle moves on a world line $z^\alpha(\tau)$ in a
region of spacetime that is empty of matter; the parameter $\tau$ is
the particle's proper time. We write the particle's stress-energy
tensor as  
\begin{equation}
\tau^{\alpha\beta}(x) = m \int u^\alpha u^\beta\, (-g)^{-1/2}  
\delta_4(x-z(\tau))\, d\tau,  
\label{6.2}
\end{equation}  
where $u^\alpha(\tau) = dz^\alpha/d\tau$ is the four-velocity and the 
$\delta$-function is a four-dimensional distribution with support on
the world line. It is easy to show that conservation of the
stress-energy tensor in the background spacetime implies geodesic
motion:     
\begin{equation}
\tau^{\alpha\beta}_{\ \ \ ;\beta} = m \int 
u^\alpha_{\ ;\beta} u^\beta\, (-g)^{-1/2} 
\delta_4(x-z(\tau))\, d\tau, 
\label{6.3}
\end{equation}
so that $\tau^{\alpha\beta}_{\ \ \ ;\beta} = 0$ leads to 
$u^\alpha_{\ ;\beta} u^\beta = 0$, the geodesic equation. Here,  
a semicolon indicates covariant differentiation in the background
metric. We will keep this notation below: a semicolon will always
refer to the metric $g_{\alpha\beta}$, which will always be used to 
raise and lower Greek indices. 

We see that the particle's stress-energy tensor involves the
background metric, through the definition of proper time and the
factor $(-g)^{-1/2}$. If we think of expressing the metric as a formal
expansion in powers of $M$, we can similarly expand the particle's
stress-energy tensor as   
\begin{equation}
\tau^{\alpha\beta} = O(m) + O(mM) + O(mM^2) + \cdots. 
\label{6.4}
\end{equation} 
To keep track of the powers of $m$ and $M$ appearing in various
quantities will extremely important in the following discussion. For 
the weak-field application considered here, we shall truncate
Eq.~(\ref{6.4}) after the first two terms.    

\subsection{Self-force} 

The picture described thus far is that of {\it test mass} moving on a
geodesic of the background spacetime: the particle has not yet been
allowed to alter the spacetime's geometry. We now incorporate this
effect by inserting $\tau^{\alpha\beta}$ on the right-hand side of the
Einstein field equations. Working linearly in the small mass $m$, we
find that this modifies the metric by a term $\gamma_{\alpha\beta}$,
so that the total metric is now $g_{\alpha\beta} +
\gamma_{\alpha\beta}$. If we impose the Lorentz-gauge condition on
$\gamma_{\alpha\beta}$, as we have done in Sec.~II B, we find that the 
metric perturbation is given by Eq.~(\ref{3.24}), 
\begin{equation}
\gamma^{\alpha\beta}(x) = 4 \int G^{\alpha\beta}_{\ \ \mu'\nu'}(x,x')
\tau^{\mu'\nu'}(x') \sqrt{-g'}\, d^4 x', 
\label{6.5}
\end{equation}
where $G^{\alpha\beta}_{\ \ \mu'\nu'}(x,x')$ is the gravitational
Green's function whose ``trace-reversed'' counterpart is a solution to 
Eq.~(\ref{3.21}) --- Green's equation for a spacetime containing
matter. If we again think of expressing the metric as a formal
expansion in powers of $M$, we can schematically write  
\begin{equation}
G^{\alpha\beta}_{\ \ \mu'\nu'} = O(1) + O(M) + O(M^2) + \cdots, 
\label{6.6}
\end{equation} 
which corresponds to Eq.~(\ref{3.29}). Expanding Eq.~(\ref{6.5}) in a
similar way, using Eqs.~(\ref{6.4}) and (\ref{6.6}), we obtain   
\begin{equation}
\gamma_{\alpha\beta} = O(m) + O(mM) + O(mM^2) + \cdots. 
\label{6.7}
\end{equation}
This expansion omits terms of order $m^2$ (and higher) that would come
from the nonlinearities of the Einstein field equations. We shall
systematically omit such terms in future expansions, and consistently
work to first order in the small mass $m$.  

The gravitational self-force corresponds to the metric perturbation
$\gamma_{\alpha\beta}$ acting on the particle. If this were the only
metric perturbation to be considered --- and we will show below that
it is not --- this action could be described by the
statement that the particle now moves on a geodesic of the full metric
$g_{\alpha\beta} + \gamma_{\alpha\beta}$. Working again to linear
order in the perturbation, we find that the new equations of motion
read   
\begin{equation} 
u^\mu_{\ ;\nu} u^\nu = 
-\frac{1}{2} \bigl( \gamma^\mu_{\ \alpha;\beta} 
+ \gamma^\mu_{\ \beta;\alpha} - \gamma_{\alpha\beta}^{\ \ \ ;\mu}  
+ u^\mu \gamma_{\alpha\beta;\gamma} u^\gamma \bigr) 
u^\alpha u^\beta, 
\label{6.8}
\end{equation} 
and the right-hand side can be identified with $f^\mu_{\rm self}/m$, 
the gravitational self-acceleration. Based on Eq.~(\ref{6.7}), we
would naively conclude that $f^\alpha_{\rm self}/m = O(m) + O(mM) +
O(mM^2) + \cdots$. 

Equation (\ref{6.8}), however, is not valid as it stands. First, we
have already indicated that we must insert a matter-mediated force on
the right-hand side. Second, we know that $\gamma_{\alpha\beta}$ is
singular on the particle's world line, so that the right-hand side of
Eq.~(\ref{6.8}) is only formally defined; this expression must be
regularized and the divergences must be removed. This renormalization
of the self-force was carefully performed by 
Mino {\it et al.}~\cite{24} and Quinn \& Wald \cite{25} for the case
of vacuum spacetimes. Fortunately, the extension of their work to 
nonvacuum spacetimes is trivial, and it leads to an identical
result. The reason is that the renormalization procedure is based on a
local analysis that is sensitive only to the immediate vicinity of the
world line. So long as the particle moves in a region of spacetime
that is empty of matter, which we assume here, the structure of the 
divergent terms in Eq.~(\ref{6.8}) is the same whether or not matter
is present {\it somewhere} in the spacetime. After renormalization,
the self-force is found to be given by Eq.~(\ref{1.3}), but it is now 
expressed in terms of a Green's function that is sensitive to the
presence of matter. 

We conclude that in a spacetime that contains matter, the
gravitational self-force is given by its usual expression (\ref{1.3}),
but that it now involves a Green's function that incorporates
information about the matter distribution; this information is encoded
in the tensor $S_{\mu\alpha\nu\beta}$ defined by Eq.~(\ref{2.14}).   

We note that the renormalization procedure removes the $O(m)$ 
part of the self-acceleration, which now admits the expansion 
\begin{equation}
f^{\alpha}_{\rm self}/m = O(mM) + O(mM^2) + \cdots.
\label{6.9}
\end{equation} 
Using this together with Eqs.~(\ref{6.3}) and (\ref{6.8}), we find
that the particle's stress-energy tensor is no longer conserved in the
background spacetime. Instead, 
\begin{equation}
\tau^{\alpha\beta}_{\ \ \ ;\beta} = O(m^2 M) + O(m^2 M^2) + \cdots. 
\label{6.10}
\end{equation} 
This is as it should be, because the motion is no longer geodesic in
the background spacetime. In a formal sense, $\tau^{\alpha\beta}$ is
now conserved in a spacetime with metric $g_{\alpha\beta} +
\gamma_{\alpha\beta}$.  

\subsection{Matter-mediated force} 

The existence of additional terms on the right-hand side of
Eq.~(\ref{6.8}) follows directly from the fact that the stress-energy
tensor $T^{\alpha\beta}[g]$ describing the matter distribution depends
on the metric. Because inserting the particle in the spacetime has the
effect of shifting the metric from $g_{\alpha\beta}$ to
$g_{\alpha\beta} + \gamma_{\alpha\beta}$, there must be a corresponding
shift in the stress-energy tensor. Physically, this corresponds to the
facts that (i) the particle raises a tide on the star and induces 
internal motions within the fluid, and (ii) the particle sets the 
star in a small motion around the system's center of mass. Both
effects are incorporated in the shift in the stress-energy tensor; as
we shall see, however, the bulk motion of the star is much more
important for our purposes. We let 
\begin{equation}
\delta T^{\alpha\beta} = T^{\alpha\beta}[g+\gamma] -
T^{\alpha\beta}[g] 
\label{6.11}
\end{equation}
be the perturbation of the star's stress-energy tensor created by the
presence of the particle. Below we will use $T^{\alpha\beta}_* \equiv 
T^{\alpha\beta}[g+\gamma]$ to designate the perturbed stress-energy
tensor of the moving star, and $T^{\alpha\beta} \equiv
T^{\alpha\beta}[g]$ to represent the background values --- the star
at rest. Because $T^{\alpha\beta} = O(M) + O(M^2) + \cdots$ and
$\gamma_{\alpha\beta}$ can be expanded as in Eq.~(\ref{6.7}), we have
that 
\begin{equation} 
\delta T^{\alpha\beta} = O(mM) + O(m M^2) + \cdots. 
\label{6.12}
\end{equation} 

The perturbation in the star's stress-energy tensor must appear,
together with $\tau^{\alpha\beta}$, on the right-hand side of the
Einstein field equations. It will contribute an additional metric
perturbation $\delta g_{\alpha\beta}$, over and above the original
perturbation $\gamma_{\alpha\beta}$ directly associated with the
particle. Physically, the new perturbation represents the difference
between the gravitational field of the moving star and the background
field of the star at rest. Because both $\tau^{\alpha\beta}$ and 
$\delta T^{\alpha\beta}$ are quantities of the first order in
$m$, the (linear) metric perturbations can simply be added; the full 
metric is therefore $g_{\alpha\beta} + \delta g_{\alpha\beta} +
\gamma_{\alpha\beta}$. 

Imposing the Lorentz-gauge condition separately on both $\delta
g_{\alpha\beta}$ and $\gamma_{\alpha\beta}$, we have that   
\begin{equation}
\delta g^{\alpha\beta}(x) = 4 \int 
G^{\alpha\beta}_{\ \ \mu'\nu'}(x,x') 
\delta T^{\mu'\nu'}(x') \sqrt{-g'}\, d^4 x'. 
\label{6.13}
\end{equation} 
Using the expansions (\ref{6.6}) and (\ref{6.12}), we see that
Eq.~(\ref{6.13}) implies 
\begin{equation} 
\delta g_{\alpha\beta} = O(mM) + O(mM^2) + \cdots.
\label{6.14}
\end{equation} 
The matter-mediated force corresponds to $\delta g_{\alpha\beta}$
acting on the particle. In the absence of $\gamma_{\alpha\beta}$, this
action can be described by the statement that the particle must move
on a geodesic of the metric $g^*_{\alpha\beta} \equiv g_{\alpha\beta}
+ \delta g_{\alpha\beta}$. Physically, this means that the particle is
now subjected to the gravitational field of the moving
star. Mathematically, the equations of motion would take the
appearance of Eq.~(\ref{6.8}), but with $\delta g_{\alpha\beta}$
taking the place of $\gamma_{\alpha\beta}$. There is no need to
renormalize this equation: $\delta g_{\alpha\beta}$ is smooth on the
world line.  

Our conclusion at this stage is that in a spacetime that contains
matter, the equations of motion of a massive particle are given by
Eq.~(\ref{6.1}), with a self-force given by Eq.~(\ref{1.3}) and a
matter-mediated force given by 
\begin{equation} 
f^\mu_{\rm mm} = 
-\frac{m}{2} \bigl( \delta g^\mu_{\ \alpha;\beta} 
+ \delta g^\mu_{\ \beta;\alpha} - \delta g_{\alpha\beta}^{\ \ \ ;\mu}   
+ u^\mu \delta g_{\alpha\beta;\gamma} u^\gamma \bigr)  
u^\alpha u^\beta. 
\label{6.15}
\end{equation}
According to Eq.~(\ref{6.14}), 
\begin{equation}
f^\alpha_{\rm mm}/m = O(mM) + O(mM^2) + \cdots, 
\label{6.16}
\end{equation}
which should be compared with Eq.~(\ref{6.9}). We see that in a
weak-field approximation, both forces contribute terms of order 
$O(mM)$ to the particle's acceleration.  

We have successfully extended the Mino-Sasaki-Tanaka-Quinn-Wald
equations of motion to spacetimes containing matter. We emphasize 
that this extension does not rely on any weak-field assumption
regarding the background spacetime. In the following subsections we
compute the matter-mediated force of Eq.~(\ref{6.15}); this
discussion will rely on a weak-field approximation. 

\subsection{Determination of $T^{\alpha\beta}_*$} 

The matter-mediated force of Eq.~(\ref{6.15}) involves the metric
perturbation $\delta g_{\alpha\beta}$ which is computed from $\delta
T^{\alpha\beta}$, the perturbation in the star's stress-energy tensor; 
the conversion is given by Eq.~(\ref{6.13}). We have already
introduced the notation $T^{\alpha\beta}_* = T^{\alpha\beta} 
+ \delta T^{\alpha\beta}$ for the stress-energy tensor 
of the moving star.   

We now need to gather information about $\delta T^{\alpha\beta}$. Our
main source will be the conservation equations for the full
stress-energy tensor $T^{\alpha\beta} + \delta T^{\alpha\beta} +
\tau^{\alpha\beta}$ in a spacetime with metric $g_{\alpha\beta} +
\delta g_{\alpha\beta} + \gamma_{\alpha\beta}$. This metric comes with
a connection $\Gamma^\mu_{\ \alpha\beta} 
+ \delta \Gamma^\mu_{\ \alpha\beta} + C^\mu_{\ \alpha\beta}$, in which 
$\Gamma^\mu_{\ \alpha\beta}$ are the Christoffel symbols constructed
from the background metric $g_{\alpha\beta}$, 
$\delta \Gamma^\mu_{\ \alpha\beta}$ the corrections generated by
$\delta g_{\alpha\beta}$, and $C^\mu_{\ \alpha\beta}$ those generated
by $\gamma_{\alpha\beta}$. For example, 
\begin{equation}
C^\mu_{\ \alpha\beta} = \frac{1}{2} \bigl( \gamma^\mu_{\ \alpha;\beta}
+ \gamma^\mu_{\ \beta;\alpha} - \gamma_{\alpha\beta}^{\ \ \ ;\mu}
\bigr). 
\label{6.17}
\end{equation}
By virtue of Eqs.~(\ref{6.7}) and (\ref{6.14}), we have  
\begin{equation}
C^\mu_{\ \alpha\beta} = O(m) + O(mM) + O(mM^2) + \cdots
\label{6.18}
\end{equation}
and 
\begin{equation}
\delta \Gamma^\mu_{\ \alpha\beta} = O(mM) + O(mM^2) + \cdots.
\label{6.19}
\end{equation}
We also have $\Gamma^\mu_{\ \alpha\beta} = O(M) + O(M^2)
+ \cdots$ and we shall use the expansions of Eqs.~(\ref{6.4}) and
(\ref{6.12}).       
 
The conservation equations can be expressed as  
\begin{equation}
T^{\alpha\beta}_{\ \ \ ;\beta} + \tau^{\alpha\beta}_{\ \ \ ;\beta}  
+ \delta T^{\alpha\beta}_{\ \ \ ,\beta} 
+ 2 C^{(\alpha}_{\ \mu\beta} T^{\beta)\mu}  
+ 2 \Gamma^{(\alpha}_{\ \mu\beta} \delta T^{\beta)\mu}  
+ 2 C^{(\alpha}_{\ \mu\beta} \bigl( \delta T^{\beta)\mu} 
        + \tau^{\beta)\mu} \bigr)   
+ 2 \delta \Gamma^{(\alpha}_{\ \mu\beta} \bigl( T^{\beta)\mu} 
        +\delta T^{\beta)\mu} + \tau^{\beta)\mu} \bigr) = 0.
\label{6.20}
\end{equation}
The first term vanishes because the background stress-energy tensor is
conserved in the background spacetime. The second term can be ignored,
because by virtue of Eq.~(\ref{6.10}), it is of second order in
$m$. The third of fourth terms have the expansions $O(mM) 
+ O(mM^2) + \cdots$. Among the remaining terms we have (in schematic
notation) $\Gamma \delta T = O(mM^2) + \cdots$, $C \delta T = O(m^2 M)
+ \cdots$, $C \tau = O(m^2) + O(m^2 M) + \cdots$, $\delta \Gamma T  
= O(m M^2) + \cdots$, $\delta \Gamma \delta T = O(m^2 M^2) +
\cdots$, and $\delta \Gamma \tau = O(m^2 M) + \cdots$. We simplify
Eq.~(\ref{6.20}) by discarding all terms that are second order in $m$.  
Incorporating at this stage our weak-field approximation, we also
neglect terms that are second order in $M$. The remaining terms are
all $O(mM)$ and we obtain  
\begin{equation} 
\delta T^{\alpha\beta}_{\ \ \ ,\beta} 
+ C^{\alpha}_{\ \beta\gamma} T^{\beta\gamma} 
+ C^{\beta}_{\ \beta\gamma} T^{\alpha\gamma} = O(mM^2).  
\label{6.21}
\end{equation}
This equation (partially) determines $\delta T^{\alpha\beta}$ in terms
of known quantities. Because we have only four equations for ten
unknowns, Eq.~(\ref{6.21}) cannot be used without first adopting a
specific phenomenology for the matter distribution. We shall come back
to this point in Sec.~VI F.  

It is important to notice that because Eq.~(\ref{6.21}) has been
truncated to order $mM$, only the $O(m)$ part of 
$C^\mu_{\ \alpha\beta}$ is involved in the determination of 
$\delta T^{\alpha\beta}$ --- cf.~Eq.~(\ref{6.18}). This can be
obtained from the $O(m)$ part of $\gamma_{\alpha\beta}$ ---
cf.~Eq.~(\ref{6.7}). By virtue of Eqs.~(\ref{3.3}), (\ref{3.29}), 
and (\ref{6.5}), this is given by 
\begin{equation}
\gamma^{\alpha\beta}_{\rm flat}(t,\bbox{x}) = 4 \int 
\frac{\bar{\tau}^{\alpha\beta}(t - |\bbox{x} 
 - \bbox{x'}|,\bbox{x'})}{|\bbox{x} - \bbox{x'}|}\, d^3 x',  
\label{6.22}
\end{equation} 
where $\bar{\tau}^{\alpha\beta} \equiv \tau^{\alpha\beta} -
\frac{1}{2} (\eta_{\mu\nu} \tau^{\mu\nu}) \eta^{\alpha\beta}$ is the
particle's ``trace-reversed'' stress-energy tensor. The subscript
``flat'' indicates that the metric perturbation is calculated as if
the particle were moving in flat spacetime (on a Newtonian trajectory
around the star); the right-hand side of Eq.~(\ref{6.22}) involves 
only the $O(m)$ part of $\tau^{\alpha\beta}$ --- cf.~Eq.~(\ref{6.4}).  

Another consequence of truncating Eq.~(\ref{6.21}) is that it can be
re-expressed as   
\begin{equation} 
T^{\alpha\beta}_{*\ \ ,\beta} 
+ C^{\alpha}_{\ \beta\gamma} T_*^{\beta\gamma}
+ C^{\beta}_{\ \beta\gamma} T_*^{\alpha\gamma} = O(mM^2), 
\label{6.23}
\end{equation}
in terms of the star's perturbed stress-energy tensor:
$T_*^{\alpha\beta} = T^{\alpha\beta}  
+ \delta T^{\alpha\beta}$. Here we use the facts that 
(i) $T^{\alpha\beta}_{\ \ \ ,\beta}$ is actually zero 
for the weak-field spacetimes considered in this 
paper --- $T^{\alpha\beta}_{\ \ \ ;\beta}$ is nonzero but is a
quantity of order $M^2$ --- and (ii) the difference between $C T_*$
and $C T$ is of second order in $m$. Equation (\ref{6.23}) implies
that $T_*^{\alpha\beta}$ is conserved in a spacetime with metric
$g_{\alpha\beta}^{\rm particle} \equiv \eta_{\alpha\beta} +
\gamma^{\rm flat}_{\alpha\beta}$, up to terms of order $mM^2$. We will
refer to this observation below.     

Let us now take stock. In order to compute the matter-mediated force 
of Eq.~(\ref{6.15}) we must first calculate $\delta g_{\alpha\beta}$
using Eq.~(\ref{6.13}). We therefore need to find 
$\delta T^{\alpha\beta}$, the perturbation of the stress-energy  
tensor associated with the star's motion around the center of
mass. For this we shall use Eq.~(\ref{6.21}), after adopting a
specific phenomenology for the matter distribution (Sec.~VI F). But
Eq.~(\ref{6.21}) involves also the connection coefficients 
$C^\mu_{\ \alpha\beta}$ computed from 
$\gamma^{\rm flat}_{\alpha\beta}$. To calculate this is our next task.  

\subsection{Computation of $\gamma^{\rm flat}_{\alpha\beta}$} 

The stress-energy tensor of the particle is given by
Eq.~(\ref{6.2}). For the purpose of computing 
$\gamma^{\rm flat}_{\alpha\beta}$ we let the particle move in flat
spacetime, so that the metric used in Eq.~(\ref{6.2}) to define proper
time $\tau$ and the determinant $g$ is $\eta_{\alpha\beta}$, the
metric of flat spacetime. After changing the variable of integration
to $z^0$ and integrating over $\delta(t-z_0)$, we obtain 
\begin{equation}
\tau^{\alpha\beta}(t,\bbox{x}) = m (dt/d\tau) v^\alpha v^\beta\, 
\delta_3(\bbox{x} - \bbox{z}), 
\label{6.24}
\end{equation}
in which $\bbox{z}(t)$ represents the trajectory of the particle,
$v^\alpha = (1,\bbox{v})$ with $\bbox{v}(t) = d\bbox{z}/dt$, and 
$dt/d\tau = (1 - \bbox{v}^2)^{-1/2}$. We recall that although the
calculation is carried out in flat spacetime, the particle is actually
moving on a geodesic of $g_{\alpha\beta}$; its motion is governed by
Newton's equation,   
\begin{equation}
\frac{d \bbox{v}}{d t} = \bbox{g} = -\bbox{\nabla} \Phi, 
\label{6.25}
\end{equation}
where the right-hand side is evaluated at $\bbox{x} = \bbox{z}(t)$.     

It will prove sufficient for our purposes to evaluate
$\tau^{\alpha\beta}$ to second order in $\bbox{v}$, the particle's
velocity. After trace reversal, we obtain   
\begin{eqnarray} 
\bar{\tau}^{00} &=& \frac{m}{2}  \biggl[1 + \frac{3}{2} \bbox{v}^2 
+ O(v^4) \biggr]\, \delta_3(\bbox{x} - \bbox{z}), \nonumber \\
\bar{\tau}^{0a} &=& m v^a \bigl[ 1 + O(v^2) \bigr]\, 
\delta_3(\bbox{x} - \bbox{z}), 
\label{6.26} \\
\bar{\tau}^{ab} &=& \frac{m}{2} \biggl\{ \delta^{ab} \biggl[1 
- \frac{1}{2} \bbox{v}^2 + O(v^4) \biggr] + 2 v^a v^b \bigl[ 1 
+ O(v^2) \bigr] \biggr\}\, \delta_3(\bbox{x} - \bbox{z}). \nonumber
\end{eqnarray}
To be consistent we also Taylor-expand the right-hand side of
Eq.~(\ref{6.22}) about the current time $t$. This gives 
\begin{equation} 
\gamma^{\alpha\beta}_{\rm flat}(t,\bbox{x}) = 4 \int  
\frac{\bar{\tau}^{\alpha\beta}(t,\bbox{x'})}{|\bbox{x} 
- \bbox{x'}|}\, d^3 x' 
- 4 \frac{\partial}{\partial t} \int 
\bar{\tau}^{\alpha\beta}(t,\bbox{x'})\, d^3 x' 
+ 2 \frac{\partial^2}{\partial t^2} \int 
\bar{\tau}^{\alpha\beta}(t,\bbox{x'}) |\bbox{x} - \bbox{x'}|\, d^3 x' 
+ \cdots. 
\label{6.27}
\end{equation} 
It is now straightforward to substitute Eqs.~(\ref{6.26}) into
Eq.~(\ref{6.27}) and perform the integrations. For example, 
\begin{equation}
\gamma^{00}_{\rm flat} = \frac{2m \lambda}{|\bbox{x}-\bbox{z}|}  
- 2 m \frac{\partial \lambda}{\partial t} 
+ m \frac{\partial^2}{\partial t^2} \lambda |\bbox{x}-\bbox{z}| 
+ \cdots, 
\label{6.28}
\end{equation}
where $\lambda = 1 + \frac{3}{2} \bbox{v}^2 + O(v^4)$. This result
simplifies considerably: When the time derivative acts on $\bbox{v}$,
it generates via Eq.~(\ref{6.25}) a term of order 
$g \sim M/|\bbox{z}|^2$ that can be neglected, because we are
interested only in the $O(m)$ part of $\gamma_{\alpha\beta}$. We are
therefore left with evaluating $\partial^2
|\bbox{x}-\bbox{z}|/\partial t^2$, which is a simple task. Subjecting
the other components to similar manipulations, we arrive at 
\begin{eqnarray}
\gamma^{00}_{\rm flat}(t,\bbox{x}) &=& 
\frac{2 m}{|\bbox{x}-\bbox{z}|}\,  
\biggl[ 1 + 2\bbox{v}^2 - \frac{1}{2} (\bbox{\hat{n}} \cdot
\bbox{v})^2 + O(v^4) \biggr] + O(mM), \nonumber \\ 
\gamma^{0a}_{\rm flat}(t,\bbox{x}) &=&  
\frac{4 m v^a}{|\bbox{x}-\bbox{z}|}\,  
\bigl[ 1 + O(v^2) \bigr] + O(mM), 
\label{6.29} \\
\gamma^{ab}_{\rm flat}(t,\bbox{x}) &=& 
\frac{2 m}{|\bbox{x}-\bbox{z}|}\,  
\biggl\{ \delta^{ab} \biggl[1 - \frac{1}{2} (\bbox{\hat{n}} \cdot
\bbox{v})^2 \biggr] + 2v^a v^b + O(v^4) \biggr\} + O(mM), \nonumber   
\end{eqnarray}
where $\bbox{\hat{n}} = 
(\bbox{x}-\bbox{z})/|\bbox{x}-\bbox{z}|$. 

\subsection{Determination of $T^{\alpha\beta}_*$ (continued)} 

As we have remarked at the end of Sec.~VI D, the determination of
$T^{\alpha\beta}_*$ via the conservation equation (\ref{6.23}) is
possible only if we adopt a specific phenomenology for the matter
distribution. We shall make the simplest choice, and describe the star
as a particle of mass $M$ moving on a trajectory $\bbox{Z}(t)$ with a
velocity $\bbox{V}(t) = d \bbox{Z}/dt$. The pointlike nature of the
moving star will not be a problem here: First, for the purposes of
this calculation the star is not subjected to its own field
because, as we have pointed out, it moves in a spacetime with metric
$g_{\alpha\beta}^{\rm particle} 
= \eta_{\alpha\beta} + \gamma^{\rm flat}_{\alpha\beta}$; second, the
singularity in $\delta g_{\alpha\beta}$ is far removed from the
particle's world line and is therefore not an obstacle in the
calculation of the matter-mediated force of Eq.~(\ref{6.15}). Notice
that in the following developments we shall formally treat both
$\bbox{Z}$ and $\bbox{V}$ as quantities of the first order in $m$.  

Following the steps leading to Eq.~(\ref{6.24}), we find that the
stress-energy tensor of the moving star takes the form 
\begin{equation}
T_*^{\alpha\beta}(t,\bbox{x}) = M 
\frac{dt/d \tau}{\sqrt{-g_{\rm particle}}}\, 
V^\alpha V^\beta \delta_3(\bbox{x} - \bbox{Z}), 
\label{6.30}
\end{equation}
where $V^\alpha = (1,\bbox{V})$ and $dt/d\tau = 
(-g_{\mu\nu}^{\rm particle} V^\mu V^\nu)^{-1/2}$. After discarding
terms of order $m^2$, this reduces to $T_*^{00} 
= M \delta_3(\bbox{x} - \bbox{Z}) + M(\gamma^{00}_{\rm flat} 
- \frac{1}{2}\delta_{ab} \gamma^{ab}_{\rm flat}) \delta_3(\bbox{x})$, 
$T_*^{0a} = M V^a \delta_3 (\bbox{x})$, and $T_*^{ab} = O(m^2 M)$. The
perturbation $\delta T^{\alpha\beta}$ can be obtained from these
results by removing the background values, 
$T^{00} = M \delta_3(\bbox{x})$ and $T^{0a} = T^{ab} = 0$.    

Substitution of our expressions for $\delta T^{\alpha\beta}$ into
Eq.~(\ref{6.21}) leads to equations of motion for the star. The
spatial components of the conservation equation imply 
\begin{equation}
\frac{d V^a}{dt} = \frac{1}{2} \frac{\partial}{\partial x^a}
\gamma^{00}_{\rm flat} + \frac{\partial}{\partial t}
\gamma^{0a}_{\rm flat} + O(mM), 
\label{6.31}
\end{equation} 
in which the right-hand side is evaluated at $\bbox{x} = 0$. We find 
that the time component of the conservation equation holds as an
identity, but this confirms the necessity of including 
$\sqrt{-g_{\rm particle}}$ in our expression for the stress-energy
tensor.  

It is a simple task to evaluate the partial derivatives of
Eq.~(\ref{6.31}) starting from the expressions listed in
Eq.~(\ref{6.29}); in this procedure we use Eq.~(\ref{6.25}) to dismiss
$d\bbox{v}/dt$ as a term of order $M/|\bbox{z}|^2$. After evaluating
the results at $\bbox{x} = 0$, we obtain
\begin{equation} 
\frac{d \bbox{V}}{dt} = \frac{m}{|\bbox{z}|^2}\, \biggl\{ 
\biggl[ 1 + 2 \bbox{v}^2 - \frac{3}{2} (\bbox{\hat{z}} \cdot
\bbox{v})^2 + O(v^4) \biggr] \bbox{\hat{z}} 
- 3 (\bbox{\hat{z}} \cdot \bbox{v}) \bbox{v} \biggr\} + O(mM), 
\label{6.32}
\end{equation}
where $\bbox{\hat{z}} = \bbox{z}/|\bbox{z}|$. The motion of the star
is now determined to a degree of accuracy sufficient for our remaining
calculations. Equation (\ref{6.32}) confirms that $\bbox{V}$, and
therefore $\bbox{Z}$, are to be treated as quantities of order $m$. We
note that in the present context in which the star has become a point
mass, the motion of the particle is still governed by
Eq.~(\ref{6.25}), but that the gravitational field $\bbox{g}$ now
takes the explicit form   
\begin{equation}
\bbox{g} = - \frac{M}{|\bbox{z}|^2}\, \bbox{\hat{z}}. 
\label{6.33}
\end{equation}
Equation (\ref{6.25}) determines the vectors $\bbox{z}(t)$ and
$\bbox{v}(t) = d\bbox{z}/dt$ that appear in Eq.~(\ref{6.32}). 

The perturbed stress-energy tensor of the moving star is now
completely determined. Its final expression is obtained by 
substituting the potentials of Eq.~(\ref{6.29}) into the results given
before. This yields 
\begin{eqnarray} 
T_*^{00}(t,\bbox{x}) &=& M\delta_3(\bbox{x} - \bbox{Z}) 
- \frac{mM}{|\bbox{z}|}\, \biggl[ 1 - 2 \bbox{v}^2 
- \frac{1}{2} (\bbox{\hat{z}} \cdot \bbox{v})^2 + O(v^4) \biggr] 
\delta_3(\bbox{x}) + O(mM^2), \nonumber \\
& & \label{6.34} \\
T_*^{0a}(t,\bbox{x}) &=& M V^a \delta_3(\bbox{x}) \nonumber, 
\end{eqnarray}
and $T_*^{ab} = O(m^2 M)$. 

\subsection{Computation of $g^*_{\alpha\beta}$} 

The metric $g^*_{\alpha\beta} = g_{\alpha\beta} +
\delta g_{\alpha\beta}$ was introduced in Sec.~VI C, in the paragraph
following Eq.~(\ref{6.14}). It represents the gravitational field of
the moving star, and the action of the matter-mediated force of
Eq.~(\ref{6.15}) can be described by the statement that the particle
must move on a geodesic of this metric. In the next subsection we will
use this statement to produce the equations of motion of the particle
under the sole action of the matter-mediated force. In this subsection
we calculate $g^*_{\alpha\beta}$. 

The metric of the moving star differs from $g_{\alpha\beta}$, the
metric of the star at rest, by $\delta g_{\alpha\beta}$, which can be
computed from Eq.~(\ref{6.13}). Keeping only the $O(mM)$ part of this
equation, which involves the $O(1)$ part of the gravitational Green's
function --- cf.~Eqs.~(\ref{3.3}) and (\ref{3.29}), we write this as 
\begin{equation}
\delta g^{\alpha\beta}(t,\bbox{x}) = 4 \int  
\frac{\delta \bar{T}^{\alpha\beta}(t - |\bbox{x} 
 - \bbox{x'}|,\bbox{x'})}{|\bbox{x} - \bbox{x'}|}\, d^3 x'
+ O(mM^2),  
\label{6.35}
\end{equation} 
where $\delta \bar{T}^{\alpha\beta} = \delta T^{\alpha\beta} - 
\frac{1}{2} (\eta_{\mu\nu} \delta T^{\mu\nu}) \eta^{\alpha\beta}$. 
At the same time, we recall from Sec.~II A that $g_{\alpha\beta}$
differs from $\eta_{\alpha\beta}$ by a quantity $h_{\alpha \beta} 
\equiv -2 \Phi \chi_{\alpha\beta}$ that can expressed as 
\begin{equation}
h^{\alpha\beta}(t,\bbox{x}) = 4 \int   
\frac{\bar{T}^{\alpha\beta}(t - |\bbox{x} 
 - \bbox{x'}|,\bbox{x'})}{|\bbox{x} - \bbox{x'}|}\, d^3 x',  
\label{6.36}
\end{equation} 
where $\bar{T}^{\alpha\beta}$ is the trace-reversed stress-energy
tensor of the background spacetime. Adding Eqs.~(\ref{6.35}) and
(\ref{6.36}), we find that 
\begin{equation}
g^*_{\alpha\beta} = \eta_{\alpha\beta} + h^*_{\alpha\beta}, 
\label{6.37}
\end{equation}
where $h^*_{\alpha\beta} \equiv h_{\alpha\beta} 
+ \delta g_{\alpha\beta}$ is given by 
\begin{equation}
h_*^{\alpha\beta}(t,\bbox{x}) = 4 \int   
\frac{\bar{T}_*^{\alpha\beta}(t - |\bbox{x} 
 - \bbox{x'}|,\bbox{x'})}{|\bbox{x} - \bbox{x'}|}\, d^3 x'  
+ O(mM^2),
\label{6.38}
\end{equation} 
and where $\bar{T}_*^{\alpha\beta} \equiv T_*^{\alpha\beta} -  
\frac{1}{2} (\eta_{\mu\nu} T_*^{\mu\nu}) \eta^{\alpha\beta}$ can be
calculated from Eq.~(\ref{6.34}). To first order in $m$, this gives 
$\bar{T}_*^{00} = \frac{1}{2} T_*^{00}$, $\bar{T}_*^{0a} = T_*^{0a}$,
and $\bar{T}_*^{ab} = \frac{1}{2} \delta^{ab} T_*^{00}$. 

We shall describe in some detail the computation of $h_*^{00}$. We
first Taylor-expand $T_*^{00}(t-|\bbox{x}-\bbox{x'}|,\bbox{x'})$ about
the current time $t$ and express $h_*^{00}$ as 
\begin{eqnarray}
h_*^{00} &=& 2 \int 
\frac{T_*^{00}(t,\bbox{x'})}{|\bbox{x} - \bbox{x'}|}\, d^3 x'  
- 2 \frac{\partial}{\partial t} \int T_*^{00}(t,\bbox{x'})\, d^3 x'  
+ \frac{\partial^2}{\partial t^2} \int 
T_*^{00}(t,\bbox{x'}) |\bbox{x} - \bbox{x'}|\, d^3 x'  
\nonumber \\ & & \mbox{} 
- \frac{1}{3} \frac{\partial^3}{\partial t^3} \int 
T_*^{00}(t,\bbox{x'}) |\bbox{x} - \bbox{x'}|^2\, d^3 x' 
+ \cdots + O(mM^2).  
\label{6.39}
\end{eqnarray} 
We will see that truncating the series after four terms gives us
sufficient accuracy; it is very important, however, to keep the 
last term involving three time derivatives. After substitution of
Eq.~(\ref{6.34}), we obtain 
\begin{eqnarray} 
h_*^{00} &=& \frac{2M}{|\bbox{x} - \bbox{Z}|} 
+ M \frac{\partial^2}{\partial t^2} |\bbox{x} - \bbox{Z}| 
- \frac{1}{3} M \frac{\partial^3}{\partial t^3} |\bbox{x} -
    \bbox{Z}|^2 + \cdots \nonumber \\ 
& & \mbox{} - \frac{2mM}{|\bbox{x}|}\, \frac{\lambda}{|\bbox{z}|}  
+ 2mM \frac{\partial}{\partial t} \frac{\lambda}{|\bbox{z}|} 
- mM |\bbox{x}| \frac{\partial^2}{\partial t^2}
                \frac{\lambda}{|\bbox{z}|} 
+ \frac{1}{3} mM |\bbox{x}|^2 \frac{\partial^3}{\partial t^3} 
                \frac{\lambda}{|\bbox{z}|} + \cdots 
+ O(mM^2),
\label{6.40}
\end{eqnarray}
where $\lambda = 1 - 2\bbox{v}^2 - \frac{1}{2} (\bbox{\hat{z}} \cdot 
\bbox{v})^2 + O(v^4)$. It is a straightforward task to evaluate the
time derivatives contained in Eq.~(\ref{6.40}); for this we involve
Eq.~(\ref{6.32}) and we make sure to discard all terms that are not 
first-order in both $m$ and $M$. For the first line ({\sc fl}) 
on the right-hand side of Eq.~(\ref{6.40}) we obtain 
\begin{equation}
\mbox{\sc fl} = \frac{2M}{|\bbox{x} - \bbox{Z}|} 
- m M \frac{ \bbox{x}  \cdot
  \bbox{z}}{|\bbox{x}||\bbox{z}|^3}\, 
  \bigl[ 1 + O(v^2) \bigr] 
- \frac{2}{3} mM \bbox{x} \cdot \frac{d \bbox{g}}{dt}\, 
  \bigl[ 1 + O(v^2) \bigr] 
+ O(mM^2), 
\label{6.41}
\end{equation}
where $\bbox{g}$ is defined by Eq.~(\ref{6.33}). In Eq.~(\ref{6.40}), 
the first set of ``$\cdots$'' terms represent a correction of order
$v^4$ to the second term of Eq.~(\ref{6.41}); these can be safely
ignored. The last member of Eq.~(\ref{6.41}) comes from the
third-derivative term in Eq.~(\ref{6.40}); this will be seen to give
rise to a radiation-damping term in the matter-mediated force. For the  
second line ({\sc sl}) on the right-hand side of Eq.~(\ref{6.40}) we
find 
\begin{equation}
\mbox{\sc sl} = - \frac{2mM}{|\bbox{x}||\bbox{z}|}\, 
    \bigl[ 1 + O(v^2) \bigr] 
- 2mM \frac{\bbox{z} \cdot \bbox{v}}{|\bbox{z}|^3}\, 
    \bigl[ 1 + O(v^2) \bigr] 
+ O(mM^2).
\label{6.42}
\end{equation}
Here the third-derivative term in Eq.~(\ref{6.40}) is seen to give
rise to the $O(v^2)$ correction in the second member of
Eq.~(\ref{6.42}), and the ``$\cdots$'' terms are responsible for
$O(v^4)$ corrections in the first member.  

Gathering the results, we arrive at 
\begin{equation}
h_*^{00}(t,\bbox{x}) = \frac{2M}{|\bbox{x} - \bbox{Z}|} 
- \frac{mM}{|\bbox{x}||\bbox{z}|}\,
  \biggl( 2 + \frac{\bbox{x} \cdot \bbox{z}}{|\bbox{z}|^2} \biggr) 
- \frac{2}{3} m \bbox{x} \cdot \frac{d \bbox{g}}{dt} 
- 2 m M \frac{\bbox{z} \cdot \bbox{v}}{|\bbox{z}|^3} 
+ O(2\mbox{\sc pn},mM^2),  
\label{6.43}
\end{equation} 
in which $O(2\mbox{\sc pn})$ regroups all the $O(v^2)$ correction
terms displayed in Eqs.~(\ref{6.41}) and (\ref{6.42}). We note that on
the right-hand side of Eq.~(\ref{6.43}), the first member is of order
$\Phi$ and is therefore designated as ``Newtonian''. The second member 
is of order $m/|\bbox{z}|$ relative to $\Phi$, and it represents a
post-Newtonian (1{\sc pn}) correction \cite{PN}. Because they involve
an additional power of $\bbox{v}$, the last two members are 
1.5{\sc pn} corrections.   

The calculation of $h_*^{0a}$ involves similar steps, but the
computations are substantially simpler. We simply quote the 
result: 
\begin{equation}
h_*^{0a}(t,\bbox{x}) = \frac{4M V^a}{|\bbox{x}|} + 4 m g^a 
+ O(1.5\mbox{\sc pn},mM^2). 
\label{6.44}
\end{equation}
Relative to the leading, ``Newtonian'' term in $h_*^{00}$, the first
term on the right-hand side of Eq.~(\ref{6.44}) represents a 
0.5{\sc pn} correction, while the second term is a 1{\sc pn}
correction. Because $h_*^{0a}$ couples to the particle's velocity
$v^a$, these post-Newtonian labels should actually be promoted to  
1{\sc pn} and 1.5{\sc pn}, respectively; the error term is then 
2{\sc pn} and $h_*^{0a}$ has been computed to a sufficient degree of 
accuracy. 

The metric of the moving star is now completely determined. We have,
from Eq.~(\ref{6.37}), 
\begin{equation}
g^*_{00} = -\bigl(1 - h_*^{00} \bigr) + O(mM^2), \qquad
g^*_{0a} = - h_*^{0a} + O(mM^2), \qquad
g^*_{ab} = \delta_{ab} \bigl(1 + h_*^{00} \bigr) + O(mM^2), 
\label{6.45}
\end{equation}
with $h_*^{00}$ and $h_*^{0a}$ given by Eqs.~(\ref{6.43}) and
(\ref{6.44}), respectively. In those expressions, $\bbox{z}(t)$
represents the Newtonian motion of the particle around the fixed star, 
$\bbox{v}(t) = d\bbox{z}/dt$, and $d\bbox{v}/dt = \bbox{g} = -M
\bbox{z}/|\bbox{z}|^3$. On the other hand, $\bbox{Z}(t)$
represents the order-$m$ motion of the star in the gravitational field
of the particle, $\bbox{V}(t) = d\bbox{Z}/dt$, and $d\bbox{V}/dt$ 
is determined by Eq.~(\ref{6.32}). 

\subsection{Motion of the particle under the matter-mediated force} 

In this subsection we calculate the motion of the particle under the
sole action of the matter-mediated force; we neglect for now the
action of the self-force, which will be incorporated later. In this 
subsection, therefore, we shall write the equations of motion as 
$u^\alpha_{\ ;\beta} u^\beta = f^{\alpha}_{\rm mm}$. The force could
be calculated directly from Eq.~(\ref{6.15}) by substituting 
$\delta g_{\alpha\beta}$ obtained from Eq.~(\ref{6.45}). It is easier,
however, to proceed in the following way. First, we recognize that the
equations of motion are equivalent to the statement that the particle
moves on a geodesic of the metric $g^*_{\alpha\beta} = g_{\alpha\beta}  
+ \delta g_{\alpha\beta}$. Second, we generate the equations of motion
by constructing a suitable Lagrangian function
$L(\bbox{x},\bbox{\dot{x}})$ which we then substitute into the
Euler-Lagrange equations. The practical advantage of this method is
that it allows us to straightforwardly switch from a world line
parameterized by $\tau$, the particle's proper time on the background
spacetime, to one parameterized by  $t$, the time coordinate. In the
following we will denote the updated trajectory of the particle by
$\bbox{x}(t)$ and its updated velocity by
$\bbox{\dot{x}}(t) = d\bbox{x}/dt$. The Newtonian values will still be 
denoted $\bbox{z}(t)$ and $\bbox{v}(t)$. Because the matter-mediated
force provides a correction of order $m$ to the Newtonian motion, we
have $\bbox{x}(t) = \bbox{z}(t) + O(m)$ and $\bbox{\dot{x}}(t) =
\bbox{v}(t) + O(m)$. In the next subsection we will abolish this
distinction of notation and use $\bbox{z}$ and $\bbox{v}$ to refer to
the updated trajectory. 

We take the Lagrangian function to be 
\begin{equation}
L(\bbox{x},\bbox{\dot{x}}) = 1 - \sqrt{ -g^*_{\alpha\beta} 
  \dot{x}^\alpha \dot{x}^\beta}, 
\label{6.46}
\end{equation}
where $\dot{x}^\alpha = (1,\bbox{\dot{x}})$; as we have indicated, we
use $t$ as a parameter on the world line. We substitute the metric of
Eq.~(\ref{6.45}) into Eq.~(\ref{6.46}), and to simplify we expand the
square root to second post-Newtonian (2{\sc pn}) order; in this
procedure we also discard terms that are second order in $M$. The 
result is  
\begin{equation}
L = \frac{1}{2} \bbox{\dot{x}}^2 + \frac{1}{2} h_*^{00} 
+ \frac{1}{8} \bbox{\dot{x}}^4 - h_*^{0a} \dot{x}^a 
+ \frac{3}{4} h_*^{00} \bbox{\dot{x}}^2 + O(2{\sc pn},M^2). 
\label{6.47}
\end{equation} 
After substitution of Eqs.~(\ref{6.43}) and (\ref{6.44}), we notice
that the Lagrangian contains a term $-mM \bbox{z} \cdot
\bbox{v}/|\bbox{z}|^3$. This is a function of time only which does not
affect the equations of motion; we therefore remove it from
the Lagrangian. The Lagrangian also contains a term $-4 m
\bbox{\dot{x}} \cdot \bbox{g}$. This differs from $4 m \bbox{x} \cdot
d\bbox{g}/dt$ by a total derivative $-4 m d(\bbox{x} \cdot
\bbox{g})/dt$ that can also be deleted from the Lagrangian. The final 
result is an effective Lagrangian 
\begin{equation} 
L_{\rm eff} = \frac{1}{2} \bbox{\dot{x}}^2 
+ \frac{M}{|\bbox{x} - \bbox{Z}|}\, \biggl( 1 + \frac{3}{2}
\bbox{\dot{x}}^2 \biggr) + \frac{1}{8} \bbox{\dot{x}}^4 
- \frac{mM}{|\bbox{x}||\bbox{z}|}\, \biggl(1 
+ \frac{\bbox{x} \cdot \bbox{z}}{2|\bbox{z}|^2} \biggr) 
- \frac{4M}{|\bbox{x}|}\, \bbox{\dot{x}} \cdot \bbox{V} 
+ \frac{11}{3} m \bbox{x} \cdot \frac{d \bbox{g}}{dt} 
+ O(2{\sc pn},M^2)
\label{6.48}
\end{equation} 
that is ready to be substituted into the Euler-Lagrange equations. 

The first member of the Euler-Lagrange equation is 
\begin{equation}
\frac{d}{dt} \frac{\partial L_{\rm eff}}{\partial \dot{x}^a} =  
\Lambda^{ab} \frac{d \dot{x}^b}{dt}  
- \frac{3M}{|\bbox{x} - \bbox{Z}|}\, 
  (\bbox{x} \cdot \bbox{\dot{x}}) \dot{x}^a
+ \frac{3M}{|\bbox{x}|^3}\, 
  (\bbox{x} \cdot \bbox{V} + \bbox{\dot{x}} \cdot \bbox{Z} ) \dot{x}^a
- \frac{4mM}{|\bbox{x}||\bbox{z}|^3}\, z^a 
+ \frac{4M}{|\bbox{x}|^3}\,
  (\bbox{x} \cdot \bbox{\dot{x}}) V^a, 
\label{6.49}
\end{equation}
where $\Lambda^{ab} = (1 + \frac{1}{2} \bbox{\dot{x}}^2 + 3M/|\bbox{x}
- \bbox{Z}|) \delta^{ab} + \dot{x}^a \dot{x}^b$. To obtain this we
have used Eq.~(\ref{6.32}), appropriately truncated at Newtonian
order, to evaluate $dV^a/dt$. The second member is 
\begin{equation}
\frac{\partial L_{\rm eff}}{\partial x^a} = 
-M \frac{x^a - Z^a}{|\bbox{x} - \bbox{Z}|^3}\, 
   \biggl( 1 + \frac{3}{2} \bbox{\dot{x}}^2 \biggr)
+ \frac{mM}{|\bbox{x}|^3|\bbox{z}|}\, \biggl(1 
  + \frac{\bbox{x} \cdot \bbox{z}}{2|\bbox{z}|^2} \biggr) x^a  
- \frac{mM}{2|\bbox{x}||\bbox{z}|^3}\, z^a 
+ \frac{4M}{|\bbox{x}|^3}\, (\bbox{\dot{x}} \cdot \bbox{V}) x^a 
+ \frac{11}{3} m \frac{dg^a}{dt}. 
\label{6.50}
\end{equation}
The equations of motion are obtained by equating Eq.~(\ref{6.49}) to
Eq.~(\ref{6.50}) and solving for $d\dot{x}^a/dt$. This requires the
inversion of $\Lambda^{ab}$, which can be performed approximately to
first post-Newtonian order. To simplify our expressions we replace
$z^a$ by $x^a + O(m)$ in the terms proportional to $mM$, and we also
replace $\bbox{x}$ by $\bbox{x} - \bbox{Z} = \bbox{x} + O(m)$ in terms
that are already linear in $m$. (It is of course important that we
make these substitutions {\it after} varying the Lagrangian, and 
not before.) 

The final result is 
\begin{eqnarray}
\frac{d \dot{x}^a}{dt} &=& - \frac{M}{|\bbox{x} - \bbox{Z}|^3}\,  
\biggl[ \biggl( 1 + \bbox{\dot{x}}^2 - \frac{5m}{|\bbox{x}-\bbox{Z}|}
- 4 \bbox{\dot{x}} \cdot \bbox{V} \biggr) \bigl(x^a-Z^a \bigr)  
- (\bbox{x} - \bbox{Z}) \cdot (4 \bbox{\dot{x}} - 3 \bbox{V})\,
\dot{x}^a + 4 (\bbox{x} - \bbox{Z}) \cdot \bbox{\dot{x}}\, V^a  
\biggr] 
\nonumber \\ & & \mbox{}
+ \frac{11}{3}\, m \frac{d g^a}{dt} + O(2{\sc pn},M^2).  
\label{6.51}
\end{eqnarray}    
We recall that $\bbox{Z}(t) = O(m)$ represents the trajectory of the 
moving star, that $\bbox{V}(t) = d\bbox{Z}/dt$ is determined by
Eq.~(\ref{6.32}), and that $\bbox{g}$, given by Eq.~(\ref{6.33}), is
the Newtonian gravitational field of the star at rest. Equation
(\ref{6.51}) governs the motion of the particle under the sole action
of the matter-mediated force. We see that the last term, which
represents a 1.5{\sc pn} correction to the Newtonian equations of
motion, gives rise to radiation damping. This term, however, is
precisely canceled out by the gravitational self-force of
Eq.~(\ref{5.25}). We conclude that the motion of the particle, under
the combined action of the self-force and the matter-mediated force,
is conservative at this level of approximation.    

\subsection{Summary and comparison with post-Newtonian theory} 

This is the end of the line, and we better summarize. A particle of
mass $m$ moves in the gravitational field of a star of mass $M$. The
particle's trajectory is denoted $\bbox{z}(t)$ and its velocity is
$\bbox{v}(t) = d\bbox{z}/dt$. The star is also moving, on a trajectory 
$\bbox{Z}(t)$ with a velocity $\bbox{V}(t) = d\bbox{Z}/dt$. We let
$\bbox{\rho} = \bbox{z} - \bbox{Z}$ be the separation between the two
objects, and we use the notation $\rho = |\bbox{z} -
\bbox{Z}|$. We let $\bbox{g} = -M \bbox{z}/|\bbox{z}|^3$ be the
Newtonian gravitational field of the star at rest, $\bbox{g}^* = -M 
\bbox{\rho}/\rho^3$ is the gravitational field of the moving star, and 
$\delta \bbox{g} = \bbox{g}^* - \bbox{g}$ is the difference between
the two. The spatial components of Eq.~(\ref{6.1}) then give 
\begin{equation}
\frac{d \bbox{v}}{dt} - \bbox{g} = \frac{\bbox{f}_{\!\rm self}}{m} +  
\frac{\bbox{f}_{\!\rm mm}}{m}   
\label{6.52}
\end{equation}
for the particle's equations of motion. The gravitational self-force
was computed in Sec.~V C and is given by 
\begin{equation} 
\frac{\bbox{f}_{\!\rm self}}{m} = - \frac{11}{3} m \frac{d \bbox{g}}{dt} 
+ O(2{\sc pn},M^2). 
\label{6.53}
\end{equation}
The matter-mediated force was computed in the preceding subsection and 
is given by 
\begin{equation}
\frac{\bbox{f}_{\!\rm mm}}{m} = \delta \bbox{g} + \bbox{g}^* \biggl(  
\bbox{v}^2 - \frac{5m}{\rho} - 4 \bbox{v} \cdot \bbox{V} \biggr)  
+ \frac{M}{\rho^3} \Bigl[ \bbox{\rho} \cdot (4 \bbox{v} - 3\bbox{V}) 
\bbox{v} - 4 (\bbox{\rho} \cdot \bbox{v}) \bbox{V} \Bigr] 
+ \frac{11}{3} m \frac{d \bbox{g}}{dt} + O(2{\sc pn},M^2). 
\label{6.54}
\end{equation} 
The final form of the equations of motion is obtained by combining
Eqs.~(\ref{6.52})--(\ref{6.54}). The result is 
\begin{equation}
\frac{d \bbox{v}}{dt} = -\frac{M}{\rho^3}\, \biggl[  
\biggl( 1 + \bbox{v}^2 - \frac{5m}{\rho} - 4 \bbox{v} \cdot \bbox{V}
\biggr) \bbox{\rho} -  \bbox{\rho} \cdot (4 \bbox{v} - 3\bbox{V}) 
\bbox{v} + 4 (\bbox{\rho} \cdot \bbox{v}) \bbox{V} \biggr]
+ O(2{\sc pn},M^2), 
\label{6.55}
\end{equation}  
and this contains no trace of a radiation-reaction force. The motion
of the star, on the other hand, is governed by Eq.~(\ref{6.32}), which
we rewrite as 
\begin{equation} 
\frac{d \bbox{V}}{dt} = \frac{m}{\rho^3}\, \Biggl\{  
\Biggl[ 1 + 2 \bbox{v}^2 - \frac{3}{2} \biggl( 
\frac{\bbox{\rho} \cdot \bbox{v}}{\rho} \biggr)^2 \Biggr] \bbox{\rho}  
- 3 (\bbox{\rho} \cdot \bbox{v}) \bbox{v} \Biggr\} 
+ O(2{\sc pn},mM).  
\label{6.56}
\end{equation}
In going from Eq.~(\ref{6.32}) to Eq.~(\ref{6.56}) we have allowed for
the fact that $\bbox{\rho} \equiv \bbox{z} - \bbox{Z} = \bbox{z} 
+ O(m)$.    

The reader familiar with post-Newtonian theory (as presented, for
example, in Ref.~\cite{55}) will have undoubtedly noticed a similarity
between the calculations presented here and the standard
post-Newtonian treatment of a system of point masses. While
similarities are surely present, the differences are important. The
most noticeable difference is concerned with self-field effects, which
are sometimes swept under the rug in the post-Newtonian treatment ---
the mass of each body is simply renormalized every time an infinite
self-field appears in the equations. By contrast, the
computations presented in this section are completely free of
infinities --- those have been taken care of, once and for all, during
the construction of the self-force. 

Equations (\ref{6.55}) and (\ref{6.56}) should be compared with the
equations derived from a standard post-Newtonian treatment --- the
Einstein-Infeld-Hoffmann equations of motion \cite{1}. These
are \cite{55.5}   
\begin{equation} 
\frac{d \bbox{v}}{dt} = -\frac{M}{\rho^3} \Biggl\{ \Biggl[ 1 
- \frac{4M}{\rho} - \frac{5m}{\rho} + \bbox{v}^2 + 2 \bbox{V}^2 
- 4 \bbox{v} \cdot \bbox{V} 
- \frac{3}{2} \biggl( \frac{\bbox{\rho} \cdot \bbox{V}}{\rho}
  \biggr)^2 \Biggr] \bbox{\rho}  
- \bbox{\rho} \cdot (4 \bbox{v} - 3\bbox{V}) (\bbox{v} - \bbox{V})
\Biggr\} + O(2{\sc pn}) 
\label{6.57}
\end{equation}
and 
\begin{equation} 
\frac{d \bbox{V}}{dt} = \frac{m}{\rho^3} \Biggl\{ \Biggl[ 1 
- \frac{4m}{\rho} - \frac{5M}{\rho} + \bbox{V}^2 + 2 \bbox{v}^2 
- 4 \bbox{v} \cdot \bbox{V} 
- \frac{3}{2} \biggl( \frac{\bbox{\rho} \cdot \bbox{v}}{\rho}
  \biggr)^2 \Biggr] \bbox{\rho}  
+ \bbox{\rho} \cdot (4 \bbox{V} - 3\bbox{v}) (\bbox{v} - \bbox{V}) 
\Biggr\} + O(2{\sc pn}). 
\label{6.58}
\end{equation}
If we eliminate all terms of order $m^2$ and $M^2$ from
Eq.~(\ref{6.57}), we recover Eq.~(\ref{6.55}). If, on the other hand,
we remove all terms of order $m^2$ and $mM$ from Eq.~(\ref{6.58}), we
recover Eq.~(\ref{6.56}). Our results are therefore perfectly  
consistent with the standard post-Newtonian treatment. 
       
\section*{Acknowledgments} 

The work presented here was supported by the Natural Sciences and
Engineering Research Council of Canada. We are grateful to Leor
Barack, Eanna Flanagan, Valeri Frolov, Amos Ori, Misao Sasaki, Bob
Wald, Alan Wiseman, and the participants of the Third and Fourth Capra
Meetings on Radiation Reaction for discussions that helped shape the
presentation of this work.   

\appendix

\section{Hadamard form of the scalar Green's function}  

In this appendix we continue our discussion of the two-point function  
$A(x,x')$. While the results derived here are not required for the
computation of the self-forces, they are interesting in their own
right. Our main objective is to show that the scalar Green's function
computed in Sec.~III B can be cast in the Hadamard form \cite{27} 
\begin{equation} 
G(x,x') = \theta(x,x') \Bigl[ u(x,x') \delta(\sigma) + v(x,x')
\theta(-\sigma) \Bigr]. 
\label{A.1}
\end{equation}
Here, $\sigma(x,x')$ is Synge's world function \cite{56}, equal to
one-half the squared geodesic distance between $x$ and $x'$; $u(x,x')$
and $v(x,x')$ are two-point functions that are smooth at $\sigma = 0$,
and $\theta(x,x')$ is a time-ordering function, equal to 1 if $x$ is
in the causal future of $x'$, and zero otherwise. The calculations
presented in this Appendix rely on methods introduced by Thorne and
Kovacs \cite{57}.  

\subsection{$A(x,x')$ on the light cone} 

We go back to Eq.~(\ref{4.6}) and seek to evaluate $A(x,x')$ in the
situation where the points $x$ and $x'$ are linked by a null geodesic
of the flat-spacetime background. In this situation $\Delta t = R$,
or $s=e$. (Recall that $\Delta t = t-t'$ and $R = |\bbox{x} -
\bbox{x'}|$.) This means that $\eta^1$ and $\eta^2$ are both zero, and
that the vector of Eq.~(\ref{4.5}) reduces to $\bbox{\eta} =
\frac{1}{2} \cos\theta \bbox{R}$. (Recall that the third axis is
oriented along $\bbox{R} = \bbox{x} - \bbox{x'}$.) Integrating over
$\phi$, we find that Eq.~(\ref{4.6}) becomes $A(x,x') = \frac{1}{2} 
\int_{-1}^{1} \Phi( \bbox{x}_0 + \bbox{\eta} )\, d\cos\theta$. After
changing the variable of integration to $\lambda = 
\frac{1}{2}(\cos\theta + 1)$, this becomes   
\begin{equation}
A_{\rm lc}(x,x') = \int_0^1 \Phi\bigl( \bbox{\xi}(\lambda) \bigr)\,
d\lambda, 
\label{A.2} 
\end{equation}
where 
\begin{equation}
\xi(\lambda) = \bbox{x'} + \lambda (\bbox{x} - \bbox{x'} ) 
\label{A.3}
\end{equation}
is a vector that interpolates linearly between $\bbox{x'}$ (when
$\lambda = 0$) and $\bbox{x}$ (when $\lambda = 1$). In Eq.~(\ref{A.2})
we have indicated with the label ``lc'' that $A(x,x')$ is evaluated on
the light cone. We see that $A_{\rm lc}(x,x')$ is the average of the
Newtonian potential $\Phi$ over the straight line joining $\bbox{x}$
and $\bbox{x'}$.  

Up to now, our view of the two-point function $A_{\rm lc}(x,x')$ has
been that it is the restriction of $A(x,x')$ on the light cone: the   
points $x$ and $x'$ must be related by a null geodesic of the
background Minkowski spacetime. The right-hand side of 
Eq.~(\ref{A.2}), however, stays meaningful even for points that are
not so related, and we take this opportunity to extend the definition
of $A_{\rm lc}(x,x')$ to arbitrarily related points. This will be our
new point of view: The two-point function $A_{\rm lc}(x,x')$ shall be 
defined for arbitrary points $x$ and $x'$ by Eq.~(\ref{A.2}). In the
case where $x$ and $x'$ are null related, then $A_{\rm lc}(x,x') =
A(x,x')$.      

We now show that $A_{\rm lc}(x,x')$, as defined for arbitrary points
by Eq.~(\ref{A.2}), is intimately related to the world function
$\sigma(x,x')$, defined by \cite{57} 
\begin{equation}
\sigma(x,x') = \frac{1}{2} \int_0^1 g_{\alpha\beta}(\xi^\mu)  
\frac{d \xi^\alpha}{d\lambda} \frac{d \xi^\beta}{d\lambda}\, d\lambda, 
\label{A.4}
\end{equation}
where $\xi^\alpha(\lambda)$ describes the (unique) geodesic linking
the points $x$ and $x'$; $\lambda$ is an affine parameter
scaled in such a way that $\xi^\alpha(0) = x^{\alpha'}$ and 
$\xi^\alpha(1) = x^\alpha$. The world function is to be calculated
with the metric of Eq.~(\ref{2.3}). 

In Eq.~(\ref{A.4}), the geodesic $\gamma$ differs from a straight line 
by a quantity of order $\Phi$. Because Eq.~(\ref{A.4}) is an action 
principle for the geodesic equation, this error of the first order in
the specification of the curve produces an error of the {\it second} 
order in $\sigma$. To calculate $\sigma$ to first order in $\Phi$, it
is sufficient to take $\gamma$ to be a straight line. The appropriate
relations are then $\xi^\alpha(\lambda) = x^{\alpha'} 
+ \lambda (x^\alpha - x^{\alpha'})$. Substituting also
$g_{\alpha\beta} = \eta_{\alpha\beta} - 2\Phi(\bbox{\xi})
\chi_{\alpha\beta}$ in Eq.~(\ref{A.4}), we obtain
\begin{equation}
\sigma(x,x') = \sigma^{\rm flat}(x,x') 
- \bigl(\Delta t^2 + R^2\bigr)  
\int_0^1 \Phi\bigl( \bbox{\xi}(\lambda) \bigr)\, d\lambda, 
\label{A.5}
\end{equation}
where $\sigma^{\rm flat}(x,x') = \frac{1}{2}(-\Delta t^2 + R^2)$.      
 
Comparing Eqs.~(\ref{A.2}) and (\ref{A.5}), we arrive at 
\begin{equation}
\sigma(x,x') = \sigma^{\rm flat}(x,x') 
- \bigl(\Delta t^2 + R^2 \bigr) A_{\rm lc}(x,x'). 
\label{A.6}
\end{equation} 
We see that the two-point function $A_{\rm lc}(x,x')$ determines by 
how much the squared geodesic interval $\sigma$ differs from its 
flat-spacetime value.  

\subsection{van Vleck determinant} 

The scalarized van Vleck determinant \cite{36,58} plays a fundamental
role in the theory of Green's functions in curved spacetime. This is
defined by  
\begin{equation}
\Delta(x,x') = - \frac{\mbox{det}\bigl[-\partial_{\alpha\beta'}
\sigma \bigr]}{\sqrt{g(x) g(x')}},  
\label{A.7}
\end{equation}
where $\sigma(x,x')$ is the world function defined in Eq.~(\ref{A.4}),
and $g(x)$ is the metric determinant evaluated at $x$. The general
theory predicts that in Eq.~(\ref{A.1}), $u(x,x')$ is the square root
of the van Vleck determinant. We will verify this result with our
weak-curvature scalar Green's function. In order to do so we must
compute $\Delta(x,x')$. 

We rewrite Eq.~(\ref{A.6}) in the form $\sigma = \sigma^{\rm flat} -
\Omega$, where 
\begin{equation}
\Omega(x,x') = \chi_{\mu\nu} (x-x')^\mu (x-x')^\nu A_{\rm lc}(x,x'). 
\label{A.8}
\end{equation} 
A straightforward computation, using $\sqrt{-g(x)} = 1 
- 2\Phi(\bbox{x})$ and $\partial_{\alpha\beta'} \sigma^{\rm flat} 
= - \eta_{\alpha\beta}$, reveals that to first order in the Newtonian 
potential, the van Vleck determinant is given by 
\begin{equation}
\Delta = 1 + 2\Phi(\bbox{x}) + 2\Phi(\bbox{x'}) 
+ \eta^{\alpha\beta} \partial_{\alpha \beta'} \Omega.  
\label{A.9}
\end{equation} 
The last term is evaluated from Eq.~(\ref{A.8}), and we obtain 
\begin{equation} 
\eta^{\alpha\beta} \partial_{\alpha \beta'} \Omega 
= \bigl( \Delta t^2 + R^2 \bigr) \delta^{ab} \partial_{ab'} A_{\rm lc}
- 2 (x-x')^a (\partial_a - \partial_{a'}) A_{\rm lc} - 4 A_{\rm lc}, 
\label{A.10}
\end{equation}
where we have used the fact that $A_{\rm lc}(x,x')$, as defined by
Eqs.~(\ref{A.2}) and (\ref{A.3}), depends only on the spatial
variables $\bbox{x}$ and $\bbox{x'}$. Calculating the derivatives is
easy, and we get $\partial_a A_{\rm lc} = \int \lambda \Phi_{,a}\,
d\lambda$, $\partial_{a'} A_{\rm lc} = \int (1-\lambda) \Phi_{,a}\,
d\lambda$, and $\partial_{ab'} A_{\rm lc} = \int \lambda(1-\lambda)
\Phi_{,ab}\, d\lambda$. Inside the integrals, the derivatives
of $\Phi$ are taken with respect to $\bbox{\xi}$, and the
various factors involving $\lambda$ come from differentiating this
vector with respect to either $\bbox{x}$ or $\bbox{x'}$. Substituting
this into Eq.~(\ref{A.10}), we have 
\begin{equation} 
\eta^{\alpha\beta} \partial_{\alpha \beta'} \Omega 
= \bigl( \Delta t^2 + R^2 \bigr) \int_0^1 \lambda(1-\lambda) 
\nabla^2 \Phi\, d\lambda - 2(x-x')^a \int_0^1 (2\lambda - 1) 
\Phi_{,a}\, d\lambda - 4 \int_0^1 \Phi\, d\lambda. 
\label{A.11}
\end{equation} 
In the first integral we replace $\nabla^2 \Phi$ by $4\pi \rho$. In
the second integral we replace $\Phi_{,a} (x-x')^a$ by
$d\Phi/d\lambda$, which allows us to integrate by parts. After
simplification, we arrive at 
\begin{equation}
\eta^{\alpha\beta} \partial_{\alpha \beta'} \Omega 
= 4\pi \bigl( \Delta t^2 + R^2 \bigr) \int_0^1 \lambda(1-\lambda)  
\rho\, d\lambda - 2\Phi(\bbox{x}) - 2 \Phi(\bbox{x'}). 
\label{A.12}
\end{equation} 
Substituting this into Eq.~(\ref{A.9}), we obtain our final expression
for the van Vleck determinant: 
\begin{equation}
\Delta(x,x') = 1 + 4\pi \bigl( \Delta t^2 + R^2 \bigr) 
\int_0^1 \lambda(1-\lambda) \rho\bigl(\bbox{\xi}(\lambda)\bigr)\, 
d\lambda. 
\label{A.13}
\end{equation} 
We see that $\Delta(x,x')$ differs from 1 if and only if the
straight line connecting the points $\bbox{x}$ and $\bbox{x'}$ passes
through the matter distribution.  

\subsection{$A(x,x')$ near the light cone} 

We have previously evaluated the two-point function $A(x,x')$ in the 
case where $x$ and $x'$ are connected by a null geodesic of the
background Minkowski spacetime. We now improve on this result, by
allowing the parameter 
\begin{equation}
\varepsilon \equiv \sqrt{s^2 - e^2} = \frac{1}{2}\,
\sqrt{\Delta t^2 - R^2} 
\label{A.14}
\end{equation}
to be nonzero. We will, however, assume that $\varepsilon \ll 1$, and
evaluate Eq.~(\ref{4.6}) in this limit. This is the near-light-cone
approximation. 

To second order in $\varepsilon$, the vector $\bbox{\eta}$ of
Eq.~(\ref{4.5}) reads $\eta^1 = \varepsilon \sin\theta \cos\phi$,
$\eta^2 = \varepsilon \sin\theta \sin\phi$, and $\eta^3 = e[1 + 
\frac{1}{2}(\varepsilon/e)^2 + O(\varepsilon^4)]\cos\theta$. We
substitute this into $\Phi(\bbox{x_0} + \bbox{\eta})$, expand to
second order in $\varepsilon$, and then integrate over the
angles. After re-introducing the parameter $\lambda =
\frac{1}{2}(\cos\theta + 1)$, we obtain  
\begin{equation}
A = A_{\rm lc} + 4\pi \varepsilon^2 \int_0^1 \lambda(1-\lambda) \rho\, 
d\lambda + \frac{\varepsilon^2}{2e} \int_0^1 (2\lambda - 1) 
\Phi_{,a} \hat{n}^a\, d\lambda - \varepsilon^2 \int_0^1
\lambda(1-\lambda) \Phi_{,ab} \hat{n}^a \hat{n}^b\, d\lambda 
+ O(\varepsilon^4).    
\label{A.15}
\end{equation}
Here, $A_{\rm lc}$ is the restriction of $A(x,x')$ on the light cone,
the quantity given by Eq.~(\ref{A.2}), and $\bbox{\hat{n}}$ is a unit
vector pointing in the direction of $\bbox{x}-\bbox{x'}$. The
quantities inside the integrals are evaluated at the point
$\bbox{\xi}(\lambda)$ defined by Eq.~(\ref{A.3}), and the derivatives
of $\Phi$ are taken with respect to this vector. To get the first
integral we have used Poisson's equation, $\nabla^2 \Phi = 4\pi \rho$.      
 
In the second and third integrals of Eq.~(\ref{A.15}), the derivatives
of $\Phi$ in the direction of $\bbox{\hat{n}}$ can be expressed as
derivatives with respect to $\lambda$: $d\Phi/d\lambda = 
2e\Phi_{,a} \hat{n}^a$ and $d^2 \Phi/d\lambda^2 
= (2e)^2 \Phi_{,ab} \hat{n}^a \hat{n}^b$. After substitution into
Eq.~(\ref{A.15}) and an integration by parts on the last term, we find
that these integrals cancel out. Our final expression for the
two-point function is therefore 
\begin{equation}
A(x,x') = A_{\rm lc}(x,x') + 4\pi \varepsilon^2 \int_0^1 
\lambda(1-\lambda) \rho\bigl(\bbox{\xi}(\lambda)\bigr)\, d\lambda 
+ O(\varepsilon^4). 
\label{A.16}
\end{equation}
We recognize here the same integral over the mass density that appears
in our previous expression for the van Vleck determinant,
Eq.~(\ref{A.13}). This allows us to rewrite Eq.~(\ref{A.16}) as 
\begin{equation}
A(x,x') = A_{\rm lc}(x,x') + \frac{\Delta t^2 - R^2}{8 R^2}\, 
\bigl[ \Delta(x,x') - 1 \bigr] + O(\varepsilon^4). 
\label{A.17}
\end{equation} 
In this calculation we have used the fact that the factor $\Delta t^2
+ R^2$ appearing on the right-hand side of Eq.~(\ref{A.13}) is equal
to $2R^2[1 + O(\varepsilon^2)]$. 

Equations (\ref{A.16}) and (\ref{A.17}) indicate that unless the
straight line connecting $\bbox{x}$ to $\bbox{x'}$ intersects the
matter distribution, $A_{\rm lc}(x,x')$ makes an excellent
approximation to $A(x,x')$. 

\subsection{Scalar Green's function} 

We now have the necessary tools to cast the scalar Green's function
calculated in Sec.~III B into the Hadamard form \cite{27} displayed in 
Eq.~(\ref{A.1}). Recall from Eq.~(\ref{3.11}) and (\ref{3.14}) that
our expression for the Green's function was 
\begin{equation} 
G(x,x) = G^{\rm flat}(x,x') - 2 \partial_{t t'} A(x,x') 
- 2\xi B(x,x'), 
\label{A.18}
\end{equation} 
where $G^{\rm flat}(x,x')$ is the retarded Green's function of flat
spacetime, given by Eq.~(\ref{3.3}). This can be re-expressed as 
\begin{equation}
G^{\rm flat}(x,x') = \theta(x,x') \delta\bigl( \sigma^{\rm flat}
\bigr), 
\label{A.19}
\end{equation}
where $\theta(x,x')$ is the time ordering function introduced in
Eq.~(\ref{A.1}), and $\sigma^{\rm flat}(x,x') = \frac{1}{2}(-\Delta
t^2 + R^2)$. 

To calculate the Green's function we need to take into account the
fact that both $A(x,x')$ and $B(x,x')$ are zero if $x$ and $x'$ are
spacelike related (in the flat-spacetime background), and if $x'$ lies 
to the future of $x$. We express this as 
\begin{equation}
A(x,x') = \theta(x,x') \theta\bigl(-\sigma^{\rm flat}\bigr)
\hat{A}(x,x'), \qquad 
B(x,x') = \theta(x,x') \theta\bigl(-\sigma^{\rm flat}\bigr)
\hat{B}(x,x'), 
\label{A.20}
\end{equation}
in which the $\theta$-functions explicitly enforce the vanishing of  
the two-point functions except when $x'$ is in the causal past of
$x$. In what follows we will keep the time-ordering function implicit,
and re-insert it at the end of the calculation.  

Differentiation of Eq.~(\ref{A.20}) gives   
\begin{equation}
\partial_{tt'} A = \theta\bigl(-\sigma^{\rm flat}\bigr) \partial_{tt'}
\hat{A} - \delta\bigl( \sigma^{\rm flat}\bigr) (1 
+ 2\Delta t \partial_t) \hat{A} 
+ \delta'\bigl( \sigma^{\rm flat}\bigr) \Delta t^2 \hat{A}, 
\label{A.21}
\end{equation}
where the prime on the $\delta$-function indicates differentiation 
with respect to $\sigma^{\rm flat}$. To calculate the time derivative
of $\hat{A}$ in the second term, we use Eq.~(\ref{A.16}) which we copy
as $\hat{A} = A_{\rm lc} - \sigma^{\rm flat} C 
+ O(\sigma^{{\rm flat}2})$, where $C = 2\pi \int \lambda(1-\lambda)  
\rho\, d\lambda$. Dropping all terms proportional to
$\sigma^{\rm flat}$, we find that the coefficient of the
$\delta$-function becomes $-A_{\rm lc} - \chi C$, where $\chi \equiv
\Delta t^2 + R^2$. Working now on the third term, we find that the
coefficient of the differentiated $\delta$-function is $\frac{1}{2}
\chi A_{\rm lc} - (A_{\rm lc} + \frac{1}{2} \chi C) \sigma^{\rm flat}
+ O(\sigma^{{\rm flat}2})$. The term that is linear in 
$\sigma^{\rm flat}$ can be transferred to the coefficient of the
$\delta$-function with the help of the distributional identity $\sigma
\delta'(\sigma) = -\delta(\sigma)$. The end result is
\begin{equation} 
\partial_{tt'} A = \theta\bigl(-\sigma^{\rm flat}\bigr)
\partial_{tt'} \hat{A} 
- \frac{1}{2}\, \chi C \delta\bigl( \sigma^{\rm flat}\bigr)
+ \frac{1}{2}\, \chi A_{\rm lc} \delta'\bigl( \sigma^{\rm flat}\bigr).  
\label{A.22}
\end{equation} 

We substitute this into Eq.~(\ref{A.18}). With the help of
Eqs.~(\ref{A.19}) and (\ref{A.20}), we obtain 
\begin{equation}
G = (1 + \chi C) \delta\bigl( \sigma^{\rm flat}\bigr) 
- \chi A_{\rm lc} \delta'\bigl( \sigma^{\rm flat}\bigr) 
+ \theta\bigl(-\sigma^{\rm flat}\bigr)\Bigl[ -2\partial_{tt'} \hat{A}
- 2\xi \hat{B} \Bigr]. 
\label{A.23}
\end{equation}
We can re-express the first three terms in the form 
$(1+\chi C)\delta(\sigma^{\rm flat} - \chi A_{\rm lc})$. From
Eq.~(\ref{A.6}) we recognize the new argument of the $\delta$-function
as the world function $\sigma(x,x')$ of the weakly curved
spacetime. And we recognize the factor $1 + \chi C$ in front as the
square root of the van Vleck determinant, calculated in
Eq.~(\ref{A.13}). Finally, because the quantity within the square
brackets in Eq.~(\ref{A.23}) is already of the first order in the
Newtonian potential, we can safely replace $\sigma^{\rm flat}$ with 
$\sigma$ as the argument of the $\theta$-function. The only remaining
task is to re-insert the time-ordering function that has been left 
out of our expressions. 

Our conclusion is that the scalar Green's function of Eq.~(\ref{A.18})
can indeed be cast in the form of Eq.~(\ref{A.1}), with $u =
\Delta^{1/2}$. For the weakly curved spacetimes considered in 
this paper, the world function $\sigma(x,x')$ is worked out in
Eq.~(\ref{A.6}), the van Vleck determinant $\Delta(x,x')$ in
Eq.~(\ref{A.13}), and 
\begin{equation}
v(x,x') = -2\partial_{tt'} \hat{A}(x,x') - 2\xi \hat{B}(x,x') 
\label{A.24}
\end{equation}
is the tail part of the Green's function. Here, $\hat{A}(x,x')$ and 
$\hat{B}(x,x')$ are the two-point functions introduced in Sec.~III A.      

\section{Two-point functions for the point-mass potential}  

In this Appendix we evaluate the two-point functions $A(x,x')$ and
$B(x,x')$ for the special case $\Phi(\bbox{x}) = -M/|\bbox{x}|$. The
results derived here were first obtained by DeWitt and DeWitt
\cite{44}; we include this discussion here for completeness. 

We begin with Eq.~(\ref{4.6}), which we rewrite as 
\begin{equation}
A(x,x') = -\frac{M}{4\pi} \int \frac{1}{|\bbox{\eta} -
\bbox{\eta}_0|}\, d\Omega, 
\label{B.1}
\end{equation}
where $\bbox{\eta}(s,\theta,\phi)$ is the vector of Eq.~(\ref{4.5}) with
$s=\frac{1}{2} \Delta t \equiv \frac{1}{2}(t-t')$, and $\bbox{\eta}_0
\equiv - \bbox{x}_0 \equiv - \frac{1}{2} (\bbox{x} + \bbox{x'})$. The
point $\bbox{\eta}_0$, at which the central mass is located (in a
coordinate system centered at $\bbox{\eta} = 0$), will be represented
by the ellipsoidal coordinates $(s_0,\theta_0,\phi_0)$. We recall that  
$e=\frac{1}{2} R = \frac{1}{2} |\bbox{x} - \bbox{x'}|$ is the
ellipticity of the coordinate system. 

To evaluate the integral of Eq.~(\ref{B.1}) we invoke the addition
theorem in ellipsoidal coordinates \cite{59},   
\begin{equation}
\frac{1}{|\bbox{\eta} - \bbox{\eta}_0|} = \frac{4\pi}{e}\,
\sum_{l=0}^\infty \sum_{m=-l}^l  
(-1)^m \frac{(l-m)!}{(l+m)!} P_l^m(s_</e) Q_l^m(s_>/e) 
Y_{lm}^*(\theta_0,\phi_0) Y_{lm}(\theta,\phi), 
\label{B.2}
\end{equation}
where $s_<$ ($s_>$) is the lesser (greater) of $s$ and $s_0$, and
$P_l^m$ and $Q_l^m$ are associated Legendre polynomials
\cite{60}. Substituting this into Eq.~(\ref{B.1}), we find that the 
integration over the spherical harmonics $Y_{lm}(\theta,\phi)$ is zero
unless $l$ and $m$ are both zero, and we obtain $A = -(M/e) P_0(s_</e)
Q_0(s_>/e)$, or   
\begin{equation}
A = -\frac{M}{R} \ln \frac{ 2 s_> + R }{ 2s_> - R }, 
\label{B.3}
\end{equation}
using the known forms for the Legendre functions of zeroth order. 

The hard part of the calculation resides with the computation of
$s_0$. Recalling that $\bbox{x}$ and $\bbox{x'}$ differ by a vector
$\bbox{R}$ pointing in the $z$ direction, we write $\bbox{x} =
(x,y,z)$, $\bbox{x'} = (x,y,z-R)$, and we have $\bbox{\eta}_0 =
(-x,-y,-z + \frac{1}{2}R)$. This and Eq.~(\ref{4.5}) give us the
equations  
\begin{equation}
\sqrt{{s_0}^2 - e^2} \sin\theta_0 \cos\phi_0 = -x, \qquad
\sqrt{{s_0}^2 - e^2} \sin\theta_0 \sin\phi_0 = -y, \qquad
s_0 \cos\theta_0 = -z + e, 
\label{B.4}
\end{equation}
which must be solved for $s_0$. It is not hard to show that this
amounts to solving the quadratic
\begin{equation}
{s_0}^4 - \frac{1}{2} \bigl(r^2 + r^{\prime 2}\bigr) {s_0}^2 +
\frac{1}{16} \bigl( r^2 - r^{\prime 2} \bigr)^2 = 0, 
\label{B.5}
\end{equation}
where $r = |\bbox{x}|$ and $r' = |\bbox{x'}|$. This finally gives $s_0
= \frac{1}{2} (r + r')$.

Going back to Eq.~(\ref{B.3}), we see that we must distinguish between
the cases $\Delta t > r + r'$, for which $2 s_> = \Delta t$, and
$\Delta t < r + r'$, for which $2 s_> = r + r'$. This gives 
\begin{equation} 
A(x,x') = -\frac{M}{R}\, \theta(\Delta t - R) 
\Biggl[ \theta(r + r' - \Delta t) 
\ln \frac{r + r' + R}{r + r' - R} + \theta(\Delta t - r - r') 
\ln \frac{\Delta t + R}{\Delta t - R} \Biggr], 
\label{B.6}
\end{equation}
where, we recall, $r = |\bbox{x}|$, $r' = |\bbox{x'}|$, $R = |\bbox{x}
- \bbox{x'}|$, and $\Delta t = t - t'$. For completeness we have
re-inserted the step function $\theta(\Delta t - R)$ that was left
implicit in Eqs.~(\ref{4.6}) and (\ref{B.1}). We see that the
two-point function undergoes a change of behavior when $\Delta t = r 
+ r'$. This time delay corresponds to a signal propagating with the 
speed of light from $\bbox{x}$ to the center (at which the Newtonian
potential is singular) and then on to $\bbox{x'}$. For shorter delays,
$A(x,x')$ is time-independent; for longer delays, $A(x,x')$ depends
explicitly on $\Delta t$. This sudden change of behavior makes the
two-point function slightly suspicious: Although our calculations are
based on the assumption that the spacetime is weakly curved
everywhere, the change of behavior is dictated by a region of
spacetime --- the center --- in which the Newtonian potential is
decidedly not small.  

The computation of $B(x,x')$ is quite simple for the potential $\Phi =
-M/r$, for which the mass density is $\rho(\bbox{x}) = M
\delta(\bbox{x})$. Substituting this into Eq.~(\ref{4.2}) gives 
\begin{equation}
B(x,x') = \frac{M}{rr'}\, \delta(\Delta t - r - r'). 
\label{B.7}
\end{equation} 
Again we notice a suspicious dependence on the conditions at the
center. 

Despite the notes of caution, the results derived here for the special
case $\Phi = -M/r$ are in complete agreement with our general results
of Sec.~IV. First, Eq.~(\ref{B.7}) is identical to Eq.~(\ref{4.17}),
which was obtained as the leading term in a multipole expansion for 
$B(x,x')$. This indicates that in fact, our result for this two-point
function is quite insensitive to the conditions near the
center. Second, Eq.~(\ref{B.6}) implies that near coincidence ($R$ 
small and smaller than $\Delta t$), $A(x,x')$ can be approximated by  
\begin{equation}
A(x,x') = - \frac{2M}{r+r'}\, \Biggl[ 1 + \frac{1}{3} \biggl
( \frac{R}{r+r'} \biggr)^2 + O\bigl(R^4\bigr) \Biggr],  
\label{B.8}
\end{equation}
which could be recast in the form of Eq.~(\ref{4.10}). In particular,
differentiation of Eq.~(\ref{B.8}) with respect to $\bbox{x}$ and
$\bbox{x'}$ confirms Eq.~(\ref{4.12}). In this case we have
$\Phi_{,ab} = (M/r^3)(\delta_{ab} - 3r_{,a} r_{,b})$, with $r_{,a} = 
x^a/r$. Third, Eq.~(\ref{B.6}) implies that for long delays 
($\Delta t$ large and larger than $R$), we have the approximation  
\begin{equation} 
A(x,x') = - \frac{2M}{\Delta t}\, \Biggl[ 1 + \frac{1}{3} \biggl
( \frac{R}{\Delta t} \biggr)^2 + O\bigl( \Delta t^{-4} \bigr)
\Biggr],  
\label{B.9}
\end{equation}
which is evidently compatible with Eq.~(\ref{4.13}). This comparison
between our general results of Sec.~IV and those of this Appendix
shows that any aspect of the two-point functions that might be
sensitive to the strong-field portion of the spacetime near the center
will not be involved in the computation of the self-forces. This
statement provides a further validation of the work of DeWitt and
DeWitt \cite{44}, which was entirely based on the special case 
$\Phi = -M/r$.


\begin{references}
\bibitem{1} A. Einstein, L. Infeld, and B. Hoffmann, Ann. Math. 
{\bf 39}, 65 (1938). 
\bibitem{2} S. Chandrasekhar and F.P. Esposito, Astrophys. J. 
{\bf 160}, 153 (1970). 
\bibitem{3} T. Damour, C. R. Acad. Sci. Paris {\bf 294}, s\'erie II, 
1355 (1982). 
\bibitem{4} G. Sch\"afer, Ann. Phys. (NY) {\bf 161}, 81 (1985). 
\bibitem{5} L.P. Grishchuk and S.M. Kopejkin, in {\it Relativity in
Celestial Mechanics and Astrometry} (IAU Symposium 114, Leningrad,
1985), edited by J. Kovalevsky and V.A. Brumberg (Reidel, Dordrecht,
1986). 
\bibitem{6} P. Jaranowski and G. Sh\"afer, Ann. Phys. (Leipzig) 
{\bf 9}, 378 (2000). 
\bibitem{6.2} T. Damour, P. Jaranowski and G. Sh\"afer, Phys. Rev. D
{\bf 63}, 044021 (2001); gr-qc/0105038 (2001).  
\bibitem{7} L. Blanchet and G. Faye, Phys. Rev. D {\bf 63}, 062005
(2001). 
\bibitem{7.2} V. de Andrade, L. Blanchet, and G. Faye,
Class. Quantum. Grav. {\bf 18}, 753 (2001). 
\bibitem{8} M.E. Pati and C.M. Will, Phys. Rev. D {\bf 62}, 124015
(2000).  
\bibitem{9} T. Damour, in {\it 300 Years of Gravitation}, edited by
S.W. Hawking and W. Israel (Cambridge University Press, Cambridge,
1987). 
\bibitem{10} P. Havas, in {\it Einstein and the History of General
Relativity}, edited by D. Howard and J. Stachel (Birkh\"auser,
Boston, 1989). 
\bibitem{11} Consider flat-spacetime electrodynamics as an
example. Here, the retarded vector potential $A^\alpha_{\rm ret}$ can
be decomposed as $\frac{1}{2}(A^\alpha_{\rm ret}+A^\alpha_{\rm adv}) + 
\frac{1}{2}(A^\alpha_{\rm ret}-A^\alpha_{\rm adv})$, where
$A^\alpha_{\rm adv}$ is the advanced potential. It can be shown that
the first part is singular at the position of the particle, but that
it does not influence its motion. The second part, on the other hand,
is well behaved at the particle's location, and it gives rise to the
Abrahams-Lorentz-Dirac radiation-reaction force. The analogous
decomposition of the vector potential in curved spacetime is much 
more delicate.    
\bibitem{12} T. Tanaka, M. Shibata, M. Sasaki, H. Tagoshi, and
T. Nakamura, Prog. Theor. Phys. {\bf 90}, 65 (1993). 
\bibitem{13} C. Cutler, D. Kennefick, and E. Poisson, Phys. Rev. D 
{\bf 50}, 3816 (1994). 
\bibitem{14} M. Shibata, Prog. Theor. Phys. {\bf 99}, 595 (1993). 
\bibitem{15} F.D. Ryan, Phys. Rev. D {\bf 53}, 3064 (1996). 
\bibitem{16} D. Kennefick and A. Ori, Phys. Rev. D {\bf 53}, 4319
(1996). 
\bibitem{17} D. Kennefick, Phys. Rev. D {\bf 58}, 064012 (1998). 
\bibitem{18} S.A. Hughes, Phys. Rev. D {\bf 61}, 084004 (2000). 
\bibitem{19} A. Ori and K.S. Thorne, Phys. Rev. D {\bf 62}, 124022
(2000).  
\bibitem{20} B. Carter, Phys. Rev. {\bf 174}, 1559 (1968). 
\bibitem{21} L.S. Finn and K.S. Thorne, Phys. Rev. D {\bf 62}, 124021
(2000).  
\bibitem{22} The LISA mission is described at
{\tt http://lisa.jpl.nasa.gov/}. A collection of relevant articles can
be found in {\it Laser Interferometer Space Antenna}, Proceedings of
the Second International LISA Symposium, AIP Conference Proceedings
456 (American Institute of Physics, Woodbury, 1998). 
\bibitem{23} The report is titled {\it Astronomy and Astrophysics in
the New Millennium}. It can be found at
{\tt http://books.nap.edu/books/0309070317/html/}.  
\bibitem{24} Y. Mino, M. Sasaki, and T. Tanaka, Phys. Rev. D {\bf 55}, 
3457 (1997). 
\bibitem{25} T.C. Quinn and R.M. Wald, Phys. Rev. D {\bf 56}, 3381
(1997). 
\bibitem{26} D.W. Sciama, P.C. Waylen, and R.C. Gilman,
Phys. Rev. {\bf 187}, 1762 (1969). 
\bibitem{27} J. Hadamard, in {\it Lectures on Cauchy's Problem in
Linear Partial Differential Equations} (Yale University Press, New
Haven, 1923), shows that the Green's function can be decomposed
into a ``direct part'' that has support on the past light cone of the
field point $x$, and a ``tail part'' that has support inside the light
cone. The direct part of the Green's function is singular, but the
tail part is well behaved. This decomposition is meaningful only in a
normal convex neighborhood of $x$, which means that $x$ and $x'$ must
be linked by a unique geodesic. 
\bibitem{28} P.L. Chrzanowski, Phys. Rev. D {\bf 11}, 2042 (1975). 
\bibitem{29} A. Ori, Phys. Lett. {\bf A202}, 347 (1995). 
\bibitem{30} L.M. Burko, Phys. Rev. Lett. {\bf 84}, 4529 (2000);
Class. Quantum Grav. {\bf 17}, 227 (2000); Am. J. Phys. {\bf 68}, 456
(2000). 
\bibitem{31} L. Barack and A. Ori, Phys. Rev. D {\bf 61}, 061502
(2000); gr-qc/0107056 (2001). 
\bibitem{32} L. Barack, Phys. Rev. D {\bf 62}, 084027 (2000);
gr-qc/0105040 (2001).  
\bibitem{33} L. Barack and L.M. Burko, Phys. Rev. D {\bf 62}, 084040
(2000). 
\bibitem{34} L.M. Burko, Y.T. Liu, and Y. Soen, Phys. Rev. D {\bf 63},
024015 (2001). 
\bibitem{34.2} L.M. Burko and Y.T. Liu, Phys. Rev. D {\bf 64}, 024006
(2001). 
\bibitem{35} C.O. Lousto, Phys. Rev. Lett. {\bf 84}, 5251 (2000). 
\bibitem{35.2} H. Nakano and M. Sasaki, Prog. Theor. Phys. {\bf 105},
197 (2001). 
\bibitem{35.4} H. Nakano, Y. Mino, and M. Sasaki, gr-qc/0104012
(2001). 
\bibitem{36} B.S. DeWitt and R.W. Brehme, Ann. Phys. (NY) {\bf 9}, 220 
(1960). 
\bibitem{37} J.M. Hobbs, Ann. Phys. (NY) {\bf 47}, 141 (1968). 
\bibitem{38} See, for example, J.D. Jackson, {\it Classical
Electrodynamics} (Wiley, New York, 1975). 
\bibitem{39} See E. Poisson, gr-qc/9912045 (1999) for a pedagogical
introduction to the Abrahams-Lorentz-Dirac equation. 
\bibitem{40} The Abrahams-Lorentz-Dirac equation is usually expressed
in terms of $\dot{a}^\alpha$, the proper-time derivative of the
particle's acceleration. Within the approximation implied by the 
point-particle description, it is consistent to replace this by
$\dot{f}^\alpha_{\rm ext}/m$, which we do throughout the paper. This
reduction-of-order of the equations of motion is reviewed
in Ref.~[44]. The procedure is explained and motivated in 
E.E. Flanagan and R.M. Wald, Phys. Rev. D {\bf 54}, 6233 (1996). 
\bibitem{41} T.C. Quinn, Phys. Rev. D {\bf 62}, 064029 (2000). 
\bibitem{42} A.G. Wiseman, Phys. Rev. D {\bf 61}, 084014 (2000). 
\bibitem{44} B.S. DeWitt and C.M. DeWitt, Physics (Long Island City,
NY) {\bf 1}, 3 (1964). 
\bibitem{45} A.G. Smith and C.M. Will, Phys. Rev. D {\bf 22}, 1276
(1980). 
\bibitem{45.5} The authors have been very slow at recognizing this
simple point, and we gratefully acknowledge relevant conversations
with Alan Wiseman and Bob Wald. Our confusion had to do with the fact 
that it ought to be possible to adopt a coordinate system in which the
star always appears to be at rest (except for internal motions which
we ignore). While undoubtedly true, this is (probably) incompatible
with the gauge conditions imposed in our calculations. Indeed, the
calculations presented in Sec.~VI give rise to a description in which
the star and the particle both move around a fixed center of mass.     
\bibitem{46} P. Havas and J.N. Goldberg, Phys. Rev. {\bf 128}, 398
(1962). 
\bibitem{PN} A correction of $n${\sc pn} order is a term of order
$\Phi^n$ relative to the leading term, which is labelled
``Newtonian''. In the post-Newtonian expansion, it is normally assumed
that $v^2 = O(\Phi)$, where $v$ is the velocity of an object moving in
the gravitational potential $\Phi$. A quantity such as $m^2 
d\bbox{g}/dt$ is 1.5{\sc pn} order relative to $m \bbox{g}$: if $r_c$
is a characteristic length scale and $t_c$ a characteristic time scale
associated with the motion, then $m^2 d\bbox{g}/dt \sim
(m/r_c)(r_c/t_c) (m \bbox{g}) = O(\Phi v) m \bbox{g}$. 
\bibitem{51} M. Carmeli, Ann. Phys. (NY) {\bf 34}, 465 (1965); 
{\bf 35}, 250 (1965). 
\bibitem{55.5} See, for example, C.W. Misner, K.S. Thorne, and
J.A. Wheeler, {\it Gravitation} (Freeman, San Francisco, 1973),
Exercise 39.15.   
\bibitem{47} A.G. Wiseman (personal communication). Wiseman's
unpublished calculations were presented at the Second Capra Meeting on
Radiation Reaction, in Dublin, Ireland (July 1999). The proceedings of
this conference, which contain Wiseman's presentation, can be found at  
{\tt http://www.lsc-group.phys.uwm.edu/\~{}patrick/ireland99/}.   
\bibitem{48} R.M. Wald (personal communication). 
\bibitem{52} A.I. Zel'nikov and V.P. Frolov, Zh. Eksp. Teor. Fiz. 
{\bf 82}, 321 (1982) [Sov. Phys. JETP {\bf 55}(2), 191 (1982)]. 
\bibitem{53} V.P. Frolov (personal communication). 
\bibitem{54} C. Morette-DeWitt and J.L. Ging, C. R. Acad. Sci. Paris
{\bf 251}, 1868 (1960). 
\bibitem{55} L.D. Landau and E.M. Lifshitz, {\it The Classical Theory
of Fields} (Pergamond, Oxford, 1962), Section 105.  
\bibitem{56} J.L. Synge {\it Relativity: The General Theory}
(North-Holland, Amsterdam, 1960). 
\bibitem{57} K.S. Thorne and S.J. Kovacs, Astrophys. J. {\bf 200}, 245
(1975). 
\bibitem{58} J.H. van Vleck, Proc. Nat. Acad. Sci. (US) {\bf 14}, 178
(1928). 
\bibitem{59} T.M. MacRobert, {\it Spherical Harmonics} (Methuen,
London, 1927), Chap. XI. 
\bibitem{60} G. Arfken, {\it Mathematical Methods for Physicists}
(Academic Press, Orlando, 1985), Chap. 12.    
\end{references}
\end{document}